\documentclass[journal]{IEEEtran}

\usepackage{cite}
\usepackage{xcolor}
\newcommand{\editrev}[1]{{\color{black}#1}}
\newcommand{\editrevr}[1]{{\color{red}#1}}

\ifCLASSINFOpdf
   \usepackage[pdftex]{graphicx}
\else
\fi
\usepackage{amsmath, amssymb, amsfonts}
\usepackage{multirow}
\usepackage{hhline}
\usepackage{comment}

\DeclareMathOperator{\Tr}{tr}
\DeclareMathOperator{\rank}{rank}

\newtheorem{define}{Definition}
\newtheorem{theorem}{Theorem}
\newtheorem{lemma}{Lemma}

\newtheorem{cor}{Corollary}

\begin{document}
%
\title{How Many Simultaneous Beamformers are Needed for Integrated Sensing and Communications?}
%
%
%

\author{Kareem M. Attiah,~\IEEEmembership{Member,~IEEE,}
and~Wei~Yu,~\IEEEmembership{Fellow,~IEEE}
\thanks{Manuscript submitted to \emph{IEEE Transactions on Information Theory} on July 20, 2025, revised on February 3, 2026 and May 12, 2026, accepted on May 18, 2026. The authors are with the Electrical and Computer Engineering Department,
University of Toronto, Toronto, ON, Canada.  
Emails: kttiah@ece.utoronto.ca, weiyu@ece.utoronto.ca.  The materials of this paper have been presented in part at the Asilomar Conference on Signals, Systems, and Computers, Pacific Grove, CA, USA, Oct. 2024~\cite{attiahbounds2024}. 
This work was supported by Natural Sciences and Engineering Research Council (NSERC) of Canada via a Discovery Grant and the Canada Research Chairs program.
}
}

\maketitle



\begin{abstract}
Consider a downlink integrated sensing and communications (ISAC) system in
which a base station employs linear beamforming to communicate to $K$ users,
while simultaneously uses sensing beams to perform a sensing task of estimating
$L$ real parameters.  How many beamformers are needed to achieve the best
performance for both sensing and communications?  This paper establishes 
bounds on the minimum number of downlink beamformers needed for ISAC, in which 
sensing performance is measured in terms of the Cram\'{e}r-Rao bound for 
parameter estimation and
communications performance is measured in terms of the
signal-to-interference-and-noise (SINR) ratios. We show that such an ISAC system
requires at most $K + \sqrt{\frac{L(L+1)}{2}}$ beamformers if the remote users
have the ability to cancel the interference caused by the sensing beams. If
cancelling interference due to the sensing beams is not possible, the bound
becomes $\sqrt{K^2 +  \frac{L(L+1)}{2}}$. 
Interestingly, in this latter case, the bound on the number of beamformers needed for joint sensing and communication is less than the sum of the bounds for each task individually. 
These results can be extended to sensing tasks for which the performance is measured as a function of $d$ quadratic terms in the beamformers. In this case, the bound becomes $K + \sqrt{d}$ and $\sqrt{K^2 + d}$, respectively. 
Specifically, for estimating complex path losses and angles-of-arrival of $N_\text{tr}$ targets while communicating to $K$ users, the bound on the minimum number of beamformers scales linearly in $K$ and in $N_\text{tr}$, assuming interference from sensing can be cancelled. 
When interference cancellation is not possible, the following exact characterization for the case of $N_\text{tr} = 1$ can be obtained: when $K=0$ or $1$, two beamformers should be used; when $K \ge 2$, exactly $K$ beamformers should be used, i.e., communication beamformers alone are already sufficient. 

\end{abstract}

%
\IEEEpeerreviewmaketitle

\section{Introduction}
Consider a downlink multiple-input multiple-output (MIMO) system where a base
station (BS) wishes to perform some sensing task, e.g., estimating $L$ real 
parameters of the environment, while simultaneously communicating to $K$ 
single-antenna users by using downlink beamforming.  What is the minimum number
of beamformers needed to realize the best possible performance for both tasks? 


The answer to such a question would have been straightforward if communication is the only goal, in which case it is clear that $K$ beamformers are sufficient. 
But if the BS also wishes to perform additional sensing tasks, it may be desirable to use additional sensing beams to facilitate the sensing operation. Then, it is not immediately obvious how many beamformers should be used. 

The preceding question is important in the emerging paradigm of integrated sensing and communication (ISAC), where the BS designs a common waveform in the downlink to convey information to the remote users, while at the same time also probing the environment to enable radar-like functionalities \cite{liutcom202, liujsac2022}. In this context, such a common waveform is typically obtained by augmenting the conventional model for communication with extra beamforming vectors for sensing \cite{Liu2020joint,LiuFCRB2022}. For implementation complexity reasons, it is desirable to keep the number of extra beamformers to a minimum, while ensuring 
good communication and sensing performance, as opposed to a system which sets the number of extra beamformers to be the maximum possible, i.e., the number of transmit antennas. Such a system would be impractical to implement when the BS is equipped with massive MIMO. 

Obtaining an exact answer to this question is challenging.
This is because the communication symbols transmitted in the
downlink can already be used for sensing, as they are
known at the BS. Augmenting the transmitted signal with extra
sensing beams can, on one hand, further improve the sensing
performance by directing additional energy toward the sensing
target, yet on the other hand, also hurt the communications performance, 
depending on whether the communication users are
capable of cancelling the interference from the sensing beams. 
Thus, sensing and communication interact in complicated ways.

Moreover, the optimal number of beamformers 
can vary widely depending on the channel realizations. For example, the optimal number of beamformers when the communication channels align with the sensing directions can be fewer than when they do not align. 

The optimal number of beamformers can also differ substantially based on the metric adopted for sensing, which depends on the nature of the sensing operation. For instance, mean-squared error (MSE), mutual information \cite{yang}, and Cram\'{e}r-Rao bound (CRB) \cite{LiuFCRB2022, Li2008range, chanisit2024} are common metrics for parameter estimation; radar signal-to-noise ratio (SNR) and signal-to-clutter-plus-noise ratio (SCNR)~\cite{mateen2023, wen2023} are often adopted for target detection; furthermore, beam pattern-based metrics \cite{soticaprobing2007, Liu2020joint, HuaOptimal2023} are often employed for probing tasks (i.e., illuminating certain spatial directions around targets). 

This paper aims toward characterizing the minimum number of beamformers in the form of theoretical upper bounds that hold across all system configurations (e.g., the number of transmit and receive antennas) and across all channel realizations. 
For the task of estimating $L$ parameters while communicating to $K$ users, these bounds are concise expressions of $K$ and $L$. For more general sensing tasks, we consider a family of sensing metrics with quadratic dependence on the beamformers, and derive bounds that are functions of the number of quadratic terms. This encompasses many of the radar metrics in \cite{yang, LiuFCRB2022, Li2008range, chanisit2024, mateen2023, wen2023, soticaprobing2007, Liu2020joint, HuaOptimal2023}.

\subsection{Prior Work}
The importance of characterizing bounds on the minimum number of beamformers for ISAC can be motivated by the following. First, the minimum number of beamformers, when expressed as a function of $K$ users and $L$ parameters, offers valuable insights into the maximum number of permissible communication users and sensing parameters that can be simultaneously served/estimated for a fixed number of transmit antennas. Second, determining 
such a minimum number can oftentimes simplify the system implementation and beamforming optimization. This is because most of the existing numerical algorithms for designing the beamformers in the ISAC literature \cite{chanisit2024, HuaOptimal2023, Liu2020joint,LiuFCRB2022, attiahactive2023} are based on semidefinite relaxation (SDR) methods that overparameterize the solution space by inherently assuming that the number of additional sensing beamformers is set to be the number of transmit antennas. In most cases of interest, 
this is unnecessary; see \cite{chanisit2024, mateen2023, salman}. 
Instead, 
by characterizing the appropriate number of beamformers ahead of time, one can forgo SDR entirely and devise low-complexity algorithms that perform the optimization in the beamforming space \cite{attiah2024ULDL, wen2023, LiuF2022conf}.


%
%
%
%
Partial characterizations of the minimum number of beamformers are available in a handful of special cases. A classical result in~\cite{Li2008range} demonstrates that the minimum number of sensing beamformers is at most twice the number of targets for estimating the angles-of-arrival (AoAs) of the targets in MIMO radar systems without communication users. This bound is derived by using the classical CRB as the sensing performance metric, which unrealistically requires the AoAs to be known ahead of time. 
More recent works \cite{chanisit2024, mateen2023, salman, HuaOptimal2023, hua2025_twc } 
examine the ISAC framework. In \cite{chanisit2024}, it is shown that two beamformers are needed in the worst-case for the simple scenario of one target and one user when the Bayesian CRB (BCRB) is considered. Meanwhile,~\cite{salman, HuaOptimal2023, mateen2023} show that the sensing beamformers are altogether unnecessary for several special cases. Specifically, \cite{salman} establishes this fact for the scenario of one communication user with radar SNR adopted as a sensing metric, whereas \cite{mateen2023} extends this result for any number of users. The result of~\cite{HuaOptimal2023} is derived under the assumptions that the communication channels follow a line-of-sight (LoS) model and the radar utilizes a beam pattern metric. \editrev{Finally, the authors of~\cite{hua2025_twc} derive a simple bound on the minimum number of beamformers needed for near-field target positioning using classical CRB and signal power metrics, based on an extension of the proof technique in~\cite{Li2008range}.}

It should be remarked that the results of\editrev{~\cite{chanisit2024, mateen2023, salman, HuaOptimal2023} and~\cite{hua2025_twc}} are derived under different assumptions. Specifically,\editrev{~\cite{chanisit2024, mateen2023} and~\cite{hua2025_twc}} consider only the scenario where the communication users do not cancel the interference caused by superimposing the sensing beam on top of the communication signal. Meanwhile,~\cite{HuaOptimal2023} and~\cite{salman} examine both situations with or without radar interference cancellation. 

Finally, the very recent work~\cite{yao2025optimal} presents bounds on the number of beamformers for estimating the AoAs of multiple targets in the special case where the path loss coefficients have zero mean and the users do not cancel the interference. These bounds can already be derived from the earlier conference version of this work \cite{attiahbounds2024}. 
This journal paper develops tighter bounds 
as compared to \cite{yao2025optimal, attiahbounds2024}
for the non-interference-cancellation scenario.
In addition, this paper also derives results for the interference cancellation scenario, which is not treated in \cite{yao2025optimal}.

\begin{table*}[ht] 
\renewcommand{\arraystretch}{1.5} 
\centering
	\caption{Upper Bounds on Minimum Number of Beamformers (BFs) for Downlink ISAC Serving $K$ Communication Users \\
	(IC denotes interference cancellation; NIC denotes no interference cancellation)}
\label{tab:results_ummary}
\resizebox{\textwidth}{!}
{\begin{tabular}{c || c|c||c|  c  c} \hline 
\multicolumn{6}{c}{Parameter Estimation Using BCRB}  \\  \hline
 &
\multicolumn{2}{c||}{$L$ Parameters} & \multicolumn{3}{c}{
 ISAC Scenario with $K$ Users and $N_\text{tr}$ LoS Targets} 
\\ \hline
& IC & NIC &  IC  & \multicolumn{2}{c}{NIC}  \\
\hline

 \multirow{7}{5em}{This Work} & \multirow{7}{10em}{$K + \left \lfloor \sqrt{L (L + 1)/2} \right \rfloor$} & \multirow{7}{10em}{$\left \lfloor \sqrt{K^2 +  L(L + 1)/2} 
 \right \rfloor$} & \multirow{2}{12em}{$K + \left \lfloor \sqrt{\frac{7}{2}N^2_\text{tr} + \frac{1}{2} N_\text{tr}}\right \rfloor$} & \multicolumn{2}{c}{\multirow{2}{10em}{$\left \lfloor  \sqrt{ K^2 + \frac{7}{2}N^2_\text{tr} + \frac{1}{2} N_\text{tr}}\right \rfloor$}} \\
& & & & & \\
 \cline{4-6} 
 & & & \multicolumn{3}{c}{Special Case I: ISAC Scenario with Arbitrary $K$ and $N_\text{tr} = 1$} \\ \cline{4-6}

 &  &   & 
  \multirow{1}{3em}{$K + 2$}  & \multicolumn{2}{c}{\multirow{2}{18em}{$\left \lfloor \sqrt{K^2 +  4}   \right \rfloor = \begin{cases} 2, & K = 0, 1 \\ K, & K \geq 2 \end{cases}$   (Tight)}}
\\  
 & & &  $2$ BFs for $K = 0$ (Tight)
& \multicolumn{2}{c}{} 
\\ \cline{4-6}
& &$K$ BFs, for $K \geq \frac{L (L + 1)}{4} $ (Tight) & \multicolumn{3}{c}{Special Case II: Sensing-Only Scenario with $K = 0$ and Arbitrary $N_\text{tr}$} \\ \cline{4-6} 
& & & \multicolumn{3}{c}{$\left \lfloor \sqrt{\frac{7}{2}N^2_\text{tr} + \frac{1}{2} N_\text{tr}}\right \rfloor = \begin{cases} 2, & N_\text{tr} = 1 \text{~~(Tight)} \\ \approx 1.871N_\text{tr}, & N_\text{tr} \gg 1 \end{cases} $} \\ \hline
\multirow{2}{5em}{Prior Work}
&  \multirow{2}{3em}{--} &  \multirow{2}{3em}{--} &   \multicolumn{3}{l}{
$2$ BFs for ISAC Scenario with NIC, $K = 1$, and $N_\text{tr} = 1$; see \cite{chanisit2024}} 
\\
& & & \multicolumn{3}{l}{$2N_\text{tr}$ BFs for Sensing-Only Scenario with $K = 0$ and Arbitrary $N_\text{tr}$; see \cite{Li2008range}}  \\
\hline 
\hline 
\multicolumn{6}{c}{Sensing Metrics Involving $d$ Quadratic Terms of Beamformers} \\
\hline 
& \multicolumn{2}{c||}{$d$ Quadratic Terms} & \multicolumn{3}{c}{Radar SNR/SCNR with $d=1$}
\\ \hline
& IC & NIC & IC & \multicolumn{2}{c}{NIC}
\\ \hline
\multirow{2}{5em}{This Work}
  & \multirow{2}{5em}{$K + \left \lfloor \sqrt{d} \right \rfloor$}  & \multirow{1}{5em}{$\left \lfloor \sqrt{K^2 + d} \right \rfloor $} & \multirow{1}{3em}{$K + 1$}
& \multicolumn{2}{c}{\multirow{2}{17em}{$\left \lfloor \sqrt{K^2 + 1} \right \rfloor = \begin{cases} 1, & K = 0 \\ K, & K > 0 \end{cases}$ (Tight)}} 
 \\  
& & $K$ BFs, for $K \geq \tfrac{d}{2}$ (Tight) &  {$1$ BF for $K = 0$ (Tight)} & \multicolumn{2}{c}{} 
\\ \hline
Prior Work
& -- & -- & $1$ BF for $K = 1$; see \cite{salman} & \multicolumn{2}{c}{$K$ BFs for $K >0 $; see \cite{mateen2023}} 
 \\
\hline
\end{tabular}}
\end{table*}

\subsection{Main Contributions}
This paper considers both the scenarios where the users can or cannot cancel the interference from the sensing beams. The derived theoretical bounds hold generically across a large family of sensing metrics.
The main contributions of this paper are as follows:
\begin{enumerate}
    \item Consider an ISAC setting with $K$ communication users and a sensing objective of estimating $L$ parameters with the BCRB as the sensing metric. We begin by 
    assuming that the users have interference cancellation capabilities. It is proved that the minimum number of beamformers is bounded by a \emph{sum bound}---that is, at most $K$ communication beamformers plus $\left \lfloor \sqrt{L (L + 1)/2} \right \rfloor$ sensing beamformers are needed.
    

    \item The situation where the users cannot cancel the radar interference is considered next, and an alternative bound $\sqrt{K^2 + {L(L+1)/2} }$ is established. Intuitively, fewer beamformers should be used in this case, because if the user cannot cancel the radar interference, a well-designed system should rely more on the communication beamformers for sensing and only use extra sensing beams that do not penalize the communication performance. 
We refer to this bound as the \emph{hypotenuse bound}. 
    Interestingly, the hypotenuse bound suggests that the total number of beamformers needed for ISAC can be strictly less than the 
    combined number of beamformers needed for each task individually. 
    We further prove that when $K \ge  {\frac{L(L+1)}{4}}$, no extra sensing beamformers are needed at all, 
		i.e., using communication beamformers alone is already sufficient. 

    \item We proceed to define a general family of metrics that depend on $d$ quadratic terms involving the beamforming matrix, and subsequently derive a similar sum bound of $K + \sqrt{d}$ and a hypotenuse bound of $\sqrt{K^2+d}$ for this family. This not only extends the analysis to other commonly utilized metrics for radar (including SNR, SCNR, and beam pattern based metrics), but also refines the existing bounds under the BCRB metric.

    \item Examples are presented to show that the developed bounds either recover or improve upon the existing results of~\cite{salman, mateen2023, chanisit2024, Li2008range}, while maintaining generality. 
    \begin{itemize}
        \item When applied to the task of estimating the LoS parameters for $N_\text{tr}$ targets, we obtain general bounds as follows: $K + \left \lfloor \sqrt{\tfrac{7}{2}N_\text{tr}^2 + \tfrac{1}{2} N_\text{tr}} \right \rfloor$ when interference cancellation is possible, and $\left \lfloor \sqrt{K^2 + \tfrac{7}{2}N_\text{tr}^2 + \tfrac{1}{2} N_\text{tr}} \right \rfloor$ when interference cancellation is not possible. 

		\item As a special case of the above, we obtain novel bounds for the case of a single target (i.e., $N_\text{tr} = 1$), while communicating to $K$ users, as follows.  
For the interference cancellation scenarios, the sum bound becomes $K + 2$, i.e., a maximum of two extra sensing beamformers are required to estimate the LoS parameters, regardless of the number of users.
For the scenarios without interference cancellation, the hypotenuse bound reduces to $\left \lfloor \sqrt{K^2 + 4} \right \rfloor$, which is tight for all $K$. This provides an exact characterization of the worst-case minimum number of beamformers as follows: $2$ beamformers when $K = 0$ or $1$, and $K$ beamformers otherwise. This generalizes the result of~\cite{chanisit2024} obtained for $K = 1$.

        \item As a second special case, we consider the sensing-only scenario with $K = 0$ and arbitrary $N_\text{tr}$ targets and obtain the bound of $\left \lfloor \sqrt{\tfrac{7}{2}N_\text{tr}^2 + \tfrac{1}{2} N_\text{tr}} \right \rfloor$ from the general result. This improves upon the  $2N_\text{tr}$ bound of~\cite{Li2008range} for all values of $N_\text{tr} > 1$, and can be shown to be tight for $N_\text{tr} = 1$. 
        
        

       \item Next, we examine an example of a target detection task using radar SNR/SCNR, which corresponds to $d = 1$ in the sum and hypotenuse bounds for $d$-quadratic metrics. For the interference cancellation scenarios, the sum bound reduces to $K + 1$, which is tight when $K = 0$. For $K = 1$, it is looser by one beamformer as compared to the true minimum in~\cite{salman}. For $K > 1$, the sum bound of this paper is also looser by one beamformer, as can be proved based on \cite{huangrank2010}.
For scenarios without interference cancellation, the general hypotenuse bound reduces to $\left \lfloor \sqrt{K^2 + 1}\right \rfloor$. This provides an exact characterization of the worst-case minimum number of beamformers as one when $K = 0$, and $K$ otherwise, which recovers the result in~\cite{mateen2023}.
    \end{itemize}
\end{enumerate}

Table~\ref{tab:results_ummary} summarizes some of the main results in the paper. Cases denoted by ``(Tight)" indicate that this minimum number of beamformers are indeed needed for some channel realizations. Additional results are provided 
in Section~\ref{sec:examples}.
    
It should be noted that the bounding techniques of this paper are rather different from those of existing \editrev{works~\cite{Li2008range, chanisit2024, salman, mateen2023, HuaOptimal2023, hua2025_twc}},
which are based on the analysis of the primal and dual semidefinite program (SDP) for the beamforming problem, but often restricted to specific channel geomertries. 
In contrast, the results in this paper are based on rank reduction techniques for SDP and for general channel models. Specifically, the derivation of this paper involve fixing certain quadratic terms associated with performance metrics for communication and sensing. For the case of sum bound for the BCRB metric, 
this strategy enables us to establish a connection between the beamforming problem for ISAC and the problem of finding a minimum-rank solution for an SDP, and to leverage constructive rank-reduction algorithms in \cite{pataki1998rank, huangrank2010} to prove the bound. For the hypotenuse bound and the subsequent analysis involving more general sensing metrics, 
novel bounding methods are developed. These bounds do not exclusively rely on the techniques in \cite{pataki1998rank, huangrank2010}.   

\subsection{Paper Organization and Notation}
The rest of this paper is organized as follows. Section~\ref{sec:mdl} presents the ISAC system model and defines the notion of minimum number of beamformers. In Section~\ref{sec:N_min_BCRB}, we derive bounds on the minimum number of beamformers assuming that the BCRB is adopted for radar estimation. Section~\ref{sec:other_metrics} extends the analysis to radar metrics that have quadratic dependence on the beamforming matrix. Section~\ref{sec:examples} presents several examples that demonstrate the utility of the developed bounds. Finally, conclusions are drawn in Section~\ref{sec:conc}. 

This paper uses lowercase, lowercase boldface, and uppercase boldface letters to denote scalars, vectors, and matrices. We adopt the notation $(\cdot)^\textsf{T}$, $(\cdot)^\textsf{H}$, $(\cdot)^{-1}$, $\Tr\left(\cdot\right)$ to represent the transpose, Hermitian, inverse, and trace of a matrix, respectively. The operators $\Re\{\cdot\}$, $\Im\{\cdot\}$ are the real and imaginary parts of a complex number. We use $\mathbf{I}_N$ to denote the $N \times N$ identity matrix and its $\ell$-th column is denoted $\mathbf{e}_\ell$. The all-zero matrix and vector are denoted by $\mathbf{0}$. The space of $n \times m$ real (complex) matrices is denoted by $\mathbb{R}^{n \times m}$ ($\mathbb{C}^{n \times m}$). The cone of $L\times L$ positive semidefinite (PSD) matrices is denoted by $\mathbb{S}_+^{L \times L}$. For random quantities, $\mathbb{E}[\cdot]$ denotes the expectation and $\mathcal{CN}(\boldsymbol{\mu}, \boldsymbol{\Sigma})$ represents the cirularly-symmetric complex Gaussian distribution with mean $\boldsymbol{\mu}$ and covariance $\boldsymbol{\Sigma}$.

\section{Downlink Beamforming for ISAC} \label{sec:mdl}
\begin{figure}
    \centering    \includegraphics[width=0.44\textwidth]{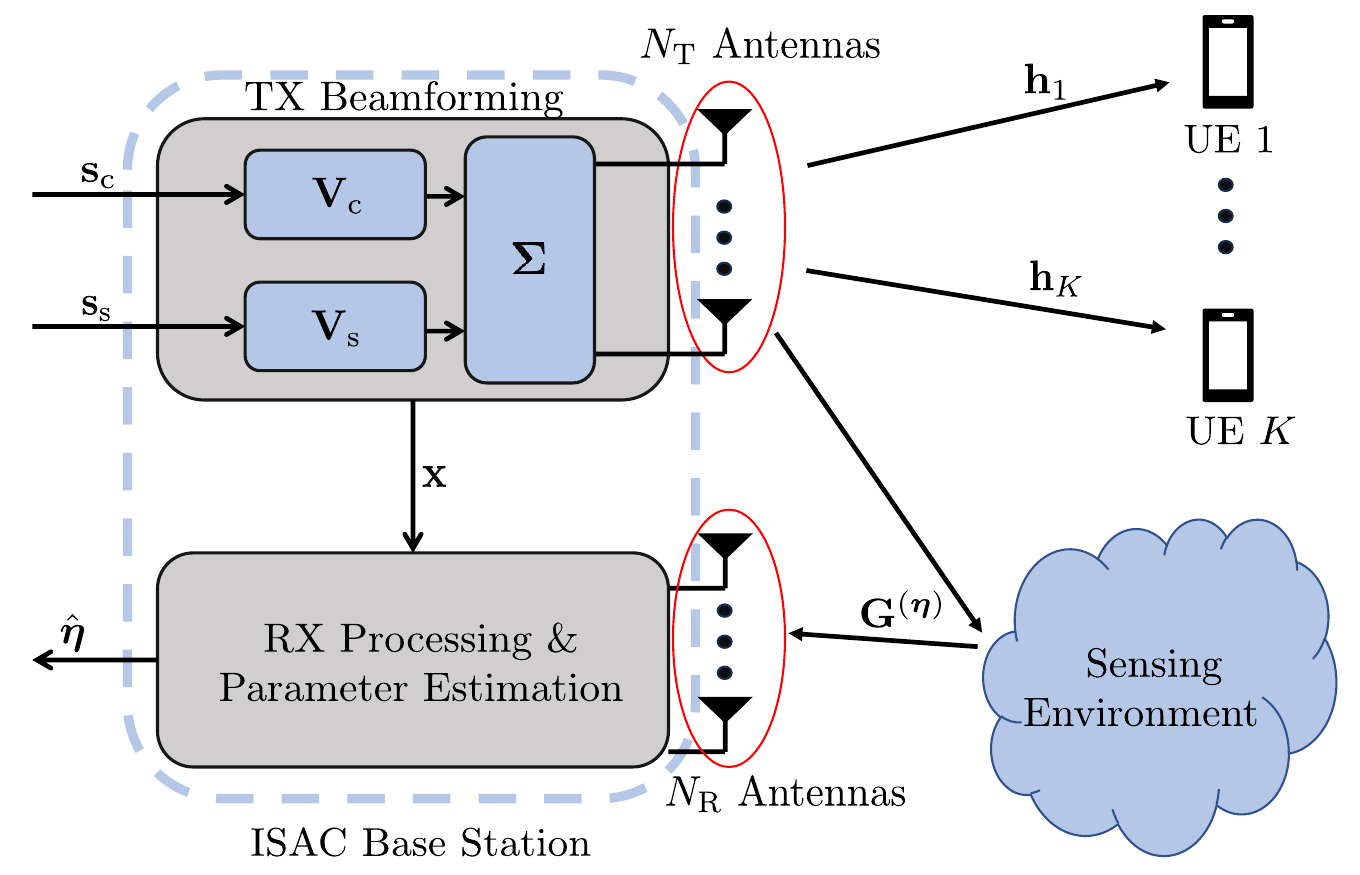}
     \caption{The ISAC downlink system where a BS serves $K$ communication users  and aims to estimate some underlying parameter of the sensing environment.}
     \label{fig:model}
\end{figure}
This paper considers a downlink ISAC system illustrated in~\figurename~\ref{fig:model} comprising a MIMO BS equipped with co-located transmit and receive antenna arrays of $N_\textup{T}$ and $N_\textup{R}$ elements, respectively. The BS performs two functions: (i) downlink communications to $K$ single-antenna users, and (ii) monostatic sensing for estimating a vector of $L$ unknown real-valued parameters $\boldsymbol{\eta} \in \mathbb{R}^{L}$ of some spatial characteristics of the environment (e.g., angle of arrival for targets) based on the received echo signal. 

We assume a block-fading model where both the communication and sensing channels remain fixed over a coherence interval spanning $\Upsilon$ transmission periods. In each transmission period, the BS transmits a joint waveform through its transmit array that conveys communication symbols to the intended users, while collecting the back-propagated echo signal through the receive array for sensing at the same time. 
Full-duplex operation is assumed here, where
the cross-array leakage 
is mitigated using appropriate self-interference cancellation techniques \cite{Sabhinband2014}. The communication and sensing may take place at different time scales. Typically, communications occur during each transmission symbol period, while parameter estimation takes place across multiple symbols within the coherence interval. 

\subsection{Signal Model} 
We consider narrowband transmission over a coherence interval of an ISAC system, where the continuous-time baseband transmit signal is given by
\begin{equation} \label{eq:cont_time_pulse}
    \mathbf{x}_\text{TX}(t) = \sum_{q = 1}^\Upsilon \mathbf{x}[q] \cdot g(t - q T) 
\end{equation}
where $g(t)$ is a unit-energy transmit pulse, e.g., a root-raised cosine waveform, shared across all the antenna elements, and $T = \tfrac{1}{B}$ is the duration of one transmission with $B$ denoting the allocated bandwidth.

The discrete-time sequence $\mathbf{x}[q] \in \mathbb{C}^{N_\text{T}}$ for $q = 1, \ldots, \Upsilon$ is used for both downlink communication to $K$ users and monostatic sensing of $L$ spatial parameters.  This work adopts a linear beamforming scheme where $\mathbf{x}[q]$ comprises dedicated components for both communication and sensing~\cite{LiuFCRB2022,
Liu2020joint}. Specifically, $\mathbf{x}[q]$  is given by a linear combination of $K$ communication beamformers and 
$N - K$ sensing beamformers:
\begin{equation}\label{eq:bf_model}
    \mathbf{x}[q]  \triangleq \left[\begin{matrix}\mathbf{V}_\text{c} & \mathbf{V}_\text{s} \end{matrix}\right] \left[\begin{matrix}\mathbf{s}_\text{c}[q] \\ \mathbf{s}_\text{s}[q] \end{matrix}\right], 
    \quad  q = 1, \ldots, \Upsilon,
\end{equation}
where $N \geq K$ is the total number of beamformers. Here, $\mathbf{V}_\text{c} \triangleq \left[ \mathbf{v}_1, \ldots, \mathbf{v}_K \right] \in \mathbb{C}^{N_\text{T} \times K}$ represents a set of communication beamformers, where $\mathbf{v}_k$ denotes the beamformer for the $k$-th user, and $\mathbf{V}_\text{s} \in \mathbb{C}^{N_\text{T} \times (N - K)}$ is a matrix of additional 
sensing beamformers. 
The vectors $\mathbf{s}_\text{c}[q] \in \mathbb{C}^{K}$ and $\mathbf{s}_\text{s}[q]\in \mathbb{C}^{N - K}$ denote the communication symbols intended for the users and the pseudo-random sequences of symbols for sensing purposes, respectively. Both are modelled as having independent and identically distributed (i.i.d.) $\mathcal{CN}(0, 1)$ entries. The use of random sequences for sensing tasks is a common assumption in the ISAC literature (e.g.,~\cite{LiuFCRB2022,
Liu2020joint}). Also, the information-bearing part of $\mathbf{x}[q]$ is already random. Randomized scheme such as this has previously been considered in the context of sensing in MIMO radar without communication; see, e.g.,~\cite{Eun2018}.

We remark here that the use of linear beamforming is without loss of generality. This is because as revealed later, for systems considered in this paper, the adopted performance metrics for communications and sensing both depend only on the covariance of $\mathbf{x}[q]$, which can be realized using an i.i.d.\ random sequence together with appropriate choices of beamformers.

It can be observed that the beamforming strategy~\eqref{eq:bf_model} naturally extends the conventional beamforming scheme for communication systems by augmenting the communication beamformers with additional beamformers for sensing. The use of additional beamformers is motivated by the fact that the communication beamformers alone may not provide sufficient degrees of freedom for serving dual purposes. This is because an ISAC system needs to steer the signal energy toward both the communication users and the directions of interest for sensing. 
The main goal of this paper is to address the question of how many beamformers are needed to efficiently carry out both sensing and communication tasks simultaneously.

Hereafter, a total power constraint is imposed on $\mathbf{x}[q]$ for $q = 1, \ldots, \Upsilon$ by enforcing 
\begin{equation}
\Tr{\left( \mathbf{V} \mathbf{V}^\mathsf{H}\right)} \leq P
\end{equation} 
on the overall beamforming matrix $\mathbf{V} \triangleq \left[\begin{matrix}\mathbf{V}_\text{c} & \mathbf{V}_\text{s} \end{matrix}\right]$. The subsequent sections detail the sensing and communication channel models.


\subsection{Sensing Model and Performance Metric}

\subsubsection{Channel Model}

We consider a model for sensing in which the round-trip channel response is represented by the sum of reflections from (a finite number of) scatterers. 
In general, the effect of the channel is described by a linear time-varying response characterized by a spatial response matrix as well as delay and Doppler shift parameters for each path, as follows: 
\begin{equation}\label{eq:time_varying_ch}
    \mathbf{G}(t) = \sum_m \mathbf{G}^{(\boldsymbol \eta)}_{\tau_m, \omega_m} \delta(t -\tau_m) \mathrm{e}^{- \jmath \omega_m t},
\end{equation}
where $\delta(\cdot)$ is the Dirac-delta function and $\mathbf{G}_{\tau, \omega}^{(\boldsymbol \eta)} \in \mathbb{C}^{N_\text{R} \times N_\text{T}}$ denotes the MIMO channel response at delay $\tau$ and Doppler frequency $\omega$. 
In this paper, we focus on estimating the spatial parameters $\boldsymbol{\eta}$, instead of estimating delay and Doppler parameters as in traditional radar applications. 

To this end, this paper makes the assumption that both the delay spread and the Doppler spread of the channel are small, (i.e., the narrowband channel model within the coherence time, as mentioned earlier). In this case, the multiple path reflections essentially overlap. Indeed, due to the assumption that sensing takes place within the coherence interval, we have $\omega_m \Upsilon T \ll 1$, for all $m$. Hence, the effect of Doppler shift can be ignored in~\eqref{eq:time_varying_ch}. Further, by the narrowband assumption, we have $\max |\tau_m - \tau_{m'}| \ll T$, and we may assume that $\tau_m \approx \tau_0$ for all $m$. Under these assumptions, the channel response can now be simplified to:
\begin{equation}
      \mathbf{G}(t) \approx \mathbf{G}^{(\boldsymbol \eta)} \delta(t-\tau_0),
\end{equation}
where $\mathbf{G}^{(\boldsymbol{\eta})} = \sum_m \mathbf{G}^{(\boldsymbol{\eta})}_{\tau_m, \omega_m} \in \mathbb{C}^{N_\text{R} \times N_\text{T}}$ models the effective channel matrix resulting from all reflections. 

The received signal at the BS due to the $q$-th transmission is described by the convolution of the channel response and the transmit signal and is given by 
    \begin{align}
        \mathbf{y}_{q}(t) 
        &=  \mathbf{G}^{(\boldsymbol \eta)} \mathbf{x}[q] \cdot g(t - q T - \tau_0) + \mathbf{z}_q(t), 
     \end{align}
where $\mathbf{z}_q(t)$ is a noise term. The effect of delay is mitigated by synchronizing the receiver using a simple matched filtering operation:
\begin{equation}
    \mathbf{y}_\text{s}[q] = \int \mathbf{y}_q(t) g^*(t - \tau)  dt,
\end{equation}
where $\tau$ is chosen to maximize the matched filter output. By setting $\tau \approx \tau_0$ and using the fact that the transmit pulse is of unit energy, we arrive at the following discrete-time model:
\begin{equation}
\mathbf{y}_\text{s}[q] = \mathbf{G}^{(\boldsymbol{\eta})} \mathbf{x}[q] + \mathbf{z}_\text{s}[q], \quad q = 1, \ldots, \Upsilon,         \label{eq:sense_mdl}
\end{equation}
where the noise term $\mathbf{z}_\text{s}[q] \in \mathbb{C}^{N_\text{R}}$ is Gaussian distributed as $\mathcal{CN}(\mathbf{0}, \sigma^2 \mathbf{I}_{N_\text{R}})$. 
This model can be alternatively derived by making the standard assumption $g(t - \tau) \approx e^{-\jmath 2 \pi f_\text{c} \tau} g(t)$ for narrowband signals, followed by sampling the received signal at $ t=q T$; see, e.g.,~\cite[Ch. 6]{stoica2005spectral}.
In the subsequent analysis, we focus on this discrete-time model~\eqref{eq:sense_mdl}. 

Furthermore, this paper makes a second crucial assumption that $\mathbf{G}^{(\boldsymbol{\eta})}$ is a deterministic function of $\boldsymbol{\eta}$, and that the functional relationship between $\mathbf{G}^{(\boldsymbol{\eta})}$ and $\boldsymbol{\eta}$ is known.
This is a reasonable assumption when the environment is relatively stationary and the only unknown parameters are that of the sensing targets.
To illustrate practical scenarios that can be handled by this setup, we present a few examples:

\begin{itemize}
    \item Consider the reflections from $N_\text{tr}$ targets in the far field, where each target has a single LoS path to the BS. The channel matrix in this case is given by  
    \begin{equation}\label{eq:multi_ath}
    \mathbf{G}^{(\boldsymbol{\eta})} = \sum_{i = 1}^{N_\text{tr}} \alpha_i \mathbf{a}_\text{R}(\theta_i) \mathbf{a}_\text{T}^\mathsf{H}(\theta_i),
    \end{equation}
    where \editrev{$\theta_i$  is azimuth AoA for the $i$-th target, and $\alpha_i$ is its associated complex scalar that includes the path loss and the radar cross section term. With a slight abuse in terminology, we refer to $\alpha_i$ as the path loss coefficient in the paper.} The vectors $\mathbf{a}_\text{T}(\cdot)$ and $\mathbf{a}_\text{R}(\cdot)$ denote the transmit and receive steering vectors when the BS employs $1$-D arrays (e.g., uniform linear arrays). In this paper, we consider both the general case where the BS seeks to estimate both the AoAs and the path loss coefficients, 
 where $\boldsymbol{\eta}$ comprises the path loss coefficients and AoAs for all targets, as well as the scenarios where the goal is to estimate the AoAs only. (See Section~\ref{sec:other_metrics} and Section~\ref{sec:examples}.)
    
\item 
A straightforward extension of the above example is when the targets are parameterized by both an azimuth and an elevation angle. Suppose that the BS is equipped with $2$-D antenna arrays (e.g., uniform planar arrays). The channel matrix is given by
 \begin{equation}
\mathbf{G}^{(\boldsymbol{\eta})} = \sum_{i = 1}^{N_\text{tr}} \alpha_i \mathbf{A}(\theta_i, \phi_i),
    \end{equation}
where $\mathbf{A}(\cdot, \cdot)$ now models the combined array response in both azimuth and elevation angels. The parameter $\boldsymbol{\eta}$ now comprises the path loss coefficients, azimuth angles, and elevation angles for all targets. 

\item When the targets are in the near field, the steering vectors depend on the positions of the targets; see~\cite{Liu2023nearfield} for the appropriate modeling of the steering vectors in this case. The vector $\boldsymbol{\eta}$ comprises the positions of all targets.
\end{itemize}

In the examples above, it is implicitly assumed that the number of targets is known prior to estimation. In practice, such knowledge can be acquired at an earlier beam sweeping stage. Furthermore, the channel model in the above examples ignore echoes from unwanted scatterers in the environment. This is a reasonable assumption when the unwanted reflections are either significantly weaker than the LoS reflections (e.g., the targets are drones in the sky), or can be mitigated in a pre-processing step at the radar receiver (e.g., reflections from large static objects in the environment can be subtracted; see~\cite{li2008mimobook, Chiriyath2015} and the references therein). 

Finally, we also consider the following most general model:

\begin{itemize}
\item Consider the problem of estimating the entire channel response matrix (i.e., $\mathbf{G}^{(\boldsymbol{\eta})}$). In this case, the parameter $\boldsymbol{\eta}$ comprises the individual entires of $\mathbf{G}^{(\boldsymbol{\eta})}$. Such an estimation task \editrev{has been considered in prior work, e.g.,~\cite{Hua_twc_2024} and} is of interest when the estimated $\mathbf{G}^{(\boldsymbol{\eta})}$ can subsequently be used to extract further information about the targets.
\end{itemize}
 

\subsubsection{Performance Metric}

We next discuss the sensing performance metric for the discrete-time channel model \eqref{eq:sense_mdl}. 
This paper adopts a Bayesian framework by assuming a prior distribution $f(\boldsymbol{\eta})$, and uses the BCRB to quantify the sensing performance. The CRB is a suitable metric for evaluating the sensing performance, because it provides a simple lower bound on the mean-squared error (MSE) for any unbiased estimator of the parameters. CRB becomes tight at high SNR or when the number of measurements is large. 
The BCRB is adopted here as it does not depend on the parameters to be estimated, unlike the classical (i.e., the non-Bayesian) CRB. 

Let $\mathbf{X} \triangleq \left[\mathbf{x}[1], \ldots, \mathbf{x}[\Upsilon]\right]$ and $\mathbf{Y}_\text{s} \triangleq [\mathbf{y}_\text{s}[1], \ldots, \mathbf{y}_\text{s}[\Upsilon]]$ denote the signals transmitted and received within a coherence interval.  For an estimator $\boldsymbol{\hat{\eta}} = \boldsymbol{\hat{\eta}}\left(\mathbf{X}, \mathbf{Y}\right)$ that satisfies certain regularity conditions \cite{van2004detection}, the MSE matrix for estimating $\boldsymbol{\eta}$, averaged over the prior $f(\boldsymbol{\eta})$, is bounded from below as
\begin{equation} \label{eq:BCRB_def}
\mathbb{E}_{\boldsymbol \eta}
\left[ 
\mathbb{E}\left[ \left.
\left( \boldsymbol{\eta} - \boldsymbol{\hat{\eta}}\right) \left( \boldsymbol{\eta} - \boldsymbol{\hat{\eta}} \right)^T 
\right| \boldsymbol{\eta}
\right] 
\right] 
\succcurlyeq \mathbf{J}_\mathbf{V}^{-1},
\end{equation}
where $\mathbf{J}_{\mathbf{V}}$ is the $L \times L$ Bayesian Fisher Information matrix (BFIM) and $\succcurlyeq$ denotes inequality with respect to the PSD matrix cone. The inner expectation in~\eqref{eq:BCRB_def} 
is taken jointly over $\mathbf{X}$ and $\mathbf{Y}_\text{s}$ since both are random. 
The $(i, j)$-th element of the BFIM is defined by
   \begin{align}\label{eq:BFIM}    \left[\mathbf{J}_{\mathbf{V}} \right]_{ij} &\triangleq -\mathbb{E} \left[ \frac{\partial^2 \log{ f\left(\mathbf{X}, \mathbf{Y}_\text{s}, \boldsymbol{\eta} \right) } }{\partial \eta_i \partial \eta_j} \right]. 
    \end{align}
In Appendix~\ref{Appen:BFIM}, we show that for the model~\eqref{eq:sense_mdl} and assuming that $\mathbf{G}^{(\boldsymbol{\eta})}$ is deterministic in $\boldsymbol{\eta}$, the BFIM can be expressed as 
\begin{equation} \label{eq:FIM_elements}
    \mathbf{J}_{\mathbf{V}} =   \mathbf{C} + \mathbf{T}_{\mathbf{V}},
    \end{equation}
where $\mathbf{C} \in \mathbb{S}_+^{L \times L}$ is a PSD matrix that depends on the prior distribution only, with elements given by
\begin{equation}~\label{eq:c_elements}
    [\mathbf{C}]_{ij} \triangleq - \mathbb{E} \left[\frac{\partial^2 \log f(\boldsymbol{\eta}) }{\partial \eta_i \partial \eta_j}\right],
\end{equation}
 and $\mathbf{T}_{\mathbf{V}} \in \mathbb{S}_+^{L \times L}$ is a PSD matrix that captures the dependence on the beamformers $\mathbf{V}$  
\begin{equation}~\label{eq:TV_elements}
	[\mathbf{T}_{\mathbf{V}}]_{ij} \triangleq \frac{\Upsilon}{\sigma^2}  \Tr\left( \tilde{\mathbf{G}}_{ij} \mathbf{V} \mathbf{V}^\textsf{H} \right),
\end{equation}
with $\tilde{\mathbf{G}}_{ij} \triangleq \mathbb{E} \left[ \dot{\mathbf{G}}_i^\mathsf{H}\dot{\mathbf{G}}_j +  \dot{\mathbf{G}}_j^\mathsf{H}\dot{\mathbf{G}}_i \right]$, $\dot{\mathbf{G}}_i \triangleq \tfrac{\partial \mathbf{G}^{\left(\boldsymbol\eta \right)} }{\partial \eta_i}$, and the expectation is taken with respect to $\boldsymbol{\eta}$. Note that the BCRB depends on \emph{both} the communication and sensing beamformers, as the reflected signals from both are used for estimating $\boldsymbol{\eta}$.
Further, the BFIM $\mathbf{J}_{\mathbf{V}}$ is a \emph{linear} function of the overall transmit covariance $\mathbf{V} \mathbf{V}^\textsf{H}$.


The BCRB presented above corresponds to the conventional bound due to Van Trees~\cite{van2004detection}, which arises by treating $(\mathbf{X}, \mathbf{Y}_\text{s})$ as a random  observation vector revealed to the estimator in each coherence interval (in a monostatic operation). 
A slightly improved bound can be obtained by deriving the inverse of the BFIM for a fixed $\mathbf{X}$ and applying an outer expectation over all possible realizations of $\mathbf{X}$; see \cite{Xiong2023}. This approach yields the so-called Miller-Chang BCRB~\cite{miller1978}.  
However, the latter bound lacks a simplified expression. 
Fortunately, it can be shown that the two bounds become equivalent for sufficiently large $\Upsilon$, which is the typical regime for which the BCRB bounds become tight for characterizing the MSE.

\subsection{Communication Model and Performance Metric}

\subsubsection{Channel Model}
The discrete-time model for communication can be derived in a manner analogous to that of~\eqref{eq:sense_mdl}. The baseband signal received at $k$-th user for $q = 1, \ldots, \Upsilon$ is given by
\begin{subequations}\label{eq:comm_mdl}
\begin{align}
    {y}_k[q] &= \mathbf{h}_k^\mathsf{H} \mathbf{x}[q] + {z}_k[q],   \\
    &= \mathbf{h}_k^\mathsf{H} \mathbf{V}_\text{c} \mathbf{s}_\text{c}[q] +  \mathbf{h}_k^\mathsf{H} \mathbf{V}_\text{s} \mathbf{s}_\text{s}[q] + z_k[q],
\end{align}
\end{subequations}
where $\mathbf{h}_k \in \mathbb{C}^{N_\text{T}}$ denotes the channel vector for the $k$-th communication user and ${z}_k[q] \in \mathbb{C}$ is the i.i.d.\ noise distributed as $\mathcal{CN}(0, \sigma^2)$. 
Here, the BS is assumed to know the communication channels $\mathbf{H} \triangleq \left[\mathbf{h}_1, \ldots, \mathbf{h}_K \right]$ perfectly within each coherence interval. Such channel knowledge can be obtained, e.g., via channel estimation using uplink pilots. 

Throughout the paper, we assume that both the communication and sensing channels are generated randomly from some non-degenerate distributions, so that it is always possible to spatially separate the sensing target and the multiple communication users. 

\subsubsection{Performance Metric}

We measure the communication performance in terms of the signal-to-interference-plus-noise ratio (SINR) at each communication receiver, which directly relates to the achievable rate of the user. Two distinct scenarios are considered in the sequel.

First, assume that the communication users are not able to cancel the interference due to the sensing component (i.e., the term $\mathbf{V}_\text{s} \mathbf{s}_\text{s}[q]$ in~\eqref{eq:bf_model}). In this case, the SINR expression for the $k$-th user is
\begin{equation}\label{eq:sinr_k_NIC}    \text{SINR}_{k, \mathbf{V}}^\text{NIC} \triangleq \frac{|\mathbf{h}_k^\mathsf{H} \mathbf{v}_k|^2 }{\sum_{i \neq k} |\mathbf{h}_k^\mathsf{H} \mathbf{v}_i|^2 + \mathbf{h}_k^\mathsf{H}\mathbf{V}_\text{s}\mathbf{V}_\text{s}^\mathsf{H}\mathbf{h}_k + \sigma^2}, 
\end{equation}
where the sensing signal is treated as additional interference by the communication users. 

In the second scenario, we assume that the sensing signal is known to the communication users (e.g., by sharing knowledge of $\mathbf{V}_\text{s}$ and $\mathbf{s}_\text{s}$) 
and the interference due to the sensing signal can be subtracted. In this case, the resulting SINR expression for the $k$-th user becomes
\begin{equation}\label{eq:sinr_k_IC}    \text{SINR}_{k, \mathbf{V}}^\text{IC} \triangleq \frac{|\mathbf{h}_k^\mathsf{H} \mathbf{v}_k|^2 }{\sum_{i \neq k} |\mathbf{h}_k^\mathsf{H} \mathbf{v}_i|^2 + \sigma^2}. 
\end{equation}
Both SINR formulations are relevant from a practical viewpoint. 
The former is more commonly adopted in the ISAC literature due to its simpler implementation, whereas the latter offers potential performance gains \cite{HuaOptimal2023, salman}. 

\subsection{Minimum Number of Beamformers}

Given that the BCRB and SINR expressions are both functions of the beamformers, it is natural to ask how many beamformers are needed for the simultaneous operation of sensing and communication. Since there are a total of $N_\text{T}$ antennas, it is easy to see that the number of additional sensing beamformers can range between $0$ and $N_\text{T}$. Hence, the total number of beamforming vectors must satisfy
$K \leq N \leq N_\text{T} + K$.

As already mentioned, setting $ N  = K$ can incur a performance loss because it may limit the ability of the BS to direct energy towards the sensing directions, while satisfying the communication SINR constraints. A common practice in ISAC literature is to instead set $N = N_\text{T} + K$; see, e.g.,~\cite{Liu2020joint, LiuFCRB2022}. However, this approach is highly inefficient since the majority of these extra sensing beamformers 
are likely to be linearly dependent or be zero when optimized. 
Indeed, several recent works \cite{chanisit2024, mateen2023, salman} show that one can often realize the same performance using much fewer beamforming vectors.

The main goal of this paper is to characterize the minimum number of beamformers that realize the same performance as the case of $N = N_\text{T} + K$. Formally, we define the minimum number of beamformers as
\begin{equation} \label{prob:N_min}
    N_\text{min} \triangleq  \arg\min ~N \quad
                \mathrm{subject \; to}~~p^*_N = p^*_{N_\text{T} + K},
\end{equation}
where $p^*_N$ is the optimal value of an ISAC beamforming problem using $N$ beamforming vectors. For instance, in scenarios where we wish to optimize the sensing performance subject to target SINRs on the communication performance, assuming that the users can cancel the interference from sensing beams, 
the beamforming optimization problem may take the following form (parameterized by $N$ the total number of beamformers at the transmitter):
\begin{subequations}\label{prob:main_problem}
\begin{align}
    \underset{\mathbf{V} \in \mathbb{C}^{N_\text{T} \times N}}{\mathrm{minimize}} ~~~&h\left({\mathbf{J}_\mathbf{V}^{-1}} \right) \label{eq:BCRB_metric} \\
    \mathrm{subject \; to}~~~ &\text{SINR}_{k, \mathbf{V}}^\text{IC} \geq \gamma_k, \quad \forall k \label{eq:SINR_const_IC} \\
    &\Tr(\mathbf{V}\mathbf{V}^\mathsf{H}) \leq P. \label{eq:power_const}
\end{align}
\end{subequations}
Here, $h(\cdot): \mathbb{S}_+^{L \times L} \rightarrow \mathbb{R}$ is a nondecreasing scalar function with respect to the cone of PSD matrices, i.e., $h(\mathbf{J}_1) \geq h(\mathbf{J}_2)$ for $\mathbf{J}_1 \succcurlyeq \mathbf{J}_2$, for quantifying sensing performance using a scalar. Scalar functions that satisfy this requirement include trace, weighted-trace, or logarithm-determinant, etc \cite{Li2008range}.

We use $N_\text{min}^\text{IC}$ to denote the minimum number 
of beamformers for the scenario in which the users can cancel
the interference from the sensing beams.
An analogous definition of the minimum number of beamformers for the scenario where the users do not cancel interference, i.e., $N_\text{min}^\text{NIC}$, is obtained by replacing $\text{SINR}_{k, \mathbf{V}}^\text{IC}$ with $\text{SINR}_{k, \mathbf{V}}^\text{NIC}$ in~\eqref{eq:SINR_const_IC}. 

Hereafter, we assume that $\gamma_k >0, \forall k$, and the constraints \eqref{eq:SINR_const_IC}-\eqref{eq:power_const} are strictly feasible to avoid degenerate cases.

\section{Upper Bounds on the Minimum Number of Beamformers for ISAC}\label{sec:N_min_BCRB}

Determining $N_\text{min}^\text{IC}$ and $N_\text{min}^\text{NIC}$ as defined in~\eqref{prob:N_min} is in general challenging as these values depend on the problem instance (e.g., the realization of the channels, nature of the parameters to be estimated, choice of prior distribution and $h(\cdot)$, etc.). In this paper, we seek concise generic upper bounds, expressed as functions of $K$ users and $L$ parameters, that hold across all problem instances.  

\subsection{ISAC Systems With Interference Cancellation} \label{sec:IC}

First, we consider systems where the communication users are able to cancel the interference from the sensing beams. The main result of this section is that $N_\text{min}^\text{IC}$ is at most $K + \sqrt{L (L+1)/2}$, which we refer to as the \emph{sum bound}. 
Note that the second term in the above expression is less than $L$. 

We prove this bound by showing the following.
Given an arbitrary pair of BFIM and SINRs achieved by a complete set of $N_\text{T} + K$ beamformers, one can always find a potentially better pair (i.e., with the same BFIM and the same or higher SINRs) which can be achievable by at most $K + \sqrt{L(L+1)/2}$ beamformers. This statement can be proved by fixing the quadratic terms associated with the BFIM-SINR pair and using a special case of the rank-reduction algorithm in~\cite{pataki1998rank, huangrank2010}. 

Consider the following set of BFIM-SINR pairs that are achievable using $K$ communication and $N_\text{T}$ sensing beams, i.e., using a full set of $\bar{N}=N_\text{T}+K$ beamformers:
\begin{equation}
\mathcal{A}^\text{IC} \triangleq 
	\left\{\left(\mathbf{J}_{\mathbf{V}}, \text{vSINR}^\text{IC}_{\mathbf{V}} \right) \Big| \mathbf{V} \in \mathbb{C}^{N_\text{T} \times \bar{N}}, \Tr(\mathbf{V}\mathbf{V}^\mathsf{H}) \leq P \right\},
\end{equation}
where $\text{vSINR}^\text{IC}_\mathbf{V} \triangleq \big[\text{SINR}_{1 , \mathbf{V}}^\text{IC}, \ldots, \text{SINR}_{K, \mathbf{V}}^\text{IC} \big]^\mathsf{T}$ denotes a vector of SINRs. Let $(\mathbf{J}, \boldsymbol{\gamma}) \in \mathcal{A}^\text{IC}$ with $\boldsymbol{\gamma} > \mathbf{0}$, and define the following family of optimization problems parameterized by $N$, the total number of beamformers at the transmitter: 
\begin{subequations}\label{prob:modified_problem}
         \begin{align}
        \mathcal{P}_N^\text{IC}: \underset{\mathbf{V} \in \mathbb{C}^{N_\text{T} \times N}}{\mathrm{minimize}}~~~&\Tr{\Big(\mathbf{V}\mathbf{V}^\mathsf{H}\Big)} \\
            {\mathrm{subject\; to}}~~&\mathbf{J}_{\mathbf{V}} = \mathbf{J}, \label{eq:BFIM_const} \\
            &\text{SINR}^\text{IC}_{k, \mathbf{V}} \geq \gamma_k, \quad \forall k. 
         \end{align}
        \end{subequations}
Observe that $\mathcal{P}_N^\text{IC}$ aims to find a set of $N$ beamforming vectors that minimize the total power, while achieving $\mathbf{J}$ and potentially improving the SINR.  Notice that $\mathcal{P}_N^\text{IC}$ may or may not be feasible depending on the value of $N$. However, when $N$ is set to $N_\text{T} + K$, a feasible solution must exist, because by assumption $(\mathbf{J}, \boldsymbol{\gamma}) \in \mathcal{A}^\text{IC}$. In the following, we begin by setting $N$ to be a value such that  \eqref{prob:modified_problem} is feasible, and denote an optimal solution of \eqref{prob:modified_problem} by $\mathbf{\hat{V}}$ and define $\boldsymbol{\gamma}'$ to be the SINRs attained by such $\mathbf{\hat{V}}$, for this value of $N$. 


Starting from $\mathbf{\hat V}$, consider the following procedure developed in ~\cite{pataki1998rank, huangrank2010} that aims to reduce the number of sensing beamformers while preserving the values of the BFIM, SINRs, and power.
Recall that $\mathbf{\hat{V}} \triangleq [\mathbf{\hat v}_1, \ldots, \mathbf{\hat v}_K, \mathbf{\hat V}_\text{s}]$ has $N$ beamformers. 
We obtain a new matrix $\mathbf{V}' \triangleq [\mathbf{v}'_1, \ldots, \mathbf{v}'_K, \mathbf{V}_\text{s}']$ with fewer beamformers $N' < N$, by scaling the communication beamformers with appropriate factors and multiplying $\mathbf{\hat{V}}_\text{s}$ by a ``tall" matrix $\mathbf{U}_\text{s}$:
\begin{align}
    \mathbf{v}_k' &= d_k \mathbf{\hat v}_k, \quad d_k \in \mathbb{C}, \quad \forall k, \label{eq:transformation_IC} \\
    \mathbf{V}'_\text{s} &= \mathbf{\hat V}_\text{s} \mathbf{U}_\text{s}, \quad \mathbf{U}_\text{s} \in \mathbb{C}^{N_\text{s} \times N_\text{s}'}, \quad  N_\text{s}' < N_\text{s}, \label{eq:transformation_IC2}
\end{align}
where $N_\text{s} = N - K$ and $N'_\text{s} = N' - K$ denote the number of sensing beamformers before and after transformation, and $\mathbf{U}_\text{s}$ is restricted to be a tall matrix to ensure that $N' < N$. The goal is to select $\{d_k\}$ and $\mathbf{U}_\text{s}$ so that $\mathbf{V}'$ achieves the same BFIM, SINR, and power as $\mathbf{\hat{V}}$, i.e.,
\begin{equation}\label{eq:BFIM_SINR}
        \mathbf{J}_{\mathbf{V}'} = \mathbf{J}_{\mathbf{\hat V}}, ~~ \text{vSINR}^\text{IC}_{\mathbf{V}'} = \boldsymbol{\gamma}', ~~ \Tr( \mathbf{V}' {\mathbf{V}'}^\mathsf{H})= \Tr( \mathbf{\hat{V}} {\mathbf{\hat{V}}}^\mathsf{H}).
\end{equation}
Observe that~\eqref{eq:BFIM_SINR} can be rearranged into a set of quadratic equations in $\{d_k\}$ and $\mathbf{U}_\text{s}$, because the BFIM elements \eqref{eq:FIM_elements}-\eqref{eq:TV_elements} and the SINRs all have quadratic dependence on $\mathbf{V}'$. 

A key question is then the following. Under what condition is it possible to find $\{d_k\}$ and a tall matrix $\mathbf{U}_\text{s}$ that would satisfy the set of quadratic equations \eqref{eq:BFIM_SINR}? 

The following lemma characterizes one such condition. 

\begin{lemma}\label{lem:quad_eq_sol_IC}
Consider $\mathcal{P}^\text{IC}_{N}$ defined in \eqref{prob:modified_problem} for sensing $L$ parameters while communicating with $K$ users, where $(\mathbf{J}, \boldsymbol{\gamma}) \in \mathcal{A}^\text{IC}$ with $\boldsymbol{\gamma} > \mathbf{0}$.  Suppose that $N$ is such that $\mathcal{P}^\text{IC}_{N}$ is feasible and has strong duality. Let $\mathbf{\hat{V}}$ be an optimal solution of $\mathcal{P}^\text{IC}_{N}$ with $K$ communication beamformers and $N_\text{s} = N - K$ sensing beamformers. Then, there exists a $\mathbf{V}' \in \mathbb{C}^{N_\text{T} \times N'}$ with $N' < N$ that achieves the same communications and sensing performance, i.e., satisfying \eqref{eq:BFIM_SINR}, if 
    \begin{equation}\label{eq:IC_condition}
            N_\text{s}^2 > \frac{L (L + 1)}{2}.
        \end{equation}    
    Furthermore, this new $\mathbf{V'}$ is an optimal solution of $\mathcal{P}^\text{IC}_{N'}$, and $\mathcal{P}^\text{IC}_{N'}$ has strong duality.
\end{lemma}

    \begin{IEEEproof}
     The proof involves rewriting the quadratic equations~\eqref{eq:BFIM_SINR} as a homogeneous linear system with $K + {N}_\text{s}^2$ unknowns and $K + L(L+1)/2$ equations 
     using a transformation of variables. Whenever~\eqref{eq:IC_condition} holds, this linear system has a nonzero solution, giving rise to $\{d_k\}$ and a tall $\mathbf{U}_\text{s}$ that yield $\mathbf{V}'$ based on \eqref{eq:transformation_IC}-\eqref{eq:transformation_IC2}. 
It can be shown that such $\mathbf{V}'$ has $N'<N$ beamformers and achieves the same BFIM, SINRs and power as $\mathbf{\hat{V}}$, and it is
an optimal solution to $\mathcal{P}^\text{IC}_{N'}$.


The details of the proof are in Appendix~\ref{appen:quad_eq_sol_IC}.
    \end{IEEEproof}

We are now ready to state the main result of this section.

\begin{theorem}\label{thm:BCRB_IC}
Consider an ISAC system for sensing $L$ parameters while communicating with $K$ users using linear beamforming, in which the communication users
can cancel the interference from the sensing beams.  Let $\mathcal{A}^\text{IC}$
denote the set of BFIM-SINR pairs achievable using a full set of 
$N_\text{T} + K$ beamformers under power constraint $P$. 
Then, for any $(\mathbf{J}, \boldsymbol{\gamma}) \in \mathcal{A}^\text{IC}$ with $ \boldsymbol{\gamma} > \mathbf{0}$, there exists a $\mathbf{V'}$ with at most $N_\text{bound}^\text{IC}$ beamformers that satisfies
    \begin{equation}
        \mathbf{J}_{\mathbf{V}'} = \mathbf{J}, ~~ \text{vSINR}^\text{IC}_{\mathbf{V}'} \geq \boldsymbol{\gamma}, ~~ \Tr( \mathbf{V}' {\mathbf{V}'}^\mathsf{H}) \leq P,
    \end{equation}
     where 
\begin{equation}\label{eq:N_bound_CRB_IC} N_\text{bound}^\text{IC} \triangleq \left \lfloor  K + \sqrt{\frac{L (L + 1)}{2}}\right \rfloor.
    \end{equation}
Thus, the minimum number of beamformers $N_\text{min}^\text{IC}$ for the ISAC system with interference cancellation 
is at most $N_\text{bound}^\text{IC}$.
\end{theorem}

\begin{IEEEproof}
Fix some arbitrary $(\mathbf{J}, \boldsymbol{\gamma}) \in \mathcal{A}^\text{IC}$ 
 with $ \boldsymbol{\gamma} > \mathbf{0}$
under power constraint $P$.  We start with a full set of beamformers, 
i.e., let $N = N_\text{T} + K$. 
By Lemma \ref{lem:P_strong_duality}, which is stated and proved in Appendix~\ref{Appen:strong_duality}, 
we have that for this $N$, $\mathcal{P}^\text{IC}_{N}$ has strong duality. 
The strong duality follows from the fact that the SDR of $\mathcal{P}^\text{IC}_{N}$,
obtained by defining the variables $\mathbf{R}_k = \mathbf{v}_k \mathbf{v}_k^\mathsf{H}$ and $\mathbf{R}_\text{s} = \mathbf{V}_\text{s}\mathbf{V}_\text{s}^\mathsf{H}$ and dropping the rank-one constraints for $\mathbf{R}_k$, is tight. 
Intuitively, this says when $N = N_\text{T} + K$, the sensing beamformers can have rank up to $N_\text{T}$, while the communication beamformers for the users are always rank-one at the optimum of $\mathcal{P}^\text{IC}_{N}$. 

Now, let $\mathbf{\hat V}$ denote an optimal solution of $\mathcal{P}^\text{IC}_{N}$. 
The idea is to apply Lemma \ref{lem:quad_eq_sol_IC} to $\mathbf{\hat V}$
to yield $\mathbf{V}'$, thereby reducing $N$ to $N'$, and then to repeat the process
iteratively, until $N \le N_\text{bound}^\text{IC}$. 

This process works, because whenever 
$N > N_\text{bound}^\text{IC}$, the number of sensing beamformers 
$N_\text{s} = N-K$ must always satisfy \eqref{eq:IC_condition}. 
Further, after each reduction step, $\mathbf{V}'$ is an optimal solution for 
$\mathcal{P}^\text{IC}_{N'}$ and $\mathcal{P}^\text{IC}_{N'}$ has strong duality, 
so the conditions of Lemma \ref{lem:quad_eq_sol_IC} are met. 
Thus, we can set $\mathbf{V'}$ as the new ${\mathbf{\hat V}}$ and $N_\text{s}'$ as the new $N_\text{s}$, and repeat the reduction step, while maintaining the BFIM, the SINRs and the power.  
The process can continue until 
\eqref{eq:IC_condition} is no longer satisfied, in which case $N_\text{s} \leq \lfloor \sqrt{L (L + 1)/2}\rfloor$ 
and $N = N_\text{s} + K$ is at most $N^\text{IC}_\text{bound}$. 

	\editrev{Based on the above, it is immediately concluded that $N_\text{min}^\text{IC} \leq N_\text{bound}^\text{IC}$. This is because the optimal solution for the original problem~\eqref{prob:main_problem} 
gives rise to a BFIM-SINR pair $(\mathbf{J}, \boldsymbol{\gamma}) \in \mathcal{A}^\text{IC}$. So, if the optimal solution has more than $N_\text{bound}^\text{IC}$ beamformers, it
can always be reduced. 
This holds even if the optimal solution of \eqref{prob:main_problem} is not unique.}
\end{IEEEproof}

The iterative procedure of Theorem~\ref{thm:BCRB_IC} is a special case of a more general rank-reduction algorithm due to Pataki \cite{pataki1998rank} for proving that a strictly feasible SDP must have a solution whose square rank is at most the number of constraints. The work of Huang and Palomar~\cite{huangrank2010} later extends this algorithm to the case of a separable SDP with $J$ SDR variables and $M$ constraints to prove the existence of a solution in which the sum of the squares of the ranks is bounded by $M$.

It is worth noting that the bounds of~\cite{huangrank2010, pataki1998rank} cannot be used to immediately recover the result of Theorem~\ref{thm:BCRB_IC} despite essentially utilizing the same algorithm. The reason is that the SDR of~\eqref{prob:modified_problem} with $\mathbf{R}_k = \mathbf{v}_k \mathbf{v}_k^\mathsf{H}$ and $\mathbf{R}_\text{s} = \mathbf{V}_\text{s}\mathbf{V}_\text{s}^\mathsf{H}$ has $J = K + 1$ variables and $M = K + \tfrac{L (L + 1)}{2}$ constraints.
Applying the result of~\cite{huangrank2010} proves the existence of a solution $\mathbf{R}^*_1, \ldots, \mathbf{R}_K^*$ and $\mathbf{R}^*_\text{s}$ satisfying 
\begin{equation}~\label{eq:huang_palomar}
\rank^2{\left(\mathbf{R}^*_\text{s}\right)} + \sum_{k}^{K} \rank^2{\left(\mathbf{R}^*_k\right)} \leq K + \frac{L (L + 1)}{2}.
\end{equation}
The above inequality does not immediately translate to a bound on the sum rank of $\mathbf{R}^*_1, \ldots, \mathbf{R}^*_K, \mathbf{R}^*_\text{s}$, (or equivalently the total number of sensing and communication beamformers). The proof of Theorem~\ref{thm:BCRB_IC} essentially shows that it is possible to find a solution satisfying~\eqref{eq:huang_palomar} with the additional property that $\rank(\mathbf{R}^*_1) = \ldots = \rank(\mathbf{R}^*_K) = 1$, and consequently, the sum rank is at most $\lfloor K + \sqrt{L (L+1)/2} \rfloor$. 

It should be remarked that the proof of Theorem~\ref{thm:BCRB_IC} reveals that the sum bound is not tied to the specific ISAC optimization formulation (i.e., \eqref{prob:main_problem}) used to define the minimum number of beamformers. For instance, we could have considered a communication-oriented optimization with a constraint on the sensing, and the same bound on $N_\text{min}^\text{IC}$ would still apply. This is so as long as the communications and sensing performances are characterized by the achievable pairs in $\mathcal{A}^\text{IC}$. 

As an observation related to the tightness of the derived bound, it can be seen that the proof is based on fixing the entire BFIM, which comprises $L(L+1)/2$ terms. This is a stronger requirement than just fixing the objective value in~\eqref{prob:main_problem}. In some cases, this can introduce slackness in the derivation. We expand on this idea in Section \ref{sec:other_metrics} to derive tighter bounds.

\subsection{ISAC Systems With No Interference Cancellation}

We now examine the scenario where the communication users cannot cancel the interference from the sensing beams. It can be verified that the proof in Theorem~\ref{thm:BCRB_IC} can be modified to show that the sum bound also holds for this scenario. In fact, the earlier conference version of this work~\cite{attiahbounds2024} establishes precisely this result. 
However, by exploiting the fact that the sensing beamformers penalize the communication in this case, it is possible to develop tighter bounds. 

Instead of the previous bound, in this section we prove that $N_\text{min}^\text{NIC} \leq \sqrt{K^2 + L(L+1)/2}$. This improved bound is referred to as the hypotenuse bound due to the following interpretation. If the effective number of beamformers for communication and sensing (i.e., $K$ and  $\sqrt{L(L+1)/2}$) correspond to the lengths of the sides in a right triangle, the minimum number of beamformers for ISAC would then be bounded by the length of the hypotenuse. 

The proof for the hypotenuse bound follows from a novel constructive algorithm that builds upon certain ideas from the previous section. Let $\mathcal{A}^\text{NIC}$ denote the set of BFIM-SINR pairs achievable by a complete set of $\bar{N}=N_\text{T}+K$ beamformers under power constraint $P$:
\begin{equation}
        \mathcal{A}^\text{NIC} \triangleq 
	\left\{\left(\mathbf{J}_{\mathbf{V}}, \text{vSINR}^\text{NIC}_{\mathbf{V}} \right) \Big| \mathbf{V} \in \mathbb{C}^{N_\text{T} \times \bar{N}}, \Tr(\mathbf{V}\mathbf{V}^\mathsf{H}) \leq P \right\},
\end{equation}
where $\text{vSINR}^\text{NIC}_{\mathbf{V}} \triangleq \big[\text{SINR}_{1, \mathbf{V}}^\text{NIC}, \ldots, \text{SINR}_{K, \mathbf{V}}^\text{NIC} \big]^\mathsf{T}$ denotes a vector of SINRs for the case of no interference cancellation. Fix an arbitrary pair $\left(\mathbf{J}, \boldsymbol{\gamma}\right) \in \mathcal{A}^\text{NIC}$ with $ \boldsymbol{\gamma} > 0$ under power constraint $P$, and define an optimization problem $\mathcal{P}_N^\text{NIC}$ analogous to that of the previous section: 
\begin{subequations}
\label{prob:NIC_formulation}
         \begin{align}
        \mathcal{P}_N^\text{NIC}: \underset{\mathbf{V} \in \mathbb{C}^{N_\text{T} \times N}}{\mathrm{minimize}}~~~&\Tr{\Big(\mathbf{V}\mathbf{V}^\textsf{H}\Big)} \\
            {\mathrm{subject\; to}}~~&\mathbf{J}_{\mathbf{V}} = \mathbf{J}, \\
            &\text{SINR}^\text{NIC}_{k, \mathbf{V}} \geq \gamma_k, \quad \forall k. \label{eq:SINR_const_NIC}
         \end{align}
        \end{subequations}
A key difference between the above optimization problem and the interference cancellation case is that it turns out $\mathcal{P}_N^\text{NIC}$ 
must have a solution whose sensing beamformers are orthogonal to the communication channels $\mathbf{h}_1, \ldots, \mathbf{h}_K$. Intuitively, this is because if the sensing beamformers had a nonzero component in the direction of
$\mathbf{h}_k$, then that component can always be moved into the communication beam $\mathbf{v}_k$, while keeping the overall transmit covariance $\mathbf{V}\mathbf{V}^\textsf{H}$ fixed. 
Doing so does not affect the BFIM or the total power, because both are functions of $\mathbf{V}\mathbf{V}^\textsf{H}$. 
On the other hand, the SINR always improves, because for user $k$,
the signal component $|\mathbf{h}_k^\mathsf{H} \mathbf{v}_k|^2$ increases, 
while the interference decreases, because 
\begin{equation}
	\sum_{i \neq k} |\mathbf{h}_k^\mathsf{H} \mathbf{v}_i|^2 + \mathbf{h}_k^\mathsf{H}\mathbf{V}_\text{s}\mathbf{V}_\text{s}^\mathsf{H}\mathbf{h}_k 
	= \mathbf{h}_k^\mathsf{H} \mathbf{V}\mathbf{V}^\textsf{H} \mathbf{h}_k - 
	|\mathbf{h}_k^\mathsf{H} \mathbf{v}_k|^2. 
\end{equation} 
For all the other users, such a move does not affect either the signal power or the interference, since $\mathbf{V}\mathbf{V}^\textsf{H}$ is fixed. Thus overall, there is an improvement.
This is a specific property of the no interference cancellation formulation.
It does not hold when interference cancellation is possible.

The following key result formalizes the above discussion.
\begin{lemma}\label{lem:Ortho_Vs_NIC}
   Fix some $\left(\mathbf{J}, \boldsymbol{\gamma}\right) \in \mathcal{A}^\text{NIC}$ with $\boldsymbol{\gamma}> \mathbf{0}$ under power constraint $P$. Suppose that the total number of beamformers $N$ is such that $\mathcal{P}_N^\text{NIC}$ is feasible. Then, $\mathcal{P}_N^\text{NIC}$ has 
 a solution 
    whose sensing beamformers satisfy
    \begin{equation}
\mathbf{h}_k^\mathsf{H} \mathbf{V}_\text{s} = \mathbf{0}, \quad \forall k.
    \end{equation}
\end{lemma}

    \begin{IEEEproof}
 See Appendix~\ref{appen:Ortho_Vs_NIC}.
    \end{IEEEproof}
\editrev{We remark that the lemma does not preclude the case $\mathbf{V}_\text{s} = \mathbf{0}$, which may happen in situations where the number of beamformers needed to attain the pair is equal to $K$.}

We now describe the rank-reduction procedure for the no interference cancellation case.
Let $\mathbf{\hat V}  \triangleq [\mathbf{\hat v}_1, \ldots, \mathbf{\hat v}_K, \mathbf{\hat V}_\text{s}] \in \mathbb{C}^{N_\text{T} \times N}$ be an optimal solution of $\mathcal{P}_{N}^\text{NIC}$ satisfying the orthogonality condition of Lemma~\ref{lem:Ortho_Vs_NIC}, so that $\mathbf{\hat{V}}_\text{s}$ is orthogonal to $\mathbf{h}_1, \ldots, \mathbf{h}_K$. Further, let $\boldsymbol{\gamma}' \geq \boldsymbol{\gamma}$ denote the SINR vector achieved by the initial $\mathbf{\hat{V}}$ with $N = N_\text{T} + K$. Similar to the previous section, the goal is to iteratively reduce the number of sensing beamformers by applying a linear transformation to $\mathbf{\hat V}$. However,
instead of relying on the construction of~\cite{pataki1998rank, huangrank2010}, we obtain the rank-reduced beamforming matrix 
$\mathbf{V}' \triangleq \left[\mathbf{v}_1', \ldots, \mathbf{v}_K', \mathbf{V}_\text{s}'\right] \in \mathbb{C}^{N_\text{T} \times N'}$ using a new transformation defined by variables $\mathbf{u}_k$ and $\mathbf{U}_\text{s}$ as follows:
\begin{align}\label{eq:lin_tran_NIC_comm}
            \mathbf{v}_k' &= \mathbf{\hat{v}}_k + \mathbf{\hat{V}}_\text{s} \mathbf{u}_k, \quad \mathbf{u}_k \in \mathbb{C}^{(N - K)}, \quad \forall k,  \\
    \mathbf{V}'_\text{s} &= \mathbf{\hat{V}}_\text{s} \mathbf{U}_\text{s}, \quad \mathbf{U}_\text{s} \in \mathbb{C}^{(N - K) \times (N' - K)},  \label{eq:lin_tran_NIC_sens} 
    \end{align}
where $N' < N$ denotes the total number of beamformers after reduction. 

Notice the difference between~\eqref{eq:lin_tran_NIC_comm} and the previous construction~\eqref{eq:transformation_IC} for the communication beamformers. Here, the new communication beamformers are given by the previous communication beamformers plus a linear transformation of the previous sensing beamformers. On the other hand, the new sensing beamformers are just a linear transformation of the previous sensing beamformers, just like in the interference cancellation case. 

We can write~\eqref{eq:lin_tran_NIC_comm}-\eqref{eq:lin_tran_NIC_sens} more compactly as follows:
\begin{equation}\label{eq:lin_transform_compact}
    \mathbf{V}'
    = \mathbf{\hat{V}} \mathbf{U}, \quad \mathbf{U} \triangleq \left[\begin{matrix} \mathbf{I}_K & \mathbf{0} \\ \mathbf{U}_\text{c} & \mathbf{U}_\text{s}  \end{matrix} \right] \in \mathbb{C}^{N \times N'}, ~ N' < N,
\end{equation}
where $\mathbf{U}_\text{c} \triangleq [\mathbf{u}_1, \ldots, \mathbf{u}_K]$, and $\mathbf{U}$ must be a tall matrix to ensure that the number of beamformers is reduced. The matrix $\mathbf{U}$ must be chosen to satisfy
\begin{equation}\label{eq:BFIM_SINR_NIC}
\mathbf{J}_{\mathbf{V}'} = \mathbf{J}_{\mathbf{\hat V}}, ~~ \text{vSINR}^\text{NIC}_{\mathbf{V}'} = \boldsymbol{\gamma}', ~~ \Tr( \mathbf{V}' {\mathbf{V}'}^\mathsf{H}) = \Tr( \mathbf{\hat{V}} {\mathbf{\hat{V}}}^\mathsf{H}).
         \end{equation}
This gives rise to a set of quadratic equations in $\mathbf{U}$. 

We ask the following question. Under what conditions can the existence of a solution to this set of quadratic equations be guaranteed? 

The next lemma provides a sufficient condition. 
\begin{lemma}\label{lem:quad_eq_sol_NIC}
Consider $\mathcal{P}^\text{NIC}_{N}$ defined in \eqref{prob:NIC_formulation} for sensing $L$ parameters while communicating with $K$ users, where $(\mathbf{J}, \boldsymbol{\gamma}) \in \mathcal{A}^\text{NIC}$ with $\boldsymbol{\gamma}>\mathbf{0}$. 
Suppose that $N$ is such that $\mathcal{P}^\text{NIC}_{N}$ is feasible and has strong duality. Let $\mathbf{\hat{V}}$ be an optimal solution of $\mathcal{P}^\text{NIC}_{N}$ with $K$ communication beamformers, and in addition $N_\text{s} = N - K$ sensing beamformers $\mathbf{\hat{V}}_\text{s}$ that are orthogonal to $\mathbf{h}_1, \ldots, \mathbf{h}_K$. Then, there exists a $\mathbf{V}' \in \mathbb{C}^{N_\text{T} \times N'}$ with $N' < N$ that achieves the same communications and sensing performance, i.e., satisfying~\eqref{eq:BFIM_SINR_NIC}, 
    if
\begin{equation}\label{eq:main_cond_NIC}
    N^2 > K^2 + \frac{L (L + 1)}{2}.
\end{equation}   
    Further, this new $\mathbf{V'}$ is an optimal solution of $\mathcal{P}^\text{NIC}_{N'}$ with $\mathbf{V}'_\text{s}$ orthogonal to $\mathbf{h}_1, \ldots, \mathbf{h}_K$, and $\mathcal{P}^\text{NIC}_{N'}$ has strong duality.
\end{lemma}

    \begin{IEEEproof}
 See Appendix~\ref{appen:quad_eq_sol_NIC}.
    \end{IEEEproof}

This leads to the following main theorem of this section. 

\begin{theorem}\label{thm:BCRB_NIC}
Consider an ISAC system with linear beamforming in which the communication users
cannot cancel the interference from the sensing beams. Let $\mathcal{A}^\text{NIC}$
denote the set of BFIM-SINR pairs achievable using a full set of 
$N_\text{T} + K$ beamformers under power constraint $P$. 
Then, for any $(\mathbf{J}, \boldsymbol{\gamma}) \in \mathcal{A}^\text{NIC}$ with  $\boldsymbol{\gamma} > 0$, there exists a $\mathbf{V'}$ with at most $N_\text{bound}^\text{NIC}$ beamformers that satisfies
    \begin{equation}
        \mathbf{J}_{\mathbf{V}'} = \mathbf{J}, ~~ \text{vSINR}^\text{IC}_{\mathbf{V}'} \geq \boldsymbol{\gamma}, ~~ \Tr( \mathbf{V}' {\mathbf{V}'}^\mathsf{H}) \leq P.
    \end{equation}
     where 
\begin{equation}\label{eq:N_bound_CRB_NIC}    
N_\text{bound}^\text{NIC} \triangleq \left \lfloor  \sqrt{K^2 + \frac{L (L + 1)}{2}}\right \rfloor.
\end{equation}
Thus, the minimum number of beamformers $N_\text{min}^\text{NIC}$ for the ISAC system without interference cancellation 
is at most $N_\text{bound}^\text{NIC}$.
\end{theorem}

\begin{IEEEproof}
The proof is similar to that of Theorem~\ref{thm:BCRB_IC}, except we now utilize the transformation~\eqref{eq:lin_transform_compact} to reduce the number of beamformers.
In the initial stage, we set $N = N_\text{T} + K$ and find the initial set of beamformers such that the sensing beamformers are orthogonal to the communication channels, which is always possible due to Lemma \ref{lem:Ortho_Vs_NIC}. Further, $\mathcal{P}^\text{NIC}_{N}$ has strong duality, which is proved in Appendix \ref{Appen:strong_duality}.

In the iteration process, whenever 
condition~\eqref{eq:main_cond_NIC} is satisfied, the rank-reduction process would be able to proceed. Throughout the process, $\mathcal{P}^\text{NIC}_{N}$ always has strong duality and the sensing beamformers can always be chosen to be orthogonal to the communication channels, so the conditions for Lemma~\ref{lem:quad_eq_sol_NIC} are satisfied. By applying  Lemma~\ref{lem:quad_eq_sol_NIC}, we can find some $\mathbf{V}'$ with reduced rank $N' < N$. This process can continue until condition~\eqref{eq:main_cond_NIC} no longer holds. Thus, the algorithm can always produce a beamforming matrix with a total number of beamformers no more than $N_\text{bound}^\text{NIC}$.
\end{IEEEproof}

The key to obtaining this tighter hypotenuse bound as compared to the sum bound 
is that the new communication beamformers can potentially absorb the sensing beamformers, as can be seen by comparing the new construction~\eqref{eq:lin_tran_NIC_comm} with the previous construction~\eqref{eq:transformation_IC} that led to the sum bound. The  construction~\eqref{eq:lin_tran_NIC_comm} is made possible by Lemma~\ref{lem:Ortho_Vs_NIC}, which shows that in the no interference cancellation scenario, the optimal sensing beamformers can be chosen to be orthogonal to the communication channels, so they do not impact the SINRs.

An interesting consequence of Lemma~\ref{lem:Ortho_Vs_NIC} is that $N_\text{min}^\text{NIC}$ is bounded by $N_\text{T}$ rather than $N_\text{T} + K$. This is true since the sensing beamformers occupy the subspace orthogonal to $\mathbf{h}_1, \ldots, \mathbf{h}_K$ as given by Lemma~\ref{lem:Ortho_Vs_NIC}. Since this orthogonal subspace is of dimension $N_\text{T} - K$, there can be at most $N_\text{T} - K$ linearly independent sensing beamformers. Thus, $N_\text{min}^\text{NIC} \leq N_\text{T}$ for the no interference cancellation case, while 
for the interference cancellation scenario, we only have $N_\text{min}^\text{IC} \leq N_\text{T}+K$.

More generally, we expect that an ISAC system where the users can cancel interference to employ more sensing beamformers (and to have better performance) than the no interference cancellation case.
This is because when interference cancellation is possible, the sensing beamformers do not penalize the communication performance regardless how they are constructed.
In contrast, when interference cancellation is not possible, the sensing beamformers must be restricted to the subspace orthogonal to the communication channels.
As a consequence, more sensing beamformers can be used and better performance can be obtained when interference cancellation is possible. 
The analysis here conforms with this intuition in the sense that $N_\text{bound}^\text{IC} \geq N_\text{bound}^\text{NIC}$. 

Theorem~\ref{thm:BCRB_NIC} reveals that with no interference cancellation, the minimum number of beamformers needed for ISAC is likely less than the sum of the number of beamformers needed for the individual tasks of communication and sensing alone. It turns out that the sensing beamformers become completely unnecessary when $K$ is larger than a threshold (which is determined by the number of sensing parameters), as the following result shows.
In essence, the communication beamformers are able to absorb all the sensing beamformers.

\begin{cor} \label{cor_Nmin=K}
Consider an ISAC system in which the communication users do not cancel the interference from the sensing beams. Suppose that the number of communication users $K$ and the number of sensing parameters $L$ are such that 
\begin{equation}\label{eq:N_min=K}
    K \geq \frac{L(L+1)}{4}.
\end{equation}
Then, for any feasible $(\mathbf{J}, \boldsymbol{\gamma}) \in \mathcal{A}^\text{NIC}$ with $\boldsymbol{\gamma} > \mathbf{0}$, it is possible to use exactly $K$ beamformers to achieve the same BFIM $\mathbf{J}$, while satisfying the SINR constraints $\boldsymbol{\gamma}$ and the power constraint $P$. 
\end{cor}

\begin{IEEEproof}
    By Theorem~\ref{thm:BCRB_NIC}, we have
    \begin{subequations}
    \begin{align}
N_\text{min}^\text{NIC} &\leq \left \lfloor  \sqrt{K^2 + \frac{L (L + 1)}{2}}\right \rfloor \\
& \leq \left \lfloor  \sqrt{K^2 + 2K} \right \rfloor \\
&< K + 1,
    \end{align}
    \end{subequations}
    where the second line follows from the assumption \eqref{eq:N_min=K}. This shows $N_\text{min}^\text{NIC} \leq K$. But, since $K$ beamformers are already needed for communications, we must have $N_\text{min}^\text{NIC} = K$. 
\end{IEEEproof}

The above result deals with the case where $K$ is large and the sensing beams become 
superfluous. In contrast, when $K$ is small, 
additional sensing beamformers are needed. In fact, Theorems \ref{thm:BCRB_IC} and 
\ref{thm:BCRB_NIC} continue to hold even in the case of $K=0$, for which the bounds of Theorems \ref{thm:BCRB_IC} and \ref{thm:BCRB_NIC} coincide
and the rank-reduction method reduces to that of \cite{pataki1998rank}. 
This sensing-only scenario is discussed in more detail in the next section.

\subsection{Implication for MIMO Radar} 

In MIMO radar with transmit beamforming, the goal is to design a continuous-time waveform $\boldsymbol{\varphi}(t) \in \mathbb{C}^{N_\text{T}}$ 
 to be transmitted over a pulse duration $T_\text{p}$ 
 for the purpose of achieving specific sensing objectives. The key difference between the signaling scheme for MIMO radar systems and ISAC is that in MIMO radar, the signals transmitted from different antennas do not share the same transmit pulse, unlike the ISAC case in~\eqref{eq:cont_time_pulse}. Furthermore, the signaling schemes adopted in MIMO radar are typically deterministic. 
Nevertheless, the results in this paper on the minimum number of beamformers can still be applied to the MIMO radar setting. 
 
A typical waveform $\boldsymbol{\varphi}(t)$ in MIMO radar is constructed based on 
a linear combination of a set of $N$ temporally-orthogonal waveforms \cite{Fried20212on}, i.e.,
\begin{equation}\label{eq:rad_decomposition}
    \boldsymbol{\varphi}(t) = \mathbf{V}_\text{rad} \mathbf{r}(t)
\end{equation}
where $\mathbf{r}(t) \triangleq \left[r_1(t), \ldots, r_{N}(t)\right]^\mathsf{T}$ are orthogonal waveforms with $\int_0^{T_\text{p}} r_i(t) r_j(t) dt = \delta_{ij}$, e.g., chirp signals with appropriate frequency shifts. 

The matrix $\mathbf{V}_\text{rad} \in \mathbb{C}^{N_\text{T} \times N}$ is a beamforming matrix with $N$ beamformers. From a design perspective, the decoupling of the temporal and spatial components in~\eqref{eq:rad_decomposition} offers significant benefits compared to designing $\boldsymbol{\varphi}(t)$ directly. The orthogonal signals $\mathbf{r}(t)$ can be selected based on their delay-Doppler properties, while the beamformers are chosen to produce a desired spatial beam pattern.
From an implementation perspective, it is desirable to use as few beamformers as possible.

Assuming a monostatic scenario, the discrete-time received signal, after delay/Doppler offset and matching filtering with the orthogonal signals, is given by~\cite{Fried20212on}
\begin{equation}~\label{eq:radar_mdl}
    \mathbf{Y} = \mathbf{G}^{(\boldsymbol{\eta})} \mathbf{V}_\text{rad} + \mathbf{Z},
\end{equation}
where $\mathbf{G}^{(\boldsymbol{\eta})} \in \mathbb{C}^{N_\text{R} \times N_\text{T}}$ is the round-trip channel assumed to be deterministic in $\boldsymbol{\eta}\in\mathbb{R}^L$, and $\mathbf{Z} \in \mathbb{C}^{N_\text{T} \times N}$ is the noise. 
Contrasting \eqref{eq:radar_mdl}, which originates from using deterministic waveforms, with the ISAC model \eqref{eq:sense_mdl} for sensing using random signals, we see that the two are closely related. In fact, it can be shown that both scenarios give rise to essentially the same BCRB structure. For the MIMO radar model~\eqref{eq:radar_mdl}, the BFIM is given by~\cite{Fried20212on, Li2008range, soticaprobing2007, huleihel2013tsp}:
\begin{equation}
    [\mathbf{J}_{\mathbf{V}}]_{ij} = [\mathbf{C}]_{ij} + \frac{1}{\sigma^2} \Tr\left(\tilde{\mathbf{G}}_{ij} \mathbf{V}_\text{rad} \mathbf{V}_\text{rad}^\mathsf{H} \right),
\end{equation}
where $\mathbf{C}$ is a prior matrix defined in~\eqref{eq:c_elements}.  If we adopt a design methodology that aims at optimizing the BCRB subject to a total power constraint~\cite{soticaprobing2007, Li2008range, huleihel2013tsp}, i.e., 
\begin{subequations}
\begin{align}
    \underset{\mathbf{V} \in \mathbb{C}^{N_\text{T} \times N}}{\mathrm{minimize}} ~~~&h\left({\mathbf{J}_\mathbf{V}^{-1}} \right) \label{eq:BCRB_metric_radar_only} \\
    \mathrm{subject \; to}~~~
    &\Tr(\mathbf{V}\mathbf{V}^\mathsf{H}) \leq P,
\end{align}
\end{subequations}
we can now ask the following question. 
What is the minimum number of beamformers needed to obtain the same performance as $N = N_\text{T}$? 
A bound can be readily established using the analysis in this paper.

\begin{cor} \label{cor_radar}
Consider an MIMO radar system that utilizes the model~\eqref{eq:rad_decomposition} with $N$ beamformers and adopts the BCRB as a performance metric for estimating $L$ real parameters of the channel. The minimum number of beamformers that achieve the same BCRB as the $N = N_\text{T}$ case is at most
\begin{equation}\label{eq:N_min_radar}
N_\text{bound}^\text{rad} \triangleq \left \lfloor \sqrt{\frac{L (L + 1)}{2}} \right \rfloor.
\end{equation}
\end{cor}

\begin{IEEEproof}
    Set $K = 0$ in either bound~\eqref{eq:N_bound_CRB_IC}  or bound~\eqref{eq:N_bound_CRB_NIC}.
\end{IEEEproof}

The main utility of Corollary~\ref{cor_radar} is that it provides an informative bound on the number of beamformers before waveform optimization. 
We remark that the bound implies $\tfrac{L}{2} \leq N_\text{bound}^\text{rad} \leq L$. This is because there are $L$ \emph{real} parameters to estimate, and each beamformer yields one complex measurement.

\section{Upper Bounds on the Minimum Number of Beamformers for General Sensing Metrics}
\label{sec:other_metrics}

Thus far, we have focused on ISAC systems that use the BCRB for parameter estimation. However, the sensing operation is generally not limited to just parameter estimation. Other tasks such as target detection along a given direction~\cite{bekkerTarget2006, mateen2023} and probing multiple spatial locations~\cite{soticaprobing2007, Fried20212on} are also of interest.
For these different objectives, the choice of appropriate sensing metric depend on the specific task. This section aims to derive bounds on the minimum number of beamformers for ISAC where the sensing task employs alternative metrics. 

\subsection{Class of $d$-Quadratic Sensing Metrics}

Instead of examining each sensing task one by one, in this paper we focus on a general class of sensing metrics for which the analysis of the previous section can be readily extended to yield useful bounds. 
We start by defining the class of sensing metrics under consideration.

\begin{define}
A real function of matrix $h_d(\mathbf{V}): \mathbb{C}^{N_\text{T} \times N} \rightarrow \mathbb{R}$ is said to be $d$-quadratic, if it has the following form:
\begin{equation}\label{eq:M_quadratic}
        h_d(\mathbf{V}) = f\left( \Tr\left(\mathbf{Q}_1 \mathbf{V}\mathbf{V}^\mathsf{H} \right), \ldots, \Tr\left(\mathbf{Q}_d \mathbf{V}\mathbf{V}^\mathsf{H} \right) \right),
    \end{equation}
where $\mathbf{Q}_1, \ldots, \mathbf{Q}_d$ are linearly independent Hermitian matrices 
and $f(\cdot)$ is a function from $\mathbb{R}^d$ to $\mathbb{R}$. 
\end{define}

In other words, a function is $d$-quadratic if it depends on the design variables (i.e., the beamforming matrix $\mathbf{V}$) only through 
$d$ distinct quadratic terms. In the ongoing discussion, we consider radar metrics that belong to the class of $d$-quadratic functions. 

There are many examples of $d$-quadratic metric that are relevant to different radar tasks. The BCRB-based metrics previously considered for parameter estimation constitute one such example. Indeed, the scalar objective of the ISAC optimization~\eqref{prob:main_problem} is $d$-quadratic with $d = L(L+1)/2$, assuming that the matrices $\big\{\tilde{\mathbf{G}}_{ij}\big\}$ are all linearly independent. More specifically, we can express the BFIM as follows:
\begin{equation}
\mathbf{J}_{\mathbf{V}} = r\left( \Tr\left(\mathbf{\tilde{G}}_{11} \mathbf{V}\mathbf{V}^\mathsf{H} \right), \ldots, \Tr\left(\mathbf{\tilde{G}}_{LL} \mathbf{V}\mathbf{V}^\mathsf{H} \right) \right),
\end{equation}
where
$r(\cdot)$ is a function that arranges its arguments as entries of an $L \times L$ symmetric matrix, with a constant matrix $\mathbf{C}$ added to it, i.e., for $\mathbf{w} = \left[w_{11}, \ldots, w_{1L}, w_{22}, \ldots w_{2L}, \ldots w_{LL} \right]^\mathsf{T} \in \mathbb{R}^{L (L + 1)/2}$, 
\begin{equation}
    r(\mathbf{w}) =  \mathbf{C}+
    \left[ \begin{matrix}\
w_{11} & w_{12} & \ldots & w_{1L} \\ 
w_{12} & w_{22} & \ldots & w_{2L}
\\
\vdots & \vdots  & \ddots & 
 \vdots \\
w_{1L} &  \ldots &  \ldots & w_{LL}
\end{matrix} \right].
\end{equation}
In this way, the objective of~\eqref{prob:main_problem} is readily seen as
a $d$-quadratic function, since it depends on the beamformers through the $L(L + 1)/2$ quadratic terms that comprise the BFIM. 

If there is dependency among $\big\{\tilde{\mathbf{G}}_{ij}\big\}$, the BCRB-based scalar metric can still be viewed as a $d$-quadratic function. However, the \emph{effective} number of quadratic terms in 
the BFIM would be less than $L(L+1)/2$. This scenario is relevant in practical parameter estimation scenarios and occurs, for example,  when some of $\big\{\tilde{\mathbf{G}}_{ij}\big\}$ are zero or repeated in different entries of the BFIM. 

An additional benefit of viewing the BCRB-based metrics as a $d$-quadratic function is that it allows us to handle cases where the goal is to estimate a strict subset of the parameters in $\boldsymbol{\eta}$, e.g., estimating the AoAs of multiple targets but not the path loss coefficients. This is achieved by simply counting the number of quadratic terms that influence the estimation of the relevant parameters, rather than counting the overall number of quadratic terms in the BFIM.

In Section~\ref{sec:examples}, we present further examples of radar metrics that belong to the family of $d$-quadratic functions beyond parameter estimation using BCRB. 

\subsection{Upper Bounds on $N_\text{min}$ for $d$-Quadratic Metrics}

The $d$-quadratic functions arise naturally as sensing performance metrics. In this section, we establish bounds on the minimum number of beamformers for ISAC systems utilizing sensing metrics in this class.

\begin{theorem}\label{thm:general_d_quad}
Consider an ISAC system with linear beamforming with $K$ communication users
and where the sensing performance is measured using a $d$-quadratic function
$h_d(\cdot)$. Let $\mathcal{\tilde{A}}^\text{IC}$
and $\mathcal{\tilde{A}}^\text{NIC}$ denote the sets of sensing metric and SINR pairs achievable with or without cancelling interference from the sensing beams at the users, respectively, using a full set of $\bar{N} = N_\text{T} + K$ beamformers under power constraint $P$: 
\begin{align}
    \mathcal{\tilde{A}}^\text{IC} &\triangleq
        \left\{\left(h_d(\mathbf{V}), \text{vSINR}^\text{IC}_{\mathbf{V}} \right) \Big| \mathbf{V} \in \mathbb{C}^{N_\text{T} \times \bar{N}}, \Tr(\mathbf{V}\mathbf{V}^\mathsf{H}) \leq P \right\}, \\
        \mathcal{\tilde{A}}^\text{NIC} &\triangleq 
        \left\{\left(h_d(\mathbf{V}), \text{vSINR}^\text{NIC}_{\mathbf{V}} \right) \Big| \mathbf{V} \in \mathbb{C}^{N_\text{T} \times \bar{N}}, \Tr(\mathbf{V}\mathbf{V}^\mathsf{H}) \leq P \right\}.
\end{align}
Then, 
\begin{enumerate}
    \item 
     For any pair $(h, \boldsymbol{\gamma}) \in  \mathcal{\tilde{A}}^\text{IC}$ with $\boldsymbol{\gamma}>0$, there exists a linear beamforming scheme with at most $\tilde{N}_\text{bound}^{\text{IC}}$ beamformers under power constraint $P$ that attains $(h, \boldsymbol{\gamma}')$ with $
     \boldsymbol{\gamma}' \geq \boldsymbol{\gamma}$, where 
\begin{equation}\label{eq:mod_sum_bound}
\tilde{N}_\text{bound}^{\text{IC}}
 \triangleq \lfloor K + \sqrt{d} \rfloor \end{equation}
 is the \emph{modified sum bound}.

\item For any pair $(h, \boldsymbol{\gamma}) \in  \mathcal{\tilde{A}}^\text{NIC}$ with $\boldsymbol{\gamma}>0$, there exists a linear beamforming scheme with at most $\tilde{N}_\text{bound}^{\text{NIC}}$ beamformers under power constraint $P$ that attains $(h, \boldsymbol{\gamma}')$ with $
     \boldsymbol{\gamma}' \geq \boldsymbol{\gamma}$, where
    \begin{equation}\label{eq:mod_hypo_bound}
\tilde{N}_\text{bound}^{\text{NIC}}
 \triangleq \lfloor \sqrt{K^2 + d} \rfloor
\end{equation} 
is the \emph{modified hypotenuse bound}.
\end{enumerate}
\end{theorem}

    \begin{IEEEproof} 
First, we note that since the radar metric $h_d(\cdot)$ is a $d$-quadratic function, by definition there are
linearly independent $\mathbf{Q}_1, \ldots, \mathbf{Q}_d$ and a function $f(\cdot)$ for which
    \begin{equation} \label{eq:c_i_def}
        h_d(\mathbf{V}) = f(c_1, \ldots, c_d), \quad c_i \triangleq \Tr\left( \mathbf{Q}_i \mathbf{V} \mathbf{V}^\mathsf{H}\right), \quad 1 \leq i \leq d.
    \end{equation} 
Since the sensing metric value $h$ is achievable, this means that there must exist $\mathbf{V}$ such that
    \begin{equation}
        h = h_d(\mathbf{V}).
    \end{equation}

To establish the modified sum bound \eqref{eq:mod_sum_bound} for the case with interference cancellation, we fix an arbitrary pair $(h, \boldsymbol{\gamma}) \in \mathcal{\tilde{A}}^\text{IC}$ and consider the optimization problem over $N$ beamformers: 
\begin{subequations}\label{prob:modified_d_quadratic}
         \begin{align}
        \mathcal{\tilde{P}}_{N}^\text{IC}: \underset{\mathbf{V} \in \mathbb{C}^{N_\text{T} \times N}}{\mathrm{minimize}}~~~&\Tr{\Big(\mathbf{V}\mathbf{V}^\textsf{H}\Big)} \\
            {\mathrm{subject\; to}}~~&  \Tr\left( \mathbf{Q}_i \mathbf{V} \mathbf{V}^\mathsf{H}\right) = c_i, \quad \forall i  \label{eq:quad_d_quad} \\
    & ~\text{SINR}^\text{IC}_{k, \mathbf{V}} \geq \gamma_k, \quad \forall k \label{eq:SINR_q_uad}
         \end{align}
        \end{subequations}
        where $\{c_i\}$ are defined in~\eqref{eq:c_i_def} for the beamformers that achieve the pair $\left(h, \boldsymbol{\gamma}\right)$. Now, start with $N_\text{T} + K$ beamformers and let $\mathbf{\hat{V}}$ denote the optimal solution of $\mathcal{\tilde{P}}_{N_\text{T} + K}^\text{IC}$. Use the iterative procedure of Theorem~\ref{thm:BCRB_IC}, with $\mathbf{\hat{V}}$ as the starting point. In each iteration, we seek to maintain the values of the $K$ quadratic SINRs equations~\eqref{eq:SINR_q_uad} and the $d$ quadratic equations~\eqref{eq:quad_d_quad}.
         By Lemma~\ref{lem:quad_eq_sol_IC}, this gives rise to a linear system with $K + d$ equations and $K + N_\text{s}^2$ unknowns. When $N_\text{s} > \sqrt{d}$, such a linear system must have a nonzero solution, thereby allowing for a reduction in the number of sensing beamformers. This procedure can continue as long as $N_\text{s} > \sqrt{d}$. This implies that the minimum number of beamformers to achieve $\left(h, \boldsymbol{\gamma}\right)$ is at most $\tilde{N}_\text{bound}^{\text{IC}} =  K + \sqrt{d}$ beamformers.
    
To establish the modified hypotenuse bound \eqref{eq:mod_hypo_bound} for the case with no interference cancellation, we follow a similar procedure as in the proof of Theorem~\ref{thm:BCRB_NIC}. 
It can be verified that at the optimum solution of the corresponding beamforming optimization problem, there must exist sensing beamformers that satisfy $\mathbf{h}_k^\mathsf{H}\mathbf{\hat{V}}_\text{s}  = \mathbf{0}$ for all $k$. Indeed, this is because the proof of Lemma~\ref{lem:Ortho_Vs_NIC} does not depend on the choice of sensing metric. 
The rest of the iterative rank-reduction process is exactly the same as in Theorem~\ref{thm:BCRB_NIC}.
    \end{IEEEproof} 

The modified bounds in Theorem~\ref{thm:general_d_quad} reduce to the sum bound in Theorem~\ref{thm:BCRB_IC} and the hypotenuse bound in Theorem~\ref{thm:BCRB_NIC}, when applied to the BCRB metric with $d = \frac{L (L + 1)}{2}$. However as mentioned earlier, these upper bounds are not limited to parameter estimation and can be applied to \emph{any} $d$-quadratic sensing metric. Furthermore, they refine the previous sum/hypotenuse bounds for the BCRB case when the set of $\{\mathbf{\tilde{G}}_{ij}\}$ is linearly dependent, or when the number of quadratic terms relevant to the estimation task is fewer than $L(L + 1)/2$. Hence, Theorem~\ref{thm:general_d_quad} encompasses Theorem~\ref{thm:BCRB_IC} and Theorem~\ref{thm:BCRB_NIC}. 

As an immediate consequence of Theorem~\ref{thm:general_d_quad}, the minimum number of beamformers for the sensing-only scenario is upper bounded by $\sqrt{d}$ for a $d$-quadratic sensing metric. This can be obtained by setting $K = 0$ in either the modified sum bound~\eqref{eq:mod_sum_bound} or hypotenuse bound~\eqref{eq:mod_hypo_bound}.

\begin{cor} \label{cor_radar_d_quadratic}
Consider a MIMO radar system that utilizes the model~\eqref{eq:rad_decomposition} with $N$ beamformers and adopts a $d$-quadratic metric for sensing. The minimum number of beamformers that achieve the same performance as the $N = N_\text{T}$ case is at most
\begin{equation}\label{eq:N_min_radar_d_quadratic}
{\tilde N}_\text{bound}^\text{rad} \triangleq \left \lfloor \sqrt{d} \right \rfloor.
\end{equation}
\end{cor}

Finally, we establish an analogous result to Corollary~\ref{cor_Nmin=K}.

\begin{cor} \label{cor_Nmin=K_d_quad}
Consider an ISAC system where the sensing performance is measured using a $d$-quadratic metric, and the communication users do not cancel the interference from the sensing beams.  Suppose that the number of communication users $K$ is such that 
\begin{equation}\label{eq:N_min=K_q_uad}
    K \geq \frac{d}{2}.
\end{equation}
Then, for any feasible $(h, \boldsymbol{\gamma}) \in \mathcal{\tilde A}^\text{NIC}$ with $\boldsymbol{\gamma} > 0$, it is possible to use exactly $K$ beamformers to achieve the same sensing metric value $h$, while satisfying the SINR constraints $\boldsymbol{\gamma}$ and the power constraint $P$. 
\end{cor}

\begin{IEEEproof}
    When \eqref{eq:N_min=K_q_uad} holds, then $N_\text{min}^\text{NIC} \leq \left \lfloor  \sqrt{K^2 + d}\right \rfloor  \leq \left \lfloor  \sqrt{K^2 + 2K} \right \rfloor < K + 1$.
    Since $N_\text{min}^\text{NIC} \geq K$, we must have $N_\text{min}^\text{NIC} = K$. 
\end{IEEEproof}


\section{Applications}\label{sec:examples}

This section presents several examples to demonstrate how the results of
this paper can be used to provide novel bounds on the minimum number of
beamformers for several ISAC scenarios of practical interest. In the
first two examples, we examine parameter estimation problems, where the
sensing performance is measured using the BCRB. The subsequent examples
consider radar functionalities, where the performance metric belongs to
the class of $d$-quadratic functions.

\subsection{Estimating the Target Channel Matrix}

\subsubsection{Sensing-Only Scenario}

We begin by considering a sensing-only task of estimating the entire channel matrix between the transmitter and the receiver. This is the conventional channel estimation problem, which we analyze it from a BCRB perspective in this paper.

First, examine the case $N_\text{R} = 1$, for which the round-trip channel is given by the row vector
\begin{equation}\label{eq:channel_row_vec}
{\mathbf{g}^{(\boldsymbol{\eta})}}^\mathsf{T} = \left[g_1, \ldots, g_{N_\text{T}} \right] \in \mathbb{C}^{1 \times N_\text{T}}.
\end{equation}
Here, there are a total of $L = 2N_\text{T}$ real parameters obtained by stacking the real and imaginary parts of different elements of $\mathbf{g}^{(\boldsymbol{\eta})}$:
\begin{equation}\label{eq:eta_full_channel_Nr1}
    \boldsymbol{\eta} = \left[\begin{matrix} \mathbf{g}_\text{re}^\mathsf{T}, & \mathbf{g}_\text{im}^\mathsf{T} \end{matrix} \right]^\mathsf{T} \in \mathbb{R}^{2N_\text{T}},
\end{equation}
where $\mathbf{g}_\text{re}$ and $\mathbf{g}_\text{im}$ denote the real and imaginary parts of $\mathbf{g}^{(\boldsymbol{\eta})}$.
A tight bound on the minimum number of beamformers for this MIMO radar application can be determined as follows. Instead of using a straightforward bound $\sqrt{L (L + 1)/2} = \sqrt{N_\text{T}(2N_\text{T} + 1)}$, we view the BCRB as a $d$-quadratic metric. The BFIM has the following structure:
\begin{equation}\label{eq:J_full_channel}
    \mathbf{J}_{\mathbf{V}} = \mathbf{C}+  \left[ \begin{matrix}\
\mathbf{T}_{\mathbf{g}_\text{re} \mathbf{g}_\text{re}} &  \mathbf{T}_{\mathbf{g}_\text{re} \mathbf{g}_\text{im}} \\
 \mathbf{T}_{\mathbf{g}_\text{im} \mathbf{g}_\text{re}} &   \mathbf{T}_{\mathbf{g}_\text{im} \mathbf{g}_\text{im}}
\end{matrix} \right] \in \mathbb{R}^{2N_\text{T} \times 2 N_\text{T}},
\end{equation}
where $\mathbf{T}_{\mathbf{g}_\text{re} \mathbf{g}_\text{re}}$ and $\mathbf{T}_{\mathbf{g}_\text{im} \mathbf{g}_\text{im}}$ are identical $N_\text{T} \times N_\text{T}$ real matrices with $(i, j)$-th elements given by
\begin{equation}\label{eq:Trr}
    \left[\mathbf{T}_{\mathbf{g}_\text{re} \mathbf{g}_\text{re}}\right]_{ij} =  \left[\mathbf{T}_{\mathbf{g}_\text{im} \mathbf{g}_\text{im}}\right]_{ij}=  \frac{\Upsilon}{\sigma^2} \Tr\left( \left(\mathbf{e}_i \mathbf{e}_j^\mathsf{H} + \mathbf{e}_j \mathbf{e}_i^\mathsf{H} \right)\mathbf{V}\mathbf{V}^\mathsf{H} \right). 
\end{equation}
The two off-diagonal terms are related by $\mathbf{T}_{\mathbf{g}_\text{re} \mathbf{g}_\text{im}} = \mathbf{T}_{\mathbf{g}_\text{im} \mathbf{g}_\text{re}}^\mathsf{T}$ with
\begin{equation}\label{eq:Tri}
  \left[\mathbf{T}_{\mathbf{g}_\text{re} \mathbf{g}_\text{im}} \right]_{ij}  = \frac{\Upsilon}{\sigma^2} \Tr\left( \jmath \left(\mathbf{e}_i \mathbf{e}_j^\mathsf{H} - \mathbf{e}_j \mathbf{e}_i^\mathsf{H} \right)\mathbf{V}\mathbf{V}^\mathsf{H} \right), ~~ \forall i, j.
\end{equation}

The above expressions
reveal the dependence among the entries of these matrices. It is readily verified that the matrices $\mathbf{T}_{\mathbf{g}_\text{re} \mathbf{g}_\text{re}}$ and $\mathbf{T}_{\mathbf{g}_\text{im} \mathbf{g}_\text{im}}$ are specified by $\tfrac{N_\text{T} (N_\text{T} + 1)}{2}$ distinct quadratic terms due to symmetry, whereas $\mathbf{T}_{\mathbf{g}_\text{re} \mathbf{g}_\text{im}}$ and $\mathbf{T}_{\mathbf{g}_\text{im} \mathbf{g}_\text{re}}$ are specified by $\tfrac{N_\text{T} (N_\text{T} - 1)}{2}$ quadratic terms since both matrices also have zero diagonal elements. A scalar function of the BCRB is therefore $d$-quadratic with $d = \tfrac{N_\text{T} (N_\text{T} + 1)}{2}  + \tfrac{N_\text{T} (N_\text{T} - 1)}{2}  = N_\text{T}^2$. The minimum number of sensing beamformers for estimating the channel matrix is then at most
\begin{equation}\label{eq:bound_sense_fullchannel}
    N^\text{rad}_\text{min} \leq \sqrt{d} = N_\text{T}.
\end{equation}
This result should not be too surprising since the quadratic matrices defining the BFIM~\eqref{eq:Trr}-\eqref{eq:Tri} correspond to the canonical basis of the space of $N_\text{T} \times N_\text{T}$ Hermitian matrices. In fact, we can equivalently express $\mathbf{J}_\mathbf{V}$ as follows:
\begin{equation}
\label{eq:Jv_sensing_full_channel}
      \mathbf{J}_{\mathbf{V}} = \mathbf{C}+ \frac{2 \Upsilon}{\sigma^2}  \left[ \begin{matrix}\
\Re\{ \mathbf{V} \mathbf{V}^\mathsf{H}\} & \Im\{ \mathbf{V} \mathbf{V}^\mathsf{H}\} \\ -\Im\{ \mathbf{V} \mathbf{V}^\mathsf{H}\} & \Re\{ \mathbf{V} \mathbf{V}^\mathsf{H}\} \end{matrix} \right],
\end{equation}
where $\Re\{ \mathbf{V} \mathbf{V}^\mathsf{H}\}$ and $\Im\{ \mathbf{V} \mathbf{V}^\mathsf{H}\}$ denote the real and imaginary part of $\mathbf{V} \mathbf{V}^\mathsf{H}$, which is a Hermitian matrix with $d = N_\text{T}^2$ degrees of freedom. Note that $\Im\{ \mathbf{V} \mathbf{V}^\mathsf{H}\}$ is skew-symmetric.

The upper bound~\eqref{eq:bound_sense_fullchannel} cannot be improved in general, because 
we can easily construct an example for which the required number of sensing beamformers achieves this bound. To see this, suppose that the elements of $\mathbf{g}^{(\boldsymbol{\eta})}$ have a prior distribution of i.i.d.\ $\mathcal{CN}(0, 2\sigma_0^2)$, so that $\mathbf{C} = \tfrac{1}{\sigma_0^2} \mathbf{I}_{2N_\text{T}}$. Consider a problem of minimizing a sensing objective of the trace-inverse of the BFIM as below:
\begin{subequations}\label{prob:ch_est_example}
\begin{align}
    \underset{\mathbf{V} \in \mathbb{C}^{N_\text{T} \times N}}{\mathrm{minimize}} & \ \Tr\left(  \mathbf{J}_\mathbf{V}^{-1} \right) = 2 \Tr\left( \left(  \tfrac{1}{\sigma_0^2} \mathbf{I}_{N_\text{T}} + \tfrac{2\Upsilon}{\sigma^2} \mathbf{V} \mathbf{V}^\mathsf{H} \right)^{-1} \right) \label{eq:obj_full_ch} \\
    \mathrm{subject \; to}
    & \ \Tr(\mathbf{V}\mathbf{V}^\mathsf{H}) \leq P.
\end{align}
\end{subequations}
The computation of $\mathbf{J}_\mathbf{V}^{-1}$ in \eqref{eq:obj_full_ch} for this case of  $\mathbf{C} = \tfrac{1}{\sigma_0^2} \mathbf{I}_{2N_\text{T}}$ involves some algebra, which is shown in Appendix \ref{app:Jv_computation}. 

Due to symmetry, this problem can be solved analytically.
For instance, when $N = N_\text{T}$, we have $\mathbf{V}^* = \sqrt{\tfrac{P}{N_\text{T}}} \mathbf{I}_{N_\text{T}}$, i.e.,   
$N_\text{T}$ beamformers are needed, which achieves the upper bound. 

This example also demonstrates the existence of problem instances for which the minimum number of beamformers can be less than the upper bound $N_\text{T}$ 
in \eqref{eq:bound_sense_fullchannel}, 
for example, when the channel coefficients have different prior distributions. Let $2\sigma_i^2$ be the variance of the $i$-th element of $\mathbf{g}^{(\boldsymbol{\eta})}$. For $N=N_\text{T}$, the solution of the trace-inverse problem is given by $\mathbf{V}^* = \mathbf{D}$ for 
diagonal matrix $\mathbf{D}$, whose entries are determined by a water-filling-like solution. Depending on the values of $\{\sigma^2_i\}$, some of the allocated powers may be zero, so only $N < N_\text{T}$ beamformers are needed. 

Finally, we examine the case where $N_\text{R} > 1$. The matrix channel now has $N_\text{R}$ rows, each of the form~\eqref{eq:channel_row_vec}:
\begin{equation}
\mathbf{G}^{(\boldsymbol{\eta})} = \left[ \begin{matrix} \mathbf{g}_1^\mathsf{H} \\ \vdots \\ \mathbf{g}_{N_\text{R}}^\mathsf{H}  \end{matrix} \right], \quad \mathbf{g}_i \in \mathbb{C}^{1 \times N_\text{T}},
\end{equation}
and the vector of parameters of interest is now obtained by stacking multiple vectors of the form in~\eqref{eq:eta_full_channel_Nr1}.
It can be verified that the BFIM can be obtained as follows:
\begin{equation}
      \mathbf{J}_{\mathbf{V}} = \mathbf{C}+  \mathbf{I}_{N_\text{R}} \otimes \frac{2 \Upsilon}{\sigma^2}  \left[ \begin{matrix}\
\Re\{ \mathbf{V} \mathbf{V}^\mathsf{H}\} & \Im\{ \mathbf{V} \mathbf{V}^\mathsf{H}\} \\ -\Im\{ \mathbf{V} \mathbf{V}^\mathsf{H}\} & \Re\{ \mathbf{V} \mathbf{V}^\mathsf{H}\} \end{matrix} \right],
\end{equation}
where $\otimes$ denotes the Kronecker product. Note that this does not change the total number of distinct quadratic terms. Thus, the minimum number of sensing beamformers is bounded by $N_\text{T}$. Again, this bound cannot be improved in general. 

The above results make intuitive sense, because to estimate an $N_\text{T} \times N_\text{R}$ 
channel, we need to transmit $N_\text{T}$ orthogonal pilots across the $N_\text{T}$ antennas, regardless of the number of receive antennas. Thus, $N_\text{T}$ beamformers are needed.
But if the prior distributions of the different channel entries have different variances, 
one may wish to allocate more power to entries with larger variances in order to reduce 
their channel estimation error, and potentially zero power to entries that are already 
quite certain, thus reducing the number of beamformers required to be less than 
$N_\text{T}$.


\subsubsection{ISAC Scenarios}
Next, we examine the case of communicating with $K$ users while estimating an $N_\text{T} \times N_\text{R}$ channel. When the communications users can cancel interference from the sensing beams, the sum bound becomes 
\begin{equation}
    N_\text{min}^\text{IC} \leq  \sqrt{d} + K =N_\text{T} + K,
\end{equation}
which is identical to the trivial bound. 

For the non-interference cancellation scenario, the hypotenuse bound is 
\begin{equation}
    N_\text{min}^\text{NIC} \leq \left \lfloor\sqrt{K^2 + N_\text{T}^2} \right\rfloor.
\end{equation}
However, recall that for the non-interference cancellation case, we always have 
$N_\text{min}^\text{NIC} \leq N_\text{T}$, which is a tighter bound. 


\subsection{Parameter Estimation for $N_\text{tr}$ Targets With LoS Paths}

Consider the task of estimating the parameters of $N_\text{tr}$ targets each with LoS paths. Assuming a $1$-D array and that the targets are in the far field, the sensing channel is given by
\begin{equation}\label{eq:multipath}
\mathbf{G}^{(\boldsymbol{\eta})} = \sum_{i = 1}^{N_\text{tr}} \alpha_i \mathbf{A}(\theta_i),
\end{equation}
where the combined (transmit and receive) array response $\mathbf{A}(\cdot) \in \mathbb{C}^{N_\text{R} \times N_\text{T}}$ depends only on the AoAs because of the far-field assumption. 
There are a
total of  $L = 3N_\text{tr}$ parameters: 
\begin{equation}
\boldsymbol{\eta} = \left[ \begin{matrix}
   \boldsymbol{\alpha}_\text{re}^\mathsf{T},  \boldsymbol{\alpha}_\text{im}^\mathsf{T}, \boldsymbol{\theta}^\mathsf{T}
\end{matrix} \right]^\mathsf{T},
\end{equation}
where $\boldsymbol{\alpha}_\text{re}^\mathsf{T} = \left[ \Re\{\alpha_1\}, \ldots, \Re\{\alpha_{N_\text{tr}}\}\right]^\mathsf{T} \in \mathbb{R}^{N_\text{tr}}$ is a vector containing the real parts of the path loss coefficients of different targets. The vectors $\boldsymbol{\alpha}_\text{im}$ and $\boldsymbol{\theta}$ are defined similarly for the imaginary parts of the path loss coefficients and for the AoAs, respectively. For convenience, we adopt the notation $\boldsymbol{\alpha} = \left[\begin{matrix} \boldsymbol{\alpha}_\text{re}^\mathsf{T} & \boldsymbol{\alpha}_\text{im}^\mathsf{T} \end{matrix} \right]^\mathsf{T}$.

Similar to the previous example, the matrices characterizing the BFIM are not all linearly independent. Therefore, viewing the sensing metric as a $d$-quadratic function can yield tighter bounds. The BFIM is given by
\begin{equation}\label{eq:BFIM_N_targets}
\mathbf{J}_{\mathbf{V}} = \left[ \begin{matrix}\
\mathbf{C}_{\boldsymbol{\alpha}, \boldsymbol{\alpha}} & \mathbf{C}_{\boldsymbol{\alpha}, \boldsymbol{\theta}} \\ \mathbf{C}_{\boldsymbol{\alpha}, \boldsymbol{\theta}}^\mathsf{T} & \mathbf{C}_{\boldsymbol{\theta}, \boldsymbol{\theta}} \end{matrix} \right]+  \left[ \begin{matrix}\
\mathbf{T}_{\boldsymbol{\alpha}, \boldsymbol{\alpha}} & \mathbf{T}_{\boldsymbol{\alpha}, \boldsymbol{\theta}} \\ \mathbf{T}_{\boldsymbol{\alpha}, \boldsymbol{\theta}}^\mathsf{T} & \mathbf{T}_{\boldsymbol{\theta}, \boldsymbol{\theta}} \end{matrix} \right].
\end{equation}
We aim to find $d$ by counting the number of distinct quadratic terms characterizing $\mathbf{T}_{\boldsymbol{\alpha}, \boldsymbol{\alpha}}, \mathbf{T}_{\boldsymbol{\alpha}, \boldsymbol{\theta}}$ and $\mathbf{T}_{\boldsymbol{\theta}, \boldsymbol{\theta}}$. The matrix $\mathbf{T}_{\boldsymbol{\alpha}, \boldsymbol{\alpha}} \in \mathbb{R}^{2N_\text{tr} \times 2N_\text{tr}}$ shares the same structure as that in~\eqref{eq:J_full_channel} for estimating the channel coefficients considered in the previous example. In particular, it can be verified that
\begin{equation}
    \mathbf{T}_{\boldsymbol{\alpha}, \boldsymbol{\alpha}} = \left[ \begin{matrix}\
\mathbf{T}_{\boldsymbol{\alpha}_\text{re} \boldsymbol{\alpha}_\text{re}} &  \mathbf{T}_{\boldsymbol{\alpha}_\text{re} \boldsymbol{\alpha}_\text{im}} \\
 \mathbf{T}_{\boldsymbol{\alpha}_\text{im} \boldsymbol{\alpha}_\text{re}} &   \mathbf{T}_{\boldsymbol{\alpha}_\text{im} \boldsymbol{\alpha}_\text{im}}
\end{matrix} \right],
\end{equation}
where $ [\mathbf{T}_{\boldsymbol{\alpha}_\text{re} \boldsymbol{\alpha}_\text{re}}]_{i,j} = [\mathbf{T}_{\boldsymbol{\alpha}_\text{im} \boldsymbol{\alpha}_\text{im}}]_{i,j} =  \Tr(\mathbf{\tilde{G}}_{ij, 1} \mathbf{V} \mathbf{V}^\mathsf{H} )$, and $\mathbf{T}_{\boldsymbol{\alpha}_\text{re} \boldsymbol{\alpha}_\text{im}} = \mathbf{T}_{\boldsymbol{\alpha}_\text{im} \boldsymbol{\alpha}_\text{re}}^\mathsf{T}$ with $\left[\mathbf{T}_{\boldsymbol{\alpha}_\text{re} \boldsymbol{\alpha}_\text{im}} \right]_{i, j} = \Tr(\mathbf{\tilde{G}}_{ij, 2} \mathbf{V} \mathbf{V}^\mathsf{H})$, with
\begin{subequations}
\begin{align}
    \mathbf{\tilde{G}}_{ij, 1} &= \frac{\Upsilon}{\sigma^2}\mathbb{E}_{\theta_i, \theta_j}\left[\mathbf{A}^\mathsf{H}(\theta_i) \mathbf{A}(\theta_j) +  \mathbf{A}^\mathsf{H}(\theta_j) \mathbf{A}(\theta_i) \right], \label{eq:G1} \\
    \mathbf{\tilde{G}}_{ij, 2} &= \frac{\Upsilon}{\sigma^2}\mathbb{E}_{\theta_i, \theta_j}\left[\jmath \left( \mathbf{A}^\mathsf{H}(\theta_i) \mathbf{A}(\theta_j) - \mathbf{A}^\mathsf{H}(\theta_j) \mathbf{A}(\theta_i) \right) \right].\label{eq:G2}
\end{align}
\end{subequations}
Notice the similarity between the previous case and this case. In the previous case, the matrices defining the BFIM correspond to the canonical basis for the space of Hermitian matrices.  For this example, the matrices correspond to a rotated coordinate system given by~\eqref{eq:G1}-\eqref{eq:G2}. It can be verified that the number of distinct defining $\mathbf{T}_{\boldsymbol{\alpha}, \boldsymbol{\alpha}}$ is $N_\text{tr}^2$. 

It can be verified that the matrix $\mathbf{T}_{\boldsymbol{\alpha}, \boldsymbol{\theta}} \in \mathbb{R}^{2N_\text{tr} \times N_\text{tr}}$ gives rise to $2N_\text{tr}^2$ distinct terms. In other words, it exhibits no particular structure in general. The matrix $\mathbf{T}_{\boldsymbol{\theta}, \boldsymbol{\theta}} \in \mathbb{R}^{N_\text{tr} \times N_\text{tr}}$ is symmetric, but also with no additional structure. It gives rise to $\tfrac{N_\text{tr} (N_\text{tr} + 1)}{2}$ terms. Summing all three terms, we obtain:
\begin{equation}
    d = N_\text{tr}^2 + 2N_\text{tr}^2 + \frac{N_\text{tr} (N_\text{tr} + 1)}{2} =
    \frac{7}{2} N_\text{tr}^2 + \frac{1}{2} N_\text{tr}.
\end{equation}

\subsubsection{Sensing-Only Scenario} 

Consider first the task of sensing the LoS channel parameters only. According to Theorem~\ref{thm:general_d_quad}, the minimum number of beamformers for estimating the parameters (path loss coefficients and AoAs) of $N_\text{tr}$ targets is bounded by
\begin{equation}\label{eq:radar_only_Ntarget}
    N_\text{min}^\text{rad} \leq \sqrt{d} = \left \lfloor \sqrt{\frac{7}{2}N_{\text{tr}}^2 + \frac{1}{2}{N_\text{tr}  }} \right \rfloor.
\end{equation}
The above result improves upon the $2N_\text{tr}$ bound in~\cite{Li2008range} derived under the special case of classical CRB. Interestingly, the bound~\eqref{eq:radar_only_Ntarget} shows that the minimum number of beamformers grows asymptotically \emph{at most} as $1.871N_\text{tr}$ for estimating $3N_\text{tr}$ parameters. 

When $N_\text{tr} = 1$, the above bound becomes $N_\text{min}^\text{rad} \leq 2$ beamformers. 
For jointly estimating the path loss coefficient and the AoA for one target, it is possible to show for some toy examples (e.g., with $N_\text{T} = 2$ and $N_\text{R} = 1$) that one sensing beamformer does not always attain the best possible performance. 
In other words, the above bound cannot be improved in general for $N_\text{tr} = 1$. 

When $N_\text{tr} > 1$, it is not clear whether the above upper bound can be improved. However, it is possible to construct examples for which at least $N_\text{tr}$ beamformers are needed.  
Consider a simple (albeit artificial) case of estimating $\boldsymbol{\alpha}$
of $N_\text{tr}$ targets when $\boldsymbol{\theta}$ is almost completely known and 
well separated.  When $N_\text{T} \gg N_\text{tr}$, the optimal beamforming scheme would 
involve $N_\text{tr}$ beamformers, one for each target direction.

\subsubsection{ISAC Scenarios} 
Next, we consider the ISAC scenario with $K$ communication users and $N_\text{tr}$ targets. For the scenario in which the interference of the sensing beams can be cancelled at the communication receivers, we have the following sum bound 
\begin{equation}\label{eq:N_traget_bound}
    N_\text{min}^\text{IC} \leq K + \left \lfloor \sqrt{\frac{7}{2}N_{\text{tr}}^2 + \frac{1}{2}{N_\text{tr}  }} \right \rfloor.
\end{equation}
This bound can yield a substantial reduction in the total number of beamformers needed as compared to the trivial bound of $K+N_\text{T}$, especially when $N_\text{tr}$ is small. For instance, if $N_\text{tr} = 1$, such a bound reduces to $ N_\text{min}^\text{IC} \leq K + 2$, i.e., at most two additional sensing beamformers are needed to achieve the same performance as a complete set of $N_\text{T} + K$ beamformers. 

For ISAC systems without interference cancellation, the bound on the minimum number of beamformers can be tightened further according to the hypotenuse bound
\begin{equation}\label{eq:N_traget_hypo_bound}
    N_\text{min}^\text{NIC} \leq \left \lfloor \sqrt{K^2 +   \frac{7}{2}N_\text{tr}^2 + \frac{1}{2} N_\text{tr} } \right \rfloor.
\end{equation}
We remark that the performance for the no interference cancellation case would be worse than the case with interference cancellation. However, as the bound for the former case
\eqref{eq:N_traget_hypo_bound} is less than that for the latter case \eqref{eq:N_traget_bound}, 
this suggests that adding beamformers is unlikely to help alleviate this performance gap. 


If we additionally assume that
\begin{equation}
    K \geq \frac{d}{2} = \frac{7}{4} N_\text{tr}^2 + \frac{1}{4} N_\text{tr},
\end{equation}
then Corollary~\ref{cor_Nmin=K_d_quad} applies. In this case, no extra sensing beamformers are needed. 

Interestingly, for the special case of one target $N_\text{tr} = 1$ and any $K$, the hypotenuse bound~\eqref{eq:N_traget_hypo_bound} provides an exact characterization of the worst-case $N_\text{min}^\text{NIC}$. In other words, there exist problem instances for which
\begin{equation}\label{eq:case_Np1}
    N_\text{min}^\text{NIC} =  \left \lfloor \sqrt{K^2 +   4} \right \rfloor = \begin{cases}
    2, & \text{if } K \in \{ 0, 1\} \\
    K, & \text{if } K \geq 2
\end{cases}.
\end{equation}
For $K = 0$ and $N_\text{tr} = 1$, it is possible to show numerically that at least two beamformers are needed for certain problem instances. 
For $K = N_\text{tr} = 1$, it is shown in \cite{chanisit2024} that $N_\text{min}^\text{NIC} = 2$ for certain problems.  Finally, by Corollary~\ref{cor_Nmin=K_d_quad}, we have $N_\text{min}^\text{NIC} = K$ for the case $K \geq 2$. 

Similar to the sensing-only case, it is unknown whether these sum bound~\eqref{eq:N_traget_bound} and hypotenuse bound~\eqref{eq:N_traget_hypo_bound} are tight for certain problem instances for all $N_\text{tr}$ and $K$. However, one can find problem instances for which the required number of beamformers is at least $\max\{K, N_\text{tr}\}$. This is because at least $K$ beamformers are needed for communication. and there are problem instances for which sensing alone requires $N_\text{tr}$ beamformers.

\subsubsection{Estimating AoA Only With Zero-Mean Path-Loss} 
The upper bounds~\eqref{eq:radar_only_Ntarget}-\eqref{eq:case_Np1} are all derived under the assumption that the goal is to estimate all path parameters. Tighter upper bounds can be developed if we further assume that: i) only the estimation of the AoAs (but not the path loss coefficients) is of interest; and ii) all parameters are independent and the path loss coefficients have zero mean. The first assumption implies that the sensing performance is measured as a function of the elements in the inverse BFIM corresponding to the estimation error in the AoAs only, i.e., it only involves the terms
\begin{equation}
    [\mathbf{J}_\mathbf{V}^{-1}]_{ii}, \quad i = 2N_\text{tr}+1, \ldots, 3N_\text{tr},
\end{equation}
where $\mathbf{J}_\mathbf{V}$ is given in~\eqref{eq:BFIM_N_targets} for the multi-target example. The second assumption implies that for $i = 2N_\text{tr}+1, \ldots, 3N_\text{tr}$, we have
\begin{equation}\label{eq:inverse_BFIM_zero_mean}
    [\mathbf{J}_\mathbf{V}^{-1}]_{ii} = \frac{1}{c_{ii} + \Tr(\mathbf{\tilde{G}}_{ii, 3} \mathbf{V} \mathbf{V}^\mathsf{H}) },
\end{equation}
where $c_{ii}$ is $i$-th diagonal entry of $\mathbf{C}$ and $\mathbf{\tilde{G}}_{ii, 3}$ is a Hermitian matrix given in the $i$-th diagonal entry of  $\mathbf{J}_\mathbf{V}$. The above relation follows from the fact that the BFIM elements sharing the same row and column as the diagonal elements at $i = 2N_\text{tr}+1, \ldots, 3N_\text{tr}$ are all zero. This holds under the assumptions that the path loss coefficients have zero mean, and all parameters are independent. 


It can be seen from~\eqref{eq:inverse_BFIM_zero_mean} that each term depends only on one quadratic term in this case. Therefore, the total number of quadratic terms relevant to estimating the AoAs is equal to $N_\text{tr}$, instead of $ \tfrac{7}{2}N_\text{tr}^2 + \tfrac{1}{2} N_\text{tr}$ for the general case. Consequently, under the above assumptions, the number of sensing beamformers for MIMO radar becomes
\begin{equation}\label{eq:sc_sensing_only}
     N_\text{min}^\text{rad} \leq \left \lfloor \sqrt{N_\text{tr}} \right \rfloor.
\end{equation}
For the ISAC setup, the sum and hypotenuse bounds become
\begin{subequations}\label{eq:sc_ISAC}
\begin{align}
     N_\text{min}^\text{IC} \leq K + \left \lfloor \sqrt{N_\text{tr}} \right \rfloor, \label{eq:IC_multitarget} \\
     N_\text{min}^\text{NIC} \leq \left \lfloor \sqrt{K ^2 + N_\text{tr}} \right \rfloor. \label{eq:NIC_multitarget}
\end{align}
\end{subequations}

The recent work \cite{yao2025optimal} derives the bound 
\eqref{eq:IC_multitarget} but for the no interference cancellation 
scenario. In this paper, we demonstrate that for the no interference 
cancellation scenario, the bound in \cite{yao2025optimal} 
can be improved to \eqref{eq:NIC_multitarget}.
Furthermore, \cite{yao2025optimal} does not treat the case 
when interference cancellation is possible. Here we show that 
\eqref{eq:IC_multitarget} is the applicable bound in this case.


We remark that the assumption of zero mean for the path loss is 
a strong one, especially when estimation can occur over multiple
stages. While the zero-mean assumption is reasonable at the initial 
stage when nothing is known about the pathloss, such assumption is
typically no longer valid as soon as some observations are made and 
the prior distribution is updated to account for these observations.
In this case, the mean is typically no longer zero and the applicable
bounds for this more general case are
\eqref{eq:N_traget_bound} and \eqref{eq:N_traget_hypo_bound}.

\subsubsection{Numerical Results}

Next, we use numerical simulations to assess the utility of the derived bounds. 
Recall that the derived bounds hold across all problem instances, a natural question to ask is whether there exist examples for which these bounds can predict the actual number of beamformers needed. The numerical simulations in this section show that there are instances where the bounds are tight, particularly when $N_\text{tr}$ is small or $K$ is large as compared to $N_\text{tr}$.

We first consider the case where both the path loss coefficients and the AoAs are to be estimated. This general case is important because estimating the path loss coefficients is essential for target identification. Furthermore, we do not make the assumption that the path loss coefficients have zero mean. As mentioned before, the zero-mean assumption may not hold in many cases, e.g., the active sensing setting \cite{active_sensing_1, active_sensing_2}, where the availability of prior information can skew the mean of the distribution.

\begin{figure}
    \centering
\includegraphics[width=0.45\textwidth]{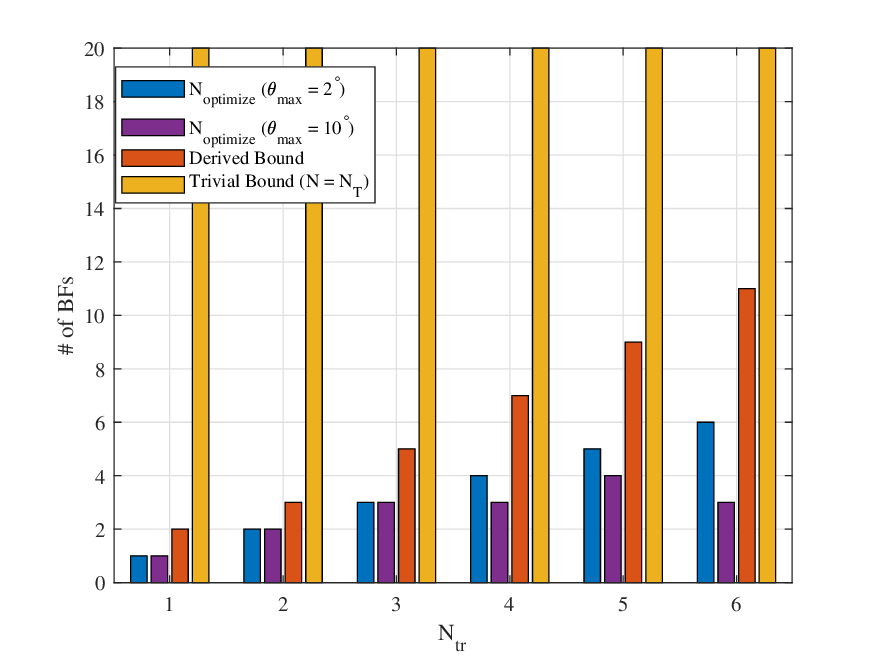}
    \caption{The number of sensing beamformers obtained by simulations with different priors versus the bound for sensing the parameters of a channel with $N_\text{tr}$ LoS paths. Here, $N_\text{T} = N_\text{R} = 20$ and $K = 0$.}
    \label{fig:bound_vs_optimization}
\end{figure}

\begin{figure*}
    \centering
\includegraphics[width=0.67\linewidth]{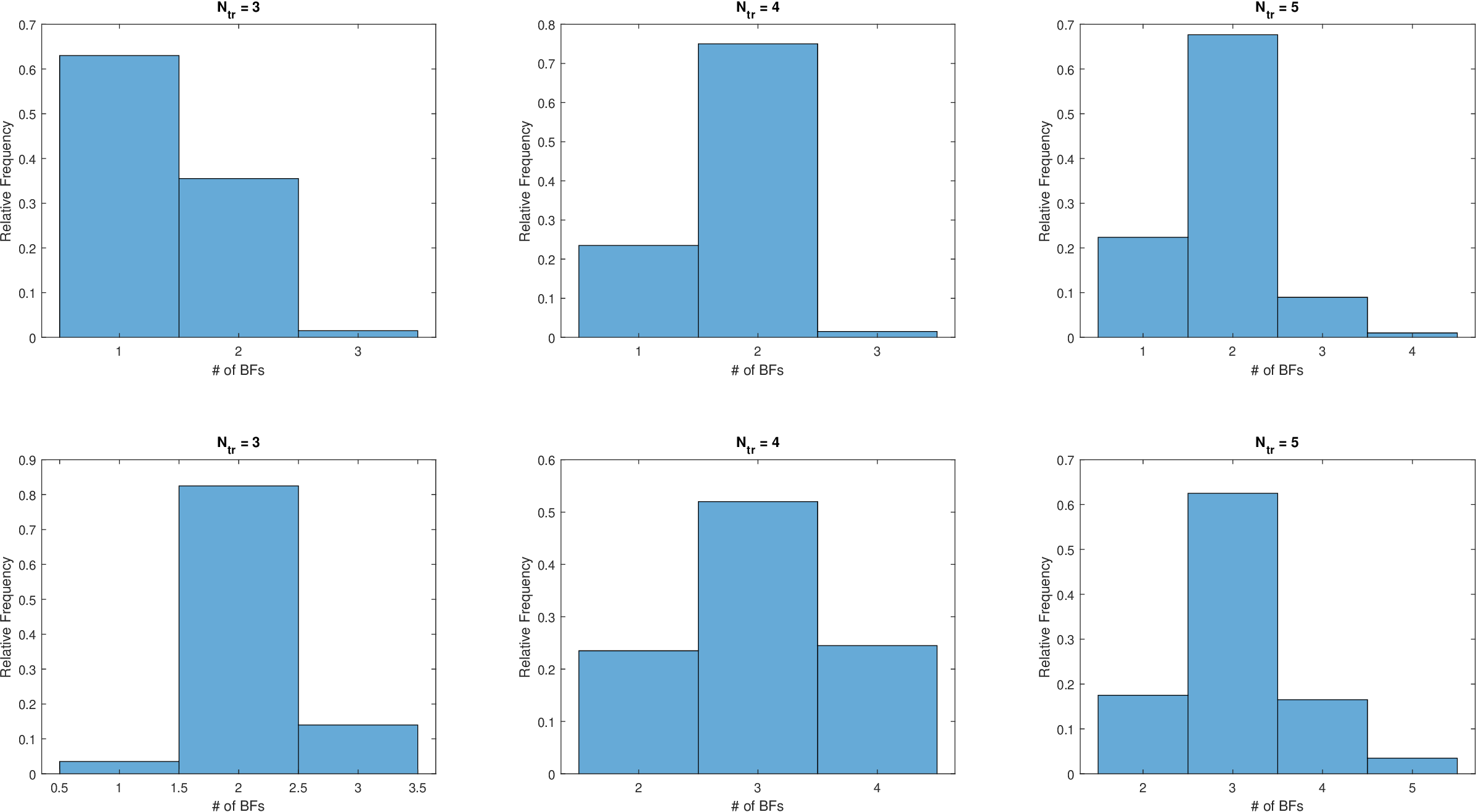}
    \caption{Histograms of the number of sensing beamformers obtained by simulation for sensing the parameters of a channel with $N_\text{tr}$ LoS paths. The top and bottom rows correspond to the different priors with $\theta_{max} = 10^\circ$ and $\theta_{max} = 2^\circ$, respectively. The number of transmit and receive antennas are $N_\text{T} = N_\text{R} = 20$ and $K = 0$.}
    \label{fig:histo}
\end{figure*}

To this end, we examine a setup where $N_\text{tr}$ targets are located at angle values selected randomly from the set of values given by $\sin^{-1}\left({\tfrac{2 j}{N_\text{T}}}\right)$ for some integer $j \in \{-\tfrac{N_\text{T}}{2} + 1, \ldots, \tfrac{N_\text{T}}{2}\}$. For uniform linear arrays, this method of selecting the angles makes the steering vectors align with Fourier transform vectors. The prior distributions of the AoAs and path loss coefficients are randomly selected from a Gaussian family. The prior distributions of the path loss coefficients are $\mathcal{CN}(\mu_i, \sigma_i^2)$, where the means and variances are selected uniformly at random. The AoAs of different targets are distributed as $\theta_i \sim \mathcal{N}\left(\sin^{-1}\left({\tfrac{2 j}{N_\text{T}}}\right), \sigma_{\theta_i}^2\right)$ for $i = 1, \ldots, N_\text{tr}$, where the standard deviations are chosen uniformly at random on the interval $[0.5^\circ, \theta_\text{max}]$.  Note that the value of $\theta_\text{max}$ controls the degree to which the prior distributions overlap, since it determines the width of the Gaussian distribution associated with each AoA.

First, consider a sensing-only scenario (i.e., with no communication users) and the objective is to minimize the maximum diagonal element in the BCRB matrix. This amounts to minimizing a lower bound on the maximum MSE in estimating all parameters (i.e., including both the path loss coefficients and the AoAs associated with the $N_\text{tr}$ targets). 
Under moderate-and-high SNRs and reasonable prior distributions for the parameters of interest, the BCRB can be a good approximation to the actual MSE for maximum a posterior estimation in this problem setting, thus justifying its use.

We obtain the minimum number of beamforming vectors required by examining the rank of the SDR solution, then applying the rank-reduction methods used to prove the theorems in Section~\ref{sec:IC} to the SDR solution when necessary. Because the number of beamformers can fluctuate based on the prior distributions, we repeat the experiment $200$ times and record the largest number of beamformers across the different Monte Carlo simulations. Such a maximum value is denoted by $N_\text{optimize}$.

\begin{figure*}[!t]
    \centering
    \begin{minipage}[b]{0.45\textwidth}
\includegraphics[width=\textwidth]{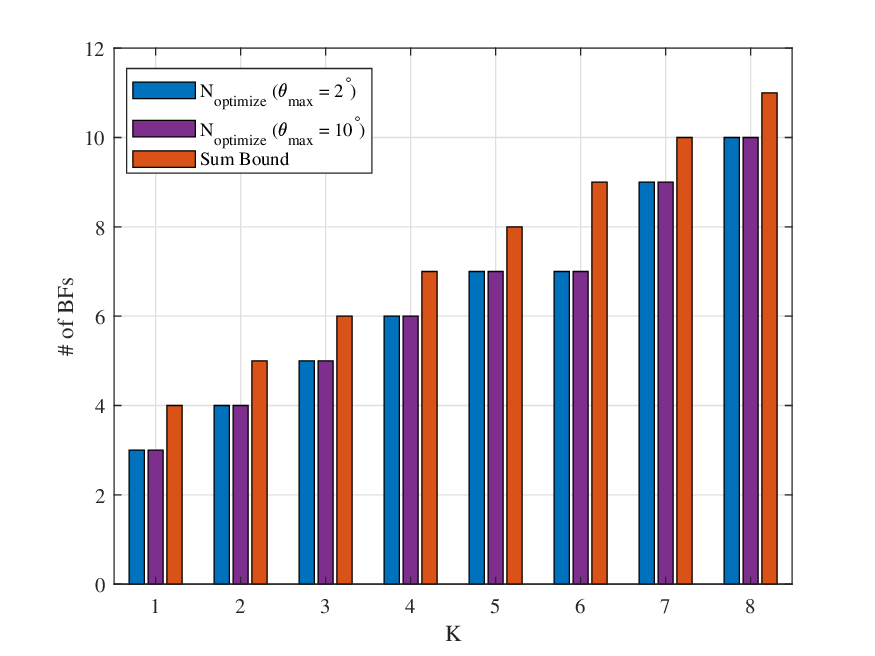}
    \caption{The number of beamformers obtained from simulations versus the upper bounds for different $K$ for an ISAC system with interference cancellation. Here, $N_\text{T} = N_\text{R} = 20$ and $N_\text{tr} = 2$.}
    \label{fig:bound_vs_optimization_diff_k_IC}
    \end{minipage}
\hfill
\begin{minipage}[b]{0.45\textwidth}
\includegraphics[width=\textwidth]{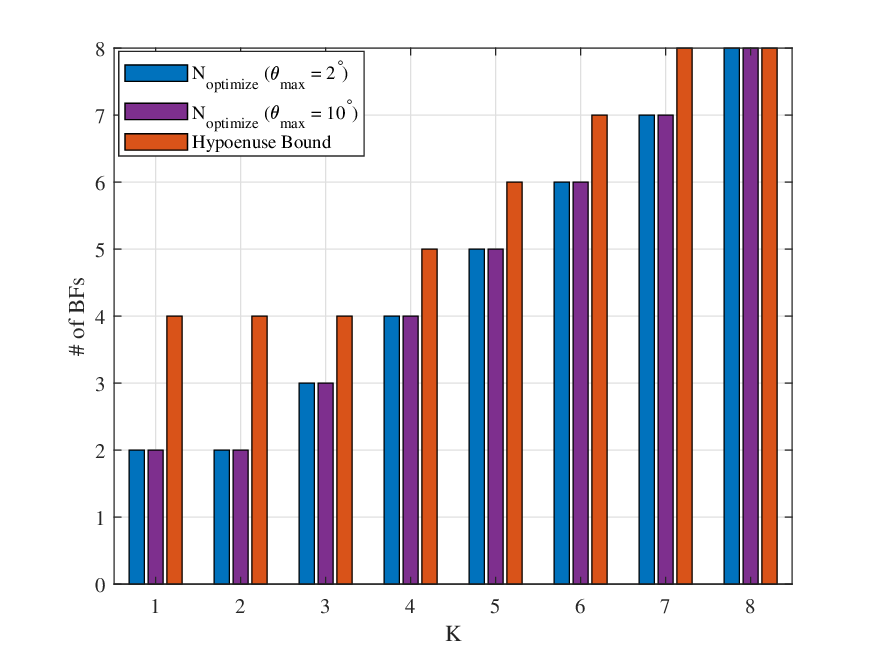}
    \caption{The number of beamformers obtained from simulations versus the upper bounds for different $K$ for an ISAC system without interference cancellation. Here, we set $N_\text{T} = N_\text{R} = 20$ and $N_\text{tr} = 2$.}
    \label{fig:bound_vs_optimization_diff_k}
\end{minipage}
\end{figure*}

\begin{figure*}[!t]
    \centering
    \begin{minipage}[b]{0.45\textwidth}
\includegraphics[width=\textwidth]{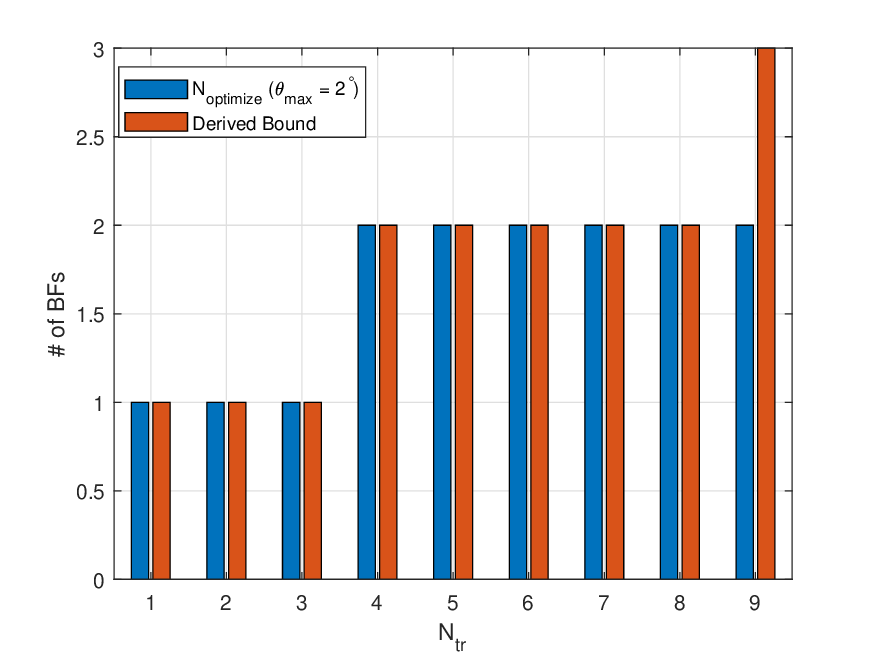}
    \caption{The number of sensing beamformers obtained from simulations versus the upper bounds for the special case of AoA estimation only. Here, $N_\text{T} = N_\text{R} = 20$ and $K = 0$.}
    \label{fig:bound_vs_optimization_diff_ntr_special_case}
    \end{minipage}
\hfill
     \begin{minipage}[b]{0.45\textwidth}
\includegraphics[width=\textwidth]{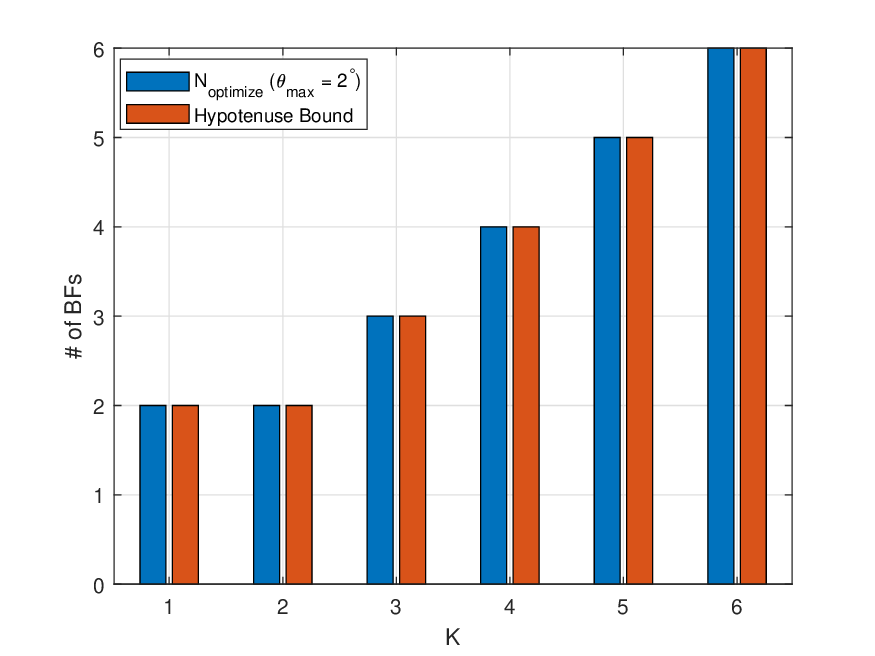}
    \caption{The number beamformers obtained from simulations versus the upper bounds for the special case of AoA estimation. Here, $N_\text{T} = N_\text{R} = 20$ and $N_\text{tr} = 4$ and assume no interference cancellation.}
\label{fig:bound_vs_optimization_diff_k_special_case}
\end{minipage}
\end{figure*}


In~\figurename~\ref{fig:bound_vs_optimization}, we compare $N_\text{optimize}$ with the bound \eqref{eq:radar_only_Ntarget} for different values of $N_\text{tr}$ for $N_\text{T} = N_\text{R} = 20$ (with $K = 0$). The transmit power is set to $P = 10$. Here, two cases are considered with $\theta_\text{max} = 2^\circ$ and $\theta_\text{max} = 10^\circ$. Note that the former case of $\theta_\text{max} = 2^\circ$ corresponds to the situation where the AoA priors have limited overlap, because the average angular spacing between two Fourier vectors is approximately $8^\circ$. 
On the other hand, there is a substantial overlap in the priors of the AoA when $\theta_\text{max} = 10^\circ$. For illustration purposes, we also include the trivial bound $N = N_\text{T}$ in the comparison. 


\figurename~\ref{fig:bound_vs_optimization} illustrates that the developed bound~\eqref{eq:radar_only_Ntarget} for jointly estimating the path loss coefficients and AoAs constitutes an upper bound on $N_\text{optimize}$ for both cases when $\theta_\text{max} = 2^\circ$ and $\theta_\text{max} = 10^\circ$. 
But, there is a gap between the bound and $N_\text{optimize}$, which indicates that the scaling of the bound is higher than the actual scaling. 
Nevertheless, the developed bound is still superior to the trivial bound of $N_\text{T}$. 

We remark here that even at $\theta_\text{max} = 2^\circ$, in which case the Fisher information of the prior distribution constitutes a significant portion of the BFIM, the observations obtained from the optimized beamformers can still contribute much more to the estimation process than the 
the prior information. 
Thus, beamforming optimization is still meaningful and can significantly refine the estimate of azimuth angle, even if the prior already provides useful information about the angle. 

\figurename~\ref{fig:bound_vs_optimization} suggests that fewer beamformers are needed when there is an overlap between the priors of different AoAs. To further confirm this, we plot the histograms of the number of beamformers required by Monte Carlo simulations in~\figurename~\ref{fig:histo}. The figure shows that the case of $\theta_\text{max} = 2^\circ$ (shown in the bottom row for different values of $N_\text{tr}$) tends to yield a larger number of beamformers more frequently than the case of $\theta_\text{max} = 10^\circ$ (top row). This behavior is understood by noting that when there is overlap between two priors, a single beamformer can be used to cover the entire region of interest. But when the AoAs have nonoverlapping priors, separate beamformers must be used to sense the individual regions in the angular domain. 

We also observe that the upper bounds derived in this paper correspond to the worst-case scenarios across all system configurations; they do not serve as an accurate predictor for the \emph{typical} number of beamformers needed, as most scenarios in Fig.~\ref{fig:histo} only require $2$-$3$ beamformers even though the upper bound is much larger.


Next, we examine how well the derived bounds predict $N_\text{optimize}$ for this example with varying $K$ and fixed $N_\text{tr} = 2$. We set the SINR targets to $5$ dB, and generate the communication channels randomly where the entries are drawn from a Gaussian distribution $\mathcal{CN}(0, 1)$. \figurename~\ref{fig:bound_vs_optimization_diff_k_IC} compares $N_\text{optimize}$ against the sum bound for the interference cancellation case. It is observed that the sum bound is a relatively tight predictor of the minimum number of beamformers over a wide range of $K$ as long as $N_\text{tr}$ is small.

In~\figurename~\ref{fig:bound_vs_optimization_diff_k}, the comparison is repeated against the hypotenuse bound for the no interference cancellation case. 
It can be observed that the derived bound becomes exact for large $K$. This is because Corollary~\ref{cor_Nmin=K_d_quad} applies for $K \geq 8$ in this case. 
Comparing \figurename~\ref{fig:bound_vs_optimization_diff_k} with  \figurename~\ref{fig:bound_vs_optimization_diff_k_IC}, 
it can be seen that indeed the no interference cancellation case
requires fewer beamformers than the interference cancellation case.

%

Finally, we study the special case of estimating the AoAs only (but not the path loss coefficients), and assume that the path loss coefficients have zero mean. We compare the bounds~\eqref{eq:sc_sensing_only}-\eqref{eq:sc_ISAC} to the number of beamformers obtained from an optimization that aims to minimize the maximum error in estimating the AoAs for all targets. The comparison is shown in for $K = 0$ and varying $N_\text{tr}$ in~\figurename~\ref{fig:bound_vs_optimization_diff_ntr_special_case} and $N_\text{tr} = 4$ and varying $K$ in~\figurename~\ref{fig:bound_vs_optimization_diff_k_special_case}. In this special case, it can be seen from the figures that the derived bounds tend to be tighter than the general case. For the case of $N_\text{tr} = 4$, the analysis predicts that hypotenuse bound becomes tight for $K \geq \frac{N_\text{tr}}{2} = 2$. In~\figurename~\ref{fig:bound_vs_optimization_diff_k_special_case}, it is shown that the bound provides a tight characterization of the worst-case scaling for all $K$.



\subsection{Target Detection with SNR and SCNR}
Consider the monostatic ISAC system of Section~\ref{sec:mdl} but now assume that the radar task is that of detecting a single target along some known direction given by $\theta_0$. If the target is present, the signal received at the radar receiver is given by
\begin{equation}
    \mathbf{y}_\text{s}[q] = \alpha \mathbf{A}(\theta_0) \mathbf{x}[q] + \mathbf{z}_\text{s}[q], \quad q = 1, \ldots, \Upsilon
\end{equation}
where $\alpha$ is an unknown path loss coefficient and $\mathbf{A}(\cdot) \in \mathbb{C}^{N_\text{T} \times N_\text{R}}$ is the combined array response. The received signal is  the noise vector if the target is absent. The radar decides whether the target is present based on the generalized likelihood ratio test (GLRT), whose detection probability is a monotone function of the radar SNR
\begin{align}
\gamma_r(\mathbf{V}) &\triangleq  \frac{\sum_q \mathbb{E} \left[ \mathbf{x}^\mathsf{H} [q]\mathbf{A}^\mathsf{H}(\theta_0) \mathbf{A}   (\theta_0) \mathbf{x}[q] \right] }{\sigma^2} \nonumber \\
&= \frac{\Upsilon\Tr\left( \mathbf{A}^\mathsf{H}(\theta_0) \mathbf{A}   (\theta_0) \mathbf{V} \mathbf{V}^\mathsf{H} \right) }{\sigma^2}. \label{eq:radar_SNR}
\end{align}
The radar SNR is a $d$-quadratic function with $d = 1$ since it is characterized in terms of a single quadratic of the beamforming matrix.

For an ISAC system with interference cancellation of the sensing 
beams at the communications users, we can apply 
Theorem~\ref{thm:general_d_quad} to obtain the sum bound
\begin{equation}\label{eq:bd_1}
    N_\text{min}^\text{IC} \leq K + 1.
\end{equation}
For $K = 0$, the above bound is tight since we need at least one beamformer. For $K = 1$, this problem has been studied in \cite{salman}, where it is shown that a single beamformer is sufficient. For $K > 1$, we can use the minimum rank bounds of~\cite{huangrank2010} to show that $N_\text{min}$ is exactly $K$. 
Consequently, the sum bound above is loose by one beamformer as compared to the true minimum for $K \geq 1$. 

If we assume that the users do not cancel the interference from sensing, we obtain the hypotenuse bound
\begin{equation}\label{eq:bd_2}
    N_\text{min}^\text{NIC} \leq \left \lfloor \sqrt{K^2 + 1} \right \rfloor  = \begin{cases}
        1, & \text{if} ~ K = 0 \\  K, & \text{if} ~ K > 0.
    \end{cases}
\end{equation}
Since $K$ beamformers are needed to communicate to $K$ users and 
at least one beamformer is always needed for sensing, it can be readily
seen that the above bound is tight. 
This result demonstrates that no extra sensing beamformers are required when the sensing performance is measured using the SNR (or any single quadratic term in $\mathbf{V}$). This is consistent with the analysis of~\cite{mateen2023}.

Next, we examine an extension to the previous scenario where clutter is present. In this case, the signal received at the radar receiver is given by
\begin{equation}
    \mathbf{y}_\text{s}[q] = 
\begin{cases}
\alpha_0 \mathbf{A}(\theta_0) \mathbf{x}[q] + \mathbf{B} \mathbf{x}[q] + \mathbf{z}_\text{s}[q] & \text{if target exists} \\
\mathbf{B} \mathbf{x}[q] + \mathbf{z}_\text{s}[q] & \text{otherwise} \\
\end{cases}
\end{equation}
where $\mathbf{B}$ is a random clutter term with known statistics.
In this case, the performance of the GLRT is determined by the SCNR \cite{xiewc2022}
\begin{equation}\label{eq:radar_SCNR}
\tilde{\gamma}_r(\mathbf{V}) \triangleq  \frac{\Upsilon \Tr\left( \mathbf{A}^\mathsf{H}(\theta_0) \mathbf{W}^\mathsf{H} \mathbf{W} \mathbf{A}   (\theta_0) \mathbf{V} \mathbf{V}^\mathsf{H} \right)}{\Upsilon \Tr\left(\mathbb{E}\left[\mathbf{B}^\mathsf{H} \mathbf{W}^\mathsf{H} \mathbf{W} \mathbf{B}\right] \mathbf{V} \mathbf{V}^\mathsf{H} \right) + \sigma^2 \Tr\left( \mathbf{W}^\mathsf{H} \mathbf{W} \right)},
\end{equation}
where $\mathbf W$ is a receiver combiner that can be optimized. The radar SCNR is a $d$-quadratic function with $d = 2$, because there is a quadratic term in the numerator and the denominator.

Even though $d = 2$ here, we can treat this case as if $d = 1$. This is because instead of fixing both quadratic terms of the SCNR separately, we can fix the SCNR itself, i.e.,
\begin{equation}
\frac{\Upsilon \Tr\left( \mathbf{A}^\mathsf{H}(\theta_0) \mathbf{W}^\mathsf{H} \mathbf{W} \mathbf{A}   (\theta_0) \mathbf{V} \mathbf{V}^\mathsf{H} \right)}{\Upsilon \Tr\left(\mathbb{E}\left[\mathbf{B}^\mathsf{H} \mathbf{W}^\mathsf{H} \mathbf{W} \mathbf{B}\right] \mathbf{V} \mathbf{V}^\mathsf{H} \right) + \sigma^2 \Tr\left( \mathbf{W}^\mathsf{H} \mathbf{W} \right)} = c,
\end{equation}
which can be rearranged into a single quadratic as follows:
\begin{equation}
    \Tr\left( \mathbf{Q} \mathbf{V} \mathbf{V}^\mathsf{H} \right) = \sigma^2 \Tr\left( \mathbf{W}^\mathsf{H} \mathbf{W} \right),
\end{equation}
with $\mathbf{Q} = \tfrac{\Upsilon}{c}\mathbf{A}^\mathsf{H}(\theta_0) \mathbf{W}^\mathsf{H} \mathbf{W} \mathbf{A}   (\theta_0) - \Upsilon \Tr\left(\mathbb{E}\left[\mathbf{B}^\mathsf{H} \mathbf{W}^\mathsf{H} \mathbf{W} \mathbf{B}\right] \right)$. 
For fixed $\mathbf{W}$, to guarantee a certain performance level, this equation can be held fixed, along with the $K$ SINR constraints, when using the rank-reduction algorithms of Theorem~\ref{thm:BCRB_IC} and Theorem~\ref{thm:BCRB_NIC}. Since this equation involves only a single quadratic, it means that the bounds~\eqref{eq:bd_1}-\eqref{eq:bd_2} also hold here. 
Now since this argument holds for arbitrary $\mathbf{W}$, it must hold for the \emph{optimal} $\mathbf{W}$ as well. 
Thus, even with the clutter term the minimum number of beamformers are upper bounded by \eqref{eq:bd_1}-\eqref{eq:bd_2}.
Again, this is consistent with the analysis of~\cite{mateen2023}.

\subsection{Probing with Beam Pattern Matching}

Consider an ISAC system where the sensing task is that of probing certain locations of interest (see~\cite{soticaprobing2007, Fried20212on, Liu2020joint}), for which the problem can be formulated as that of finding a set of beamformers to approximate a desired transmit beam pattern. \editrev{Assuming $K$ communication users with radar interference cancellation, this problem can be expressed as \cite{HuaOptimal2023, Liu2020joint}:
\begin{subequations}\label{prob:beam_pattern}
\begin{align}
	\underset{\mathbf{V} \in \mathbb C^{N_\text{T}\times N}, ~\xi }{\mathrm{minimize}} ~~~ &\sum_{n = 1}^{N_\textup{g}} |\mathbf{a}_\textup{T}^\mathsf{H}(\theta_n) \mathbf{V} \mathbf{V}^\mathsf{H} \mathbf{a}_\textup{T}(\theta_n) - \xi t_n|^2 \\
	\mathrm{subject\ to} ~~ & \text{SINR}^\text{IC}_{k, \mathbf{V}} \geq \gamma_k, \\
    &\Tr\left( \mathbf{V}\mathbf{V}^\mathsf{H} \right) = P.
\end{align}
\end{subequations}
where the objective function corresponds to   
matching some desired beam pattern given by $t_1, \ldots, t_{N_\textup{g}}$ up to some scaling factor $\xi$, with $\theta_1, \ldots, \theta_{N_\textup{g}}$ denoting grid angles and $\mathbf{a}_\textup{T}(\cdot)$ denoting the transmit array steering vector. The scaling factor $\xi$ is to be optimized, because typically the goal is to approximate only the shape of the desired beam pattern and not the exact magnitude \cite{soticaprobing2007}. 
The values of $\{t_n\}$ and $\{\theta_n\}$ are known in advance. 

The beam pattern matching problem~\eqref{prob:beam_pattern} differs from the previous examples in two aspects. First, $\mathbf{V}$ is not the only optimization variable; $\xi$ is also to be optimized. Second, the power constraint is enforced with an equality to ensure that all available power is utilized for illuminating the directions of interest (otherwise, a power constraint with inequality would give rise to a trivial solution $\mathbf V = \mathbf 0$ and $\xi=0$). Despite these differences, we show next that the analysis in this paper remains useful for deriving bounds on the minimum number of beamformers for this type of problems.


First, note that for every fixed $\xi$, the objective of \eqref{prob:beam_pattern} is $d$-quadratic, as it depends on the beamformers only through $N_\text{g}$ quadratic terms $\Tr\left(\mathbf{a}(\theta_1) \mathbf{a}^\mathsf{H}(\theta_1) \mathbf{V} \mathbf{V}^\mathsf{H}\right)$, $\cdots$ $\Tr\left(\mathbf{a}(\theta_{N_\textup{g}}) \mathbf{a}^\mathsf{H}(\theta_{N_\textup{g}}) \mathbf{V} \mathbf{V}^\mathsf{H}\right)$. 
Note that the set of matrices $\{\mathbf{a}(\theta_n) \mathbf{a}^\mathsf{H}(\theta_n)\}$ is not necessarily linearly independent. 
This is because the matrix of type $\mathbf{a}(\theta_n)\mathbf{a}^\mathsf{H}(\theta_n) \in \mathbb C^{N_\text{T} \times N_\text{T}}$ has a rank of one, so the maximum cardinality of the set of linearly independent such matrices is at most $N_\text{T}$. Thus, we have $d \le \min\{N_\text{T}, N_\text{g}\}$.


Now, let $(\mathbf{V}^*, \xi^*)$ be an optimal solution of \eqref{prob:beam_pattern}
for $N= N_\text{T}+K$. Such an optimal solution can be obtained by solving its SDR (see~\cite{HuaOptimal2023, Liu2020joint}). We can perform the rank reduction procedure on the optimal solution, while maintaining the same transmit power by requiring
\begin{equation}\label{eq:pw_beampattern}
\Tr\left(\mathbf{\hat V}\mathbf{\hat V}^\mathsf{H} \right) = P,
\end{equation}
and the same optimal objective value by fixing the $N_g$ quadratic terms, i.e.,
\begin{align}\label{eq:beampattern}
    \Tr\left(\mathbf{a}(\theta_n) \mathbf{a}^\mathsf{H}(\theta_n) \mathbf{\hat{V}} \mathbf{\hat{V}}^\mathsf{H}\right) = \Tr\left(\mathbf{a}(\theta_n) \mathbf{a}^\mathsf{H}(\theta_n) \mathbf{V}^* {\mathbf{V}^*}^\mathsf{H}\right), \notag \\ \forall n=1,\cdots, N_g
\end{align}
where the right-hand side of the above consists of scalars that can be computed from $\mathbf{V}^*$, and $\mathbf{\hat V}$ is the new beamforming matrix with potentially fewer sensing beamformers.
Note that when the set of matrices $\{\mathbf{a}(\theta_n) \mathbf{a}^\mathsf{H}(\theta_n)\}$ is linearly dependent, some of the equations can be removed from \eqref{eq:beampattern}. In other words, \eqref{eq:beampattern} is equivalent to a set of $d$ independent equations.

Together, \eqref{eq:pw_beampattern}-\eqref{eq:beampattern}
correspond to $d + 1$ quadratic equations that must be held fixed \emph{in addition} to the $K$ SINR constraints. By iteratively applying the transformation~\eqref{eq:transformation_IC}-\eqref{eq:transformation_IC2} to reduce the number of sensing beamformers, it is easy to see that the following bound holds on the minimum number of beamformers for the interference cancellation case:
\begin{equation}
     N_\text{min}^\text{IC} \leq \left \lfloor K + \editrev{\sqrt{d + 1}} \right \rfloor,
\end{equation}
where the extra plus one inside the square root (cf. \eqref{eq:mod_sum_bound}) is due to the extra power constraint \eqref{eq:pw_beampattern}.
 
For the no interference cancellation cases, we can define an analogous problem as \eqref{prob:beam_pattern} and use the transformation \eqref{eq:lin_tran_NIC_comm}-\eqref{eq:lin_tran_NIC_sens} iteratively to reduce the number of beamformers. This yields the following bound:
\begin{equation}
N_\text{min}^\text{NIC} \leq \left \lfloor  \editrev{\sqrt{K^2 + d + 1}} \right \rfloor,
\end{equation}
where again the extra plus one inside the square root (cf. \eqref{eq:mod_hypo_bound}) is due to the power constraint \eqref{eq:pw_beampattern}.
}

To see the usefulness of the above bounds, consider an example with $N_\text{T} = 256$ antennas and $N_\text{g} = 120$ (e.g., for $1^\circ$ separation over $[-60^\circ, 60^\circ]$). 
The trivial bound requires optimizing over $K + 256$ beamformers---many of which are likely to be zero after optimization. 
In contrast, the bounds developed herein demonstrate that at most $K + \lfloor  \editrev{\sqrt{N_\text{g} + 1}} \rfloor = K + \editrev{11}$ beamformers are needed for probing in ISAC systems with interference cancellation. For ISAC systems without radar interference cancellation, the minimum number of beamformers is bounded by $\left \lfloor  \sqrt{K^2 + \editrev{121}} \right \rfloor$. 


\section{Conclusion}\label{sec:conc}
This paper derives concise and useful bounds on the minimum number of beamformers required for integrated sensing and communication. For the parameter estimation scenario where the users can cancel the interference caused by the sensing beams, we prove that the minimum number of beamformers is roughly bounded by the number of communication users plus a function linear in the number of sensing parameters. A tighter bound can be developed if we further assume that the users cannot cancel the interference from sensing. The theoretical analysis of this paper is not limited to parameter estimation; it can be applied to a large family of sensing metrics that have quadratic dependence on the beamformers. Several examples demonstrate that the developed bounds are tight in several cases of practical interest.

\appendices

\section{Derivation of the BFIM expression}
\label{Appen:BFIM}
First, note that we may factor $f(\mathbf{X}, \mathbf{Y}_\text{s}, \boldsymbol{\eta})$ as follows
\begin{equation}\label{eq:pdf_factorization}
    f(\mathbf{X}, \mathbf{Y}_\text{s}, \boldsymbol{\eta}) = f(\boldsymbol{\eta}) \prod_{q = 1}^{\Upsilon} f(\mathbf{x}[q]) f\left(\mathbf{y}_\text{s}[q] \big{|} \mathbf{x}[q],  \boldsymbol{\eta}\right),
\end{equation}
where the above relation follows directly by noting that the random vectors $\boldsymbol{\eta}, \mathbf{x}[1], \ldots, \mathbf{x}[\Upsilon]$ are independent of each other, and that $\mathbf{y}_\text{s}[q]$ depends only on the current $\mathbf{x}[q]$ and $\boldsymbol{\eta}$. Using~\eqref{eq:pdf_factorization}, we can write the BFIM elements in~\eqref{eq:BFIM} as the following sum $[\mathbf{J}_{\mathbf{V}}]_{ij} = [\mathbf{C}]_{ij} + [\mathbf{T}_{\mathbf{V}}]_{ij}$, where $\mathbf{C}$ and $\mathbf{T}_{\mathbf{V}}$ are chosen as follows: 
\begin{equation}~\label{eq:C_elements}
        [\mathbf{C}]_{ij} = - \mathbb{E}\left[ \frac{\partial^2 \log\left( f(\boldsymbol{\eta})\cdot  \prod_{q = 1}^{\Upsilon} f(\mathbf{x}[q]) \right) }{\partial \eta_i \partial \eta_j }\right]
\end{equation}
and
\begin{equation}~\label{eq:T_elements}
[\mathbf{T}_\mathbf{V}]_{ij} = -  \mathbb{E}\left[ \frac{\partial^2 \log \prod_{q = 1}^\Upsilon f(\mathbf{y}_\text{s}[q]\big{|} \mathbf{x}[q], \boldsymbol{\eta}) }{\partial \eta_i \partial \eta_j }\right]. 
\end{equation}
Since $\mathbf{x}[q]$ is independent of $\boldsymbol{\eta}$, we can simplify~\eqref{eq:C_elements} as follows
\begin{subequations}
    \begin{align}
        [\mathbf{C}]_{ij} &= - \mathbb{E}\left[ \frac{\partial^2 \log f(\boldsymbol{\eta}) }{\partial \eta_i \partial \eta_j }\right] - \sum_{q = 1}^{\Upsilon} \mathbb{E}\left[ \frac{\partial^2 \log f( \mathbf{x}[q] ) }{\partial \eta_i \partial \eta_j }\right] \label{eq:C_linea}  \\
        &= - \mathbb{E}\left[ \frac{\partial^2 \log f(\boldsymbol{\eta}) }{\partial \eta_i \partial \eta_j }\right],
       \end{align} \end{subequations}
since the second term in~\eqref{eq:C_linea} is zero due to the independence of $\boldsymbol{\mathbf{x}[q]}$ and $\boldsymbol{\eta}$. This shows that $\mathbf{C}$ has the desired form. For $\mathbf{T}_{\mathbf{V}}$ in~\eqref{eq:T_elements}, it is easy to see that $f(\mathbf{y}_\text{s}[q]\big{|} \mathbf{x}[q], \boldsymbol{\eta})$ is a Gaussian distribution $\mathcal{CN}\left(\mathbf{G}^{(\boldsymbol{\eta})} \mathbf{x}[q], \sigma^2 \mathbf{I}_{N_\text{R}}\right)$. Using the law of total expectation, we can write
\begin{subequations}
\begin{align}
    [\mathbf{T}_{\mathbf{V}}]_{ij} &{=} \sum_{q = 1}^\Upsilon \mathbb{E}_{\mathbf{x}[q], \boldsymbol{\eta}} \Biggl[ \mathbb{E}\Biggl[  \frac{-\partial^2 \log f(\mathbf{y}_\text{s}[q]\big{|} \mathbf{x}[q], \boldsymbol{\eta}) }{\partial \eta_i \partial \eta_j } \Biggr| \mathbf{x}[q], \boldsymbol{\eta} \Biggr] \Biggr] \\ 
	&{=} \frac{2}{\sigma^2} \sum_{q = 1}^\Upsilon \mathbb{E}_{\mathbf{x}[q], \boldsymbol{\eta}} \Re\biggl\{\Tr\left( \dot{\mathbf{G}}_i^\textsf{H} \dot{\mathbf{G}}_j \mathbf{x}[q] \mathbf{x}^\textsf{H}[q]\right) \biggr\} \label{eq:T_V_line_b} \\
    &= \frac{\Upsilon}{\sigma^2} \Tr\left( \mathbb{E}_{\boldsymbol{\eta}} \left[ \dot{\mathbf{G}}_i^\textsf{H} \dot{\mathbf{G}}_j + \dot{\mathbf{G}}_j^\textsf{H} \dot{\mathbf{G}}_i\right] \mathbf{V}\mathbf{V}^\textsf{H}\right) \label{eq:T_V_line_c}
\end{align}
\end{subequations}
where~\eqref{eq:T_V_line_b} follows from the BFIM of Gaussian signals~\cite{kay1993fundamentals} and~\eqref{eq:T_V_line_c} uses the fact that $\mathbb{E}\left[\mathbf{x}[q] {\mathbf{x}[q]}^\textsf{H}\right] = \mathbf{V} \mathbf{V}^\textsf{H}$.

\section{Proof of Lemma~\ref{lem:quad_eq_sol_IC}}\label{appen:quad_eq_sol_IC} 

The first step is to express the BFIM and SINR constraints in~\eqref{eq:BFIM_SINR} as quadratic equations in $\left\{d_k\right\}$ and $\mathbf{U}_\text{s}$. For the BFIM constraint, we have the following $L(L + 1)/2$ quadratic equations in $\mathbf{V}'$ due to the symmetric nature of $\mathbf{J}$
\begin{equation}\label{eq:reform_BFIM}        \mathbf{J}_{\mathbf{V}'} = \mathbf{J} ~ \Leftrightarrow \Tr{\left(\mathbf{\tilde{G}}_{ij} \mathbf{V}' {\mathbf{V}'}^\mathsf{H}\right)} = t_{ij}, ~ 1 \leq i \leq j \leq L, 
        \end{equation}
        with $t_{ij} \triangleq \tfrac{\sigma^2}{\Upsilon}\left([\mathbf{J}]_{ij} - [\mathbf{C}]_{ij}\right)$. Further,  we have the following $K$ quadratic equations for the SINR constraints 
\begin{equation}\label{eq:reform_SINR}        \text{vSINR}^\text{IC}_{\mathbf{V}'} = \boldsymbol{\gamma}' \Leftrightarrow \frac{|\mathbf{h}_k^\mathsf{H}\mathbf{v}'_k|^2}{\gamma_k'} - \sum_{n \neq k} |\mathbf{h}_k^\mathsf{H} \mathbf{v}'_{n} |^2  = \sigma^2, \quad \forall k.
        \end{equation}
Plugging the rank reduction structure \eqref{eq:transformation_IC}-\eqref{eq:transformation_IC2} into \eqref{eq:reform_BFIM}-\eqref{eq:reform_SINR}, we obtain the following set of quadratic equations 
in $\left\{d_k\right\}$ and $ \mathbf{U}_\text{s}$
\begin{align}
            \sum_k |d_k|^2 {\mathbf{\hat{v}}_k}^\mathsf{H} \tilde{\mathbf{G}}_{ij}\mathbf{\hat{v}}_k {+} \Tr\Big(  \tilde{\mathbf{G}}_{ij}\mathbf{\hat{V}}_\text{s} \mathbf{U}_\text{s} \mathbf{U}_\text{s}^\mathsf{H}{\mathbf{\hat{V}}_\text{s}}^\mathsf{H} \Big) & = t_{ij}, \forall i, j \label{eq:BFIM2} \\
            \frac{ |d_k\mathbf{h}_k^\mathsf{H} \mathbf{\hat{v}}_k|^2}{\gamma_k'}  {-} \sum_{n \neq k} |d_n\mathbf{h}_k^\mathsf{H} \mathbf{\hat{v}}_n|^2 
             & = \sigma^2,~ \forall k. \label{eq:SINR2} 
\end{align}
In addition to~\eqref{eq:BFIM2}-\eqref{eq:SINR2}, we also require $\mathbf{U}_\text{s}$ to be a tall matrix to ensure that $\mathbf{V}'$ has fewer sensing beamformers. 

Next, we aim to prove the existence of $\{d_k\}$ and tall matrix $\mathbf{U}_\text{s}$ that solve~\eqref{eq:BFIM2}-\eqref{eq:SINR2}, provided that condition~\eqref{eq:IC_condition} holds. Subsequently, we prove that the $\mathbf{V'}$ obtained by such $\{d_k\}$ and $\mathbf{U}_\text{s}$ also satisfies the power constraint.

Define $a_k = 1 - |d_k|^2$ and $\mathbf{M}_\text{s} = \mathbf{I}_{N_\text{s}} - \mathbf{U}_\text{s} \mathbf{U}_\text{s}^\mathsf{H}$. Since $\mathbf{\hat{V}}$ also achieves $\left(\mathbf{J}, \boldsymbol{\gamma}'\right)$, 
        we can express \eqref{eq:BFIM2}-\eqref{eq:SINR2} in terms of the new variables $\{a_k\}$ and $\mathbf{M}_\text{s}$ as follows:
\begin{align}
\sum_k a_k {\mathbf{\hat{v}}_k}^\mathsf{H} \tilde{\mathbf{G}}_{ij}\mathbf{\hat{v}}_k + \Tr\Big(  \tilde{\mathbf{G}}_{ij}\mathbf{\hat{V}}_\text{s} \mathbf{M}_\text{s} {\mathbf{\hat{V}}_\text{s}}^\mathsf{H} \Big) & = 0, \forall i, j \label{eq:linear_BFIM} \\
\frac{a_k|\mathbf{h}_k^\mathsf{H} \mathbf{\hat{v}}_k|^2}{\gamma_k'} - \sum_{n \neq k}a_n |\mathbf{h}_k^\mathsf{H} \mathbf{\hat{v}}_n|^2 
             & = 0, ~\forall k. \label{eq:linear_SINR}
\end{align}
This is a system of linear equations.  In addition, we have the following conditions due to the definitions of $\{a_k\}$ and $\mathbf{M}_\text{s}$:
\begin{equation}\label{eq:extra_constraints}
            \mathbf{I}_{N_\text{s}} - \mathbf{M}_\text{s} \succcurlyeq \mathbf{0}, ~~{\mathbf{I}_{N_\text{s}} - \mathbf{M}_\text{s} ~ \text{singular}}, ~~ 1 - a_k \geq 0, ~~ \forall k. 
        \end{equation}
Note that $\mathbf{I}_{N_\text{s}} - \mathbf{M}_\text{s} = \mathbf{U}_\text{s} \mathbf{U}_\text{s}^\mathsf{H}$ must be singular since $\mathbf{U}_\text{s}$ is restricted to be a tall matrix.
        
The set of equations \eqref{eq:linear_BFIM}-\eqref{eq:linear_SINR} comprise a linear homogeneous system with $K + L(L+1)/2$ equations and $K + N_\text{s}^2$ real unknowns. There are $K$ equations from the SINR constraints and $L(L+1)/2$ equations from the BFIM constraint (due to symmetry), and there are $K$ real variables $a_1, \ldots, a_K$ and $N_\text{s}^2$ real variables from the Hermitian matrix $\mathbf{M}_\text{s}$. 
So if~\eqref{eq:IC_condition} holds, the number of unknowns exceeds the number of equations and such system must have a solution $a'_1, \ldots, a'_K, \mathbf{M}'_\text{s}$, which are not all zero. This solution can be scaled as follows
\begin{equation}~\label{eq:scaling}
\mathbf{M}_\text{s} = \frac{1}{\delta} \mathbf{M}_\text{s}', \quad a_k = \frac{a_k'}{\delta}, \quad \forall k,
\end{equation}
to additionally satisfy \eqref{eq:extra_constraints}. Here, $\delta$ is chosen to satisfy 
\begin{equation}\label{eq:delta}
|\delta| = \max\{|a_1'|, \ldots, |a_K'|, |\delta_1'|, \ldots, |\delta_{\hat{m}}'|\},
\end{equation}
so either $\delta = a_k'$ for some $k$ or $\delta=\delta_m'$ for some $m$,
where $\delta_1', \ldots, \delta_{\hat{N}_\text{s}}'$ are the eigenvalues of $\mathbf{M}'_\text{s}$. 
Note that $\delta \neq 0$ since $\{a'_k\}$ and $\{\delta_m'\}$ are not all zero.

It is straightforward to verify that $\{a_k\}$ and $\mathbf{M}_\text{s}$ defined in~\eqref{eq:scaling} must satisfy $\mathbf{I}_{N_\text{s}}- \mathbf{M}_\text{s} \succcurlyeq \mathbf{0}$ and $1 - a_k \geq 0, \forall k$. The fact that $\mathbf{I}_{N_\text{s}} - \mathbf{M}_\text{s}$ is singular is established by contradiction. Suppose that $\mathbf{I}_{N_\text{s}} - \mathbf{M}_\text{s}$ is nonsingular, then by~\eqref{eq:scaling}, we must have $1 - \tfrac{\delta_m}{\delta} > 0$ for all $m =\{1, \ldots, N_\text{s}\}$, which implies that $\delta = a_i$ for some $i$. In this case, the corresponding $d_i$ is zero and $\mathbf{v}_i'$ in~\eqref{eq:transformation_IC} is the all-zero vector. However, this cannot happen since it violates the SINR constraint, because $\gamma_k > 0, \forall k$ by assumption.

Given a solution of~\eqref{eq:linear_BFIM}-\eqref{eq:extra_constraints}, we can recover $\{d_k\}$ and $\mathbf{U}_\text{s}$ and subsequently find $\mathbf{V}'$ that attains $\left(\mathbf{J}, \boldsymbol{\gamma}'\right)$. We claim that such $\mathbf{V}'$ automatically 
has the same power as $\mathbf{\hat{V}}$. 
To see why, recall that $\mathbf{\hat{V}}$ is a solution to the problem
$\mathcal{P}_{N}^\text{IC}$, which has strong duality by assumption. The dual problem of  $\mathcal{P}_{N}^\text{IC}$ is the following
\begin{subequations}\label{prob:dual_IC}
\begin{align}
    \underset{\{\nu_{i, j}\}, \{\mu_k\}}{\mathrm{maximize}} ~~ &\sum_{i \le j} \nu_{i,j} t_{i, j} + \sigma^2 \sum_{k} \mu_k \\
    \mathrm{subject \; to} ~~& \mathbf{I}_N \succcurlyeq \sum_{i \le j} \nu_{i, j} \tilde{\mathbf{G}}_{i, j} \\
    &\mathbf{I}_N \succcurlyeq \sum_{i \le j} \nu_{i, j} \tilde{\mathbf{G}}_{i, j}  {+} \frac{\mu_k}{\gamma_k} \mathbf{h}_k \mathbf{h}_k^\mathsf{H}  {-} \sum_{n \neq k} {\mu_n} \mathbf{h}_n \mathbf{h}_n^\mathsf{H} \\
    & \mu_k \geq 0, \quad \forall k,
\end{align}
\end{subequations}
where $\{\nu_{i, j}\}_{1 \leq i \le j \leq L}$ and $\{\mu_k\}_{1\leq k \leq K}$ are the dual variables associated with the BFIM  and SINR constraints, respectively. By strong duality, it follows that there exist optimal dual variables $\{\nu^*_{i, j}\}_{1 \leq i \le j \leq L}$ and $\{\mu^*_k\}_{1\leq k \leq K}$ such that the optimality conditions as written below are satisfied:
\begin{align}
\mathbf{\hat{V}}_\text{s} &= \left(\sum_{i \le j} \nu^*_{i, j} \tilde{\mathbf{G}}_{i, j} \right)  \mathbf{\hat{V}}_\text{s} \label{eq:CS_vs} \\
\mathbf{\hat{v}}_k & = \left(\sum_{i \le j} \nu^*_{i, j} \tilde{\mathbf{G}}_{i, j}  + \frac{\mu^*_k}{\gamma_k} \mathbf{h}_k \mathbf{h}_k^\mathsf{H}  - \sum_{n \neq k} \mu^*_n \mathbf{h}_n \mathbf{h}_n^\mathsf{H}\right) \mathbf{\hat{v}}_k, \forall k \label{eq:CS_vk}
\end{align}
and that the primal optimum is equal to the dual optimum
\begin{equation}
\sum_{i \le j} \nu^*_{i,j} t_{i, j} + \sigma^2 \sum_{k} \mu^*_k = \Tr\big(\mathbf{ \hat V} \mathbf{\hat V}^\mathsf{H}\big).
\end{equation} 
Multiplying both sides of \eqref{eq:CS_vk} from the right by $|d_k|^2 \mathbf{\hat{v}}_k^\mathsf{H}$ and both sides of \eqref{eq:CS_vs} from the right by $\mathbf{\hat{V}}^\mathsf{H}_\text{s}\mathbf{U}^\mathsf{H}\mathbf{U}$, and applying a trace operation, followed by summing the resulting equations, we obtain 
\begin{align}
\Tr\left({\mathbf{V}'}^\mathsf{H} {\mathbf{V}'}\right) & = \sum_{i \le j} \nu^*_{i, j} \Tr\left( {\mathbf{V}'}^\mathsf{H}\tilde{\mathbf{G}}_{i,j} {\mathbf{V}'}  \right) 
\nonumber\\
        & \qquad +  
 \sum_k \mu^*_k \left( \frac{|\mathbf{h}_k^\mathsf{H} \mathbf{v}'_k|^2}{\gamma_k}  - \sum_{n \neq k} |\mathbf{h}_k^\mathsf{H} \mathbf{v}'_n|^2 \right) \nonumber \\
 & = \sum_{i \le j} \nu^*_{i, j} t_{i, j}  \nonumber \\
 & \qquad +  
 \sum_k \mu^*_k \left( \frac{|\mathbf{h}_k^\mathsf{H} \mathbf{v}'_k|^2}{\gamma_k'}  - \sum_{n \neq k} |\mathbf{h}_k^\mathsf{H} \mathbf{v}'_n|^2 \right) \nonumber \\
            & = \sum_{i \le j} \nu^*_{i, j} t_{i, j} +  \sigma^2 \sum_k \mu^*_k \nonumber \\
        &=\Tr\left({\mathbf{\hat{V}}}^\mathsf{H} {\mathbf{\hat{V}}}\right),
\end{align}
where we make use of \eqref{eq:reform_BFIM} and \eqref{eq:reform_SINR}. Note that in the second equality we replace $\gamma_k$ with $\gamma_k'$ since $\mu_k^*$ must be zero whenever $\gamma_k' \neq \gamma_k$ due to complementary slackness. 
To summarize, provided that the number of sensing beamformers satisfies \eqref{eq:IC_condition}, it is always possible to reduce the number of sensing beamformers while satisfying \eqref{eq:BFIM_SINR}.
        
Finally, we prove that i) $\mathbf{V}'$ is an optimal
solution of $\mathcal{P}_{N'}^\text{IC}$; and ii) $\mathcal{P}_{N'}^\text{IC}$ must also have strong duality. The former holds because the optimal value of $\mathcal{P}_{N'}^\text{IC}$
must be greater than or equal to that of $\mathcal{P}_{N}^\text{IC}$ (since $N'
 < N$), but 
$\mathbf{V'}$ achieves the the same power as $\mathbf{\hat{V}}$, which is the optimal value of $\mathcal{P}_{N}^\text{IC}$.
So it must be an optimal solution of $\mathcal{P}_{N'}^\text{IC}$. The latter holds because $\mathcal{P}_{N'}^\text{IC}$ and $\mathcal{P}_{N}^\text{IC}$ give rise to 
the same dual problem~\eqref{prob:dual_IC} and both attain the same minimum. 
Since $\mathcal{P}_{N}^\text{IC}$ has strong duality, $\mathcal{P}_{N'}^\text{IC}$ must also have strong duality.

\section{Proof of Strong Duality for $\mathcal{P}^\text{IC}_{N_\text{T} + K}$ and $\mathcal{P}^\text{NIC}_{N_\text{T} + K}$}
\label{Appen:strong_duality}

In this appendix, we establish that $\mathcal{P}^\text{IC}_{N_\text{T} + K}$ and
$\mathcal{P}^\text{NIC}_{N_\text{T} + K}$ have strong duality by showing that their SDR is tight. 
The proof follows from a standard trick used to establish strong duality for ISAC problems with a similar structure; see~\cite{LiuFCRB2022, HuaOptimal2023}, and \cite[Theorem 1]{Liu2020joint}. We include the proof here to keep this paper self-contained. 

\begin{lemma}\label{lem:P_strong_duality}
The optimization problems $\mathcal{P}^\text{IC}_{N_\text{T} + K}$ as stated in \eqref{prob:modified_problem} and 
$\mathcal{P}^\text{NIC}_{N_\text{T} + K}$  as stated in \eqref{prob:NIC_formulation} have strong duality. Further, their SDRs, obtained by defining $\mathbf{R} = \mathbf{V} \mathbf{V}^\mathsf{H} = \mathbf{V}_\text{s} \mathbf{V}_\text{s}^\mathsf{H} + \sum_k \mathbf{v}_k \mathbf{v}_k^\mathsf{H}$, and $\mathbf{R}_k = \mathbf{v}_k \mathbf{v}_k^\mathsf{H}$, and dropping the rank-one constraints on $\mathbf{R}_k$ are tight. In other words, 
\begin{subequations} \label{prob:P_SDR}
         \begin{align}
\mathcal{R}^\text{IC}: \underset{\mathbf{R}, \mathbf{R}_1, \ldots, \mathbf{R}_K}{\mathrm{minimize}}~~&\Tr{\left(\mathbf{R}\right)} \\
            {\mathrm{subject\; to}}~~&\mathbf{J}_{\mathbf{R}} = \mathbf{J}, \label{eq:BFIM_SDR} \\
            &\frac{1}{\gamma_k} \mathbf{h}_k^\mathsf{H}\mathbf{R}_k \mathbf{h}_k - \sum_{n \neq k } \mathbf{h}_k^\mathsf{H}\mathbf{R}_n \mathbf{h}_k \geq \sigma^2, \forall k \label{eq:SDR_SINR_IC}\\
            & \mathbf{R} \succcurlyeq \sum_k \mathbf{R}_k, \quad   \mathbf{R}_k \succcurlyeq \mathbf{0}, \quad \forall k,
         \end{align}
        \end{subequations}
and 
\begin{subequations} 
         \begin{align}
\mathcal{R}^\text{NIC}: \underset{\mathbf{R}, \mathbf{R}_1, \ldots, \mathbf{R}_K}{\mathrm{minimize}}~&\Tr{\left(\mathbf{R}\right)} \\
            {\mathrm{subject\; to}}~&\mathbf{J}_{\mathbf{R}} = \mathbf{J}, \\
            &\left( 1+ \frac{1}{\gamma_k}\right)\mathbf{h}_k^\mathsf{H}\mathbf{R}_k \mathbf{h}_k -\mathbf{h}_k^\mathsf{H}\mathbf{R} \mathbf{h}_k \geq \sigma^2, \forall k \label{eq:SDR_SINR_NIC} \\
            & \mathbf{R} \succcurlyeq \sum_k \mathbf{R}_k, \quad   \mathbf{R}_k \succcurlyeq \mathbf{0}, \quad \forall k,
\end{align}
\end{subequations}
have optimal solutions where $\mathbf{R}_k$'s have rank one. Here, $\mathbf{J}_\mathbf{R}$ denotes, with a slight abuse of notation, the BFIM matrix expressed in $\mathbf{R}$ (i.e., with $\mathbf{V}\mathbf{V}^\textsf{H}$ replacing $\mathbf{R}$ in~\eqref{eq:TV_elements}). 
\end{lemma}

\begin{IEEEproof}
First, consider the SDR of $\mathcal{P}^\text{IC}_{N_\text{T} + K}$, as given by~$\mathcal{R}^\text{IC}$.  To prove that this SDR is tight, we aim to prove the existence of a solution $\mathbf{\tilde R}, \mathbf{\tilde{R}}_1, \ldots, \mathbf{\tilde{R}}_K$, for which $\rank (\mathbf{\tilde{R}}_k) = 1$ for all $k$.  Let $\mathbf{\hat{R}}, \mathbf{\hat{R}}_1, \ldots, \mathbf{\hat{R}}_K$ be an arbitrary high-rank solution. We construct $\mathbf{\tilde R}, \mathbf{\tilde{R}}_1, \ldots, \mathbf{\tilde{R}}_K$ as follows:
\begin{equation}\label{eq:rank_one_solution}
    \mathbf{\tilde R} = \mathbf{\hat R}, \quad \mathbf{\tilde{R}}_k = \frac{\mathbf{\hat{R}}_k \mathbf{h}_k \mathbf{h}_k^\mathsf{H} \mathbf{\hat{R}}_k}{\mathbf{h}_k^\mathsf{H} \mathbf{\hat{R}}_k \mathbf{h}_k }, \quad \forall k. 
\end{equation}
Clearly, $\mathbf{\tilde{R}}_1, \ldots, \mathbf{\tilde{R}}_K$ as given by~\eqref{eq:rank_one_solution} are all rank-one PSD matrices. The new solution attains the same power and BFIM as $\mathbf{\hat{R}}, \mathbf{\hat{R}}_1, \ldots, \mathbf{\hat{R}}_K$ since the overall covariance is kept fixed, i.e., $ \mathbf{\tilde R} = \mathbf{\hat R}$.  Further, we note that the following holds:
\begin{align}\label{eq:rank_one_conditions2}
\mathbf{h}_k^\mathsf{H}\mathbf{\hat{R}}_k \mathbf{h}_k  = \mathbf{h}_k^\mathsf{H}\mathbf{\tilde{R}}_k \mathbf{h}_k, \quad    \mathbf{\hat{R}}_k \succcurlyeq \mathbf{\tilde{R}}_k, \quad \forall k.
\end{align}
This is proved as follows. First, the equality is immediate from~\eqref{eq:rank_one_solution}. Let $\mathbf{w} \in \mathbb{C}^{N_\text{T}}$ be an arbitrary vector. Then,
\begin{equation}
    \mathbf{w}^\mathsf{H} \left( \mathbf{\hat{R}}_k - \mathbf{\tilde{R}}_k\right) \mathbf{w} = \mathbf{w}^\mathsf{H} \mathbf{\hat{R}}_k \mathbf{w} - \frac{\left| \mathbf{h}_k^\mathsf{H} \mathbf{\hat{R}}_k \mathbf{w} \right|^2}{\mathbf{h}_k^\mathsf{H} \mathbf{\hat{R}}_k \mathbf{h}_k } \geq 0,
\end{equation}
where the last inequality follows from the Cauchy-Schwarz inequality. This shows that $\mathbf{w}^\mathsf{H} \mathbf{\hat{R}}_k\mathbf{w} \geq \mathbf{w}^\mathsf{H}   \mathbf{\tilde{R}}_k\mathbf{w}$, for all $\mathbf{w}$. Equivalently, $\mathbf{\hat R}_k \succcurlyeq \mathbf{\tilde{R}}_k$. This implies $\mathbf{R}  \succcurlyeq \sum_k \mathbf{\hat{R}}_k \succcurlyeq \sum_k \mathbf{\tilde R}_k$.

Notice that~\eqref{eq:rank_one_conditions2} implies that, for the $k$-th SINR constraint in~\eqref{eq:SDR_SINR_IC}, the useful power term  remains the same when $\mathbf{\hat R}_k$ is replaced with $\mathbf{\tilde R}_k$. But the interference powers are reduced. Hence, the SINR constraints are also satisfied for the new solution. Specifically,
\begin{align}
\frac{1}{\gamma_k} \mathbf{h}_k^\mathsf{H}\mathbf{\tilde R}_k \mathbf{h}_k - \sum_{n \neq k } \mathbf{h}_k^\mathsf{H}\mathbf{\tilde R}_n \mathbf{h}_k &= \frac{1}{\gamma_k} \mathbf{h}_k^\mathsf{H}\mathbf{\hat R}_k \mathbf{h}_k - \sum_{n \neq k } \mathbf{h}_k^\mathsf{H}\mathbf{\tilde R}_n \mathbf{h}_k \nonumber \\
&\geq \frac{1}{\gamma_k} \mathbf{h}_k^\mathsf{H}\mathbf{\hat R}_k \mathbf{h}_k - \sum_{n \neq k } \mathbf{h}_k^\mathsf{H}\mathbf{\hat R}_n \mathbf{h}_k \nonumber \\
&\geq \sigma^2.
\end{align}
Thus, $\mathbf{\tilde{R}}, \mathbf{\tilde{R}}_1, \ldots, \mathbf{\tilde{R}}_K$ is an optimal solution. Such a solution can always be decomposed as $\mathbf{\tilde R} = \mathbf{V}\mathbf{V}^\mathsf{H}$ and $\mathbf{\tilde R}_k = \mathbf{v}_k\mathbf{v}_k^\mathsf{H}$ for some $\mathbf{V} = \left[ \mathbf{v}_1, \ldots, \mathbf{v}_K, \mathbf{V}_\text{s} \right] \in \mathbb{C}^{N_\text{T} \times (N_\text{T} + K)}$. Hence, $\mathcal{R}^\text{IC}$ is tight relaxation of $\mathcal{P}^\text{IC}_{N_\text{T} + K}$. Notice that $\mathcal{R}^\text{IC}$ is a convex problem since the BFIM constraint~\eqref{eq:BFIM_SDR} is linear in $\mathbf{R}$.

The argument that $\mathcal{P}^\text{IC}_{N_\text{T} + K}$ has strong duality can be proved as follows. 
First, it can be verified that the dual of $\mathcal{P}_{N_\text{T} +K}^\text{IC}$ as written out explicitly in \eqref{prob:dual_IC} (with $N=N_\text{T}+K$) has a dual given by the SDR $\mathcal{R}^\text{IC}$ (i.e., the relaxation is the dual of the dual problem). Now, the dual problem~\eqref{prob:dual_IC} is convex and strictly feasible (so the Slater's condition is satisfied). Therefore, the dual problem~\eqref{prob:dual_IC} and $\mathcal{R}^\text{IC}$ both enjoy strong duality (they are the dual of each other). Now, since the SDR attains the same optimal value as $\mathcal{P}_{N_\text{T} + K}^\text{IC}$, we have that the primal and the dual of  $\mathcal{P}_{N_\text{T} + K}^\text{IC}$ achieve the same optimal value, i.e.,  $\mathcal{P}_{N_\text{T} + K}^\text{IC}$ has strong duality.

Next, we examine the no interference cancellation case. Note that $\mathcal{R}^\text{IC}$ and $\mathcal{R}^\text{NIC}$ only differ in the SINR constraints. In the former, the interference is only due to the sum $\sum_{n \neq k} \mathbf{R}_n$, whereas in the latter, the interference is due to the sum of $\sum_{n \neq k} \mathbf{R}_n$ and $\mathbf{R}_\text{s} = \mathbf{V}_\text{s} \mathbf{V}_\text{s}^\mathsf{H}$ (or, equivalently, the matrix $\mathbf{R} - \mathbf{R}_k$ since $\mathbf{R} = \mathbf{R}_\text{s} + \sum_n \mathbf{R}_n$). This gives rise to an SINR expression as shown in~\eqref{eq:SDR_SINR_NIC} for the no interference cancellation case.

Despite the difference, the same proof remains applicable. To prove that the relaxation $\mathcal{R}^\text{NIC}$ is tight, we let $\mathbf{\hat{R}}, \mathbf{\hat{R}}_1, \ldots, \mathbf{\hat{R}}_K$ be an arbitrary high-rank solution and show that the matrices defined in~\eqref{eq:rank_one_solution} constitute a solution with $\rank(\mathbf{\tilde{R}_k}) = 1$. Indeed, the power and BFIM remain the same, and the SINR constraints are satisfied, because
\begin{multline}
\left(1 + \frac{1}{\gamma_k} \right) \mathbf{h}_k^\mathsf{H}\mathbf{\tilde R}_k \mathbf{h}_k -\mathbf{h}_k^\mathsf{H}\mathbf{\tilde R} \mathbf{h}_k  \\
= \left(1 + \frac{1}{\gamma_k} \right) \mathbf{h}_k^\mathsf{H}\mathbf{\hat R}_k \mathbf{h}_k - \mathbf{h}_k^\mathsf{H}\mathbf{\hat R} \mathbf{h}_k \geq \sigma^2.
\end{multline}
The proof that $\mathcal{P}_{N_\text{T} +K}^\text{NIC}$ has strong duality follows the same reasoning as that of $\mathcal{P}_{N_\text{T} +K}^\text{IC}$. 
\end{IEEEproof}


\section{Proof of Lemma~\ref{lem:Ortho_Vs_NIC}}
\label{appen:Ortho_Vs_NIC}

      Since problem $\mathcal{P}_N^\text{NIC}$ is feasible by assumption, there exists at least one solution. Let $\mathbf{\hat{V}} = [\mathbf{\hat{v}}_1, \ldots,  \mathbf{\hat{v}}_K, \mathbf{\hat{V}}_\text{s}] \in \mathbb{C}^{N_\text{T} \times N}$ denote one such solution. Note that $\mathbf{\hat V}$ defines a particular covariance matrix $\mathbf{\hat R} = \mathbf{\hat V} \mathbf{ \hat V}^\mathsf{H}$. The idea of the proof is to find a new decomposition of $\mathbf{\hat R}$, i.e.,
      \begin{equation}
          \mathbf{\hat R} = \mathbf{V}' {\mathbf{V}'}^\mathsf{H}, \quad \mathbf{V}' = \left[\mathbf{v}'_1, \ldots, \mathbf{v}_K', \mathbf{V}_\text{s}' \right] \in \mathbb{C}^{N_\text{T} \times N},
      \end{equation}
      so that the projection of $\mathbf{V}_\text{s}'$ in the direction of the communication channels is zero, i.e., $ \mathbf{h}_k^\mathsf{H} \mathbf{V}_\text{s}'= \mathbf{0}$ for all $k$. Intuitively, we seek to move the components of $\mathbf{\hat V}_\text{s}$ (i.e., the initial sensing beamforming matrix) in the direction of $\mathbf{h}_1, \ldots, \mathbf{h}_K$ to the communication beamformers, while keeping the overall covariance the same.

      We construct $\mathbf{V}'$ from $\mathbf{\hat{V}}$ as follows. First, find a set of communication beamformers ${\mathbf{v}'_1, \ldots, \mathbf{v}'_K}$ by maximizing some weighted minimum power of the beaformers in the direction of $\mathbf{h}_1, \ldots, \mathbf{h}_K$,  where the weights $w_1, \ldots, w_K$ are to be determined later: 
\begin{subequations} \label{prob:max_min_pw}
\begin{align} \label{eq:max_min_obj} \underset{{\mathbf{v}_1, \ldots, \mathbf{v}_K}}{\mathrm{maximize}}  ~~&\min_k ~~ w_k |\mathbf{h}_k^\mathsf{H} \mathbf{v}_k|^2 \\
    \mathrm{subject \; to} ~~& \sum_k \mathbf{v}_k \mathbf{v}_k^\mathsf{H}  \preccurlyeq \mathbf{\hat R}.\label{eq:max_min_const_a}
\end{align}
  \end{subequations}
Let  ${\mathbf{v}'_1, \ldots, \mathbf{v}'_K}$  be a solution to the above optimization problem.
 The sensing beamformers are formed by decomposing the difference between $\mathbf{\hat R}$ and the covariance of the communication beamformers, i.e.,
 \begin{equation}\label{eq:V_s_prime}
    \mathbf{V}_\text{s}' {\mathbf{V}_\text{s}'}^\mathsf{H} = \mathbf{\hat{R}} - \sum_k \mathbf{v}_k' {\mathbf{v}_k'}^\mathsf{H}.
    \end{equation}
The optimal solution to \eqref{prob:max_min_pw} is not necessarily unique.
We would like to show that for positive weights $w_1, \ldots, w_K$, there exists one set of solutions ${\mathbf{v}'_1, \ldots, \mathbf{v}'_K}$ to \eqref{prob:max_min_pw} such that the $\mathbf{V}_\text{s}'$ given by the above is orthogonal to $\mathbf{h}_1, \ldots, \mathbf{h}_K$. This makes sense intuitively, because $\mathbf{V}_\text{s}'$ is obtained from the residual component of the communication beamformers that have maximal projection in the direction of the communication channels. A formal proof of this fact is provided later. 
    
      
    We now select $w_1. \ldots. w_K$ so that $\mathbf{V}'$ satisfies the SINR constraints as given in~\eqref{eq:SINR_const_NIC}.
    Note that the total power and the BFIM attained by $\mathbf{V}'$ are already equal to those attained by $\mathbf{\hat V}$, since the overall transmit covariance $\mathbf{\hat R}$ is the same. Therefore, as long as $\mathbf{V}'$ meets the SINR constraints, such a beamforming matrix would be an optimal solution of $\mathcal{P}_N^\text{NIC}$.
    The following choice of $w_1, \ldots, w_K$ works: 
    \begin{equation}\label{eq:w_choice}
        w_k \triangleq \left(1 + \frac{1}{\gamma_k}\right) \frac{1}{\mathbf{h}_k^\mathsf{H}\mathbf{\hat R} \mathbf{h}_k +  \sigma^2}.
    \end{equation}
      This can be verified as follows. 

First, with this set of weights, we have the following lower bound on the weighted projected power:
    \begin{subequations}
        \begin{align}
            \min_k  w_k |\mathbf{h}_k^\mathsf{H} \mathbf{v}_k'|^2 & \geq \min_k w_k |\mathbf{h}_k^\mathsf{H} \mathbf{\hat{v}}_k|^2 \\
            &= \min_k  \left(1 + \frac{1}{\gamma_k} \right) \frac{|\mathbf{h}_k^\mathsf{H}\mathbf{\hat{v}}_k|^2}{\mathbf{h}_k^\mathsf{H}\mathbf{\hat{V}}\mathbf{\hat{V}}^\mathsf{H} \mathbf{h}_k +  \sigma^2}  \label{eq:121_second_line}\\
            &= \min_k  \left(1 + \frac{1}{\gamma_k} \right) \left(1 +  \frac{1}{\text{SINR}^\text{NIC}_{k, \mathbf{\hat{V}}}} \right)^{-1} \\
            & \geq \min_k ~ \left(1 + \frac{1}{\gamma_k}\right) \left(1 +  \frac{1}{\gamma_k} \right)^{-1} = 1,
        \end{align}
        \end{subequations}
        where in \eqref{eq:121_second_line} we use~\eqref{eq:w_choice} and the fact that $\mathbf{\hat R} = \mathbf{\hat{V}}\mathbf{\hat{V}}^\mathsf{H}$. The last line follows because $\mathbf{\hat{V}}$ satisfies $\text{SINR}^\text{NIC}_{k, \mathbf{\hat{V}}} \geq \gamma_k$. 

The above inequality implies 
    \begin{subequations}
        \begin{align}
            w_k |\mathbf{h}_k^\mathsf{H} \mathbf{v}_k'|^2 \geq 1 &\Leftrightarrow \left(1 + \frac{1}{\gamma_k}\right)  \geq \frac{\mathbf{h}_k^\mathsf{H}\mathbf{\hat R} \mathbf{h}_k +  \sigma^2}{|\mathbf{h}_k^\mathsf{H}\mathbf{v}'_k|^2} \\
            &\Leftrightarrow \left(1 + \frac{1}{\gamma_k} \right)  \geq \frac{\mathbf{h}_k^\mathsf{H}\mathbf{V}'{\mathbf{V}'}^\mathsf{H} \mathbf{h}_k +  \sigma^2}{|\mathbf{h}_k^\mathsf{H}\mathbf{v}'_k|^2} \label{eq:SINR_proof_line_b}\\& \Leftrightarrow \left(1 + \frac{1}{\gamma_k}\right)  \geq 1 + \frac{1}{\text{SINR}^\text{NIC}_{k, \mathbf{V}'}} \\
            & \Leftrightarrow \text{SINR}^\text{NIC}_{k, \mathbf{V}'} \geq \gamma_k,
        \end{align}
        \end{subequations}
        where in~\eqref{eq:SINR_proof_line_b}
        we use $\mathbf{\hat R} = \mathbf{V}'{\mathbf{V}'}^\mathsf{H}$. 
        Hence, $\mathbf{V}'$ obtained by the weights in~\eqref{eq:w_choice} is an optimal solution of $\mathcal{P}_N^\text{NIC}$.

To complete the proof, it remains to prove the existence of one optimal $\mathbf{V}'_\text{s}$ that satisfies $\mathbf{h}_k^\mathsf{H}\mathbf{V}'_\text{s}= \mathbf{0}$ for all $k$. We provide a proof based on the examination of the Karush-Kuhn-Tucker (KKT) conditions of problem~\eqref{prob:max_min_pw}. First, the KKT conditions are necessary at the optimum, because \eqref{prob:max_min_pw} has strong duality. The strong duality can be shown using the same trick as in Appendix~\ref{Appen:strong_duality}, which involves proving that the SDR of~\eqref{prob:max_min_pw} is tight. In details, the following problem, obtained by defining $\mathbf{R}_k = \mathbf{v}_k \mathbf{v}_k^\mathsf{H}$ and dropping the rank-one constraints, has a rank-one solution
\begin{subequations} \label{prob:max_min_pw_SDR}
\begin{align}
\underset{{\mathbf{R}_1 \succcurlyeq \mathbf{0}, \ldots, \mathbf{R}_K \succcurlyeq \mathbf{0}, t}}{\mathrm{maximize}}  ~~& t \\
    \mathrm{subject \; to} ~~~~~& \sum_k \mathbf{R}_k  \preccurlyeq \mathbf{\hat R}, \label{eq:SDR_const_a} \\
    &  t \leq  w_k \Tr\left(\mathbf{h}_k^\mathsf{H}\mathbf{R}_k \mathbf{h}_k \right), ~~\forall k. \label{eq:SDR_const_b}
\end{align}
\end{subequations}
Let $\mathbf{\hat{R}}_1, \ldots, \mathbf{\hat{R}}_K$ be an arbitrary high-rank solution of the SDR~\eqref{prob:max_min_pw_SDR}. Then, the following rank-one matrices
\begin{align}\label{eq:rank_one_solution2}
    \mathbf{\tilde{R}}_k = \frac{\mathbf{\hat{R}}_k \mathbf{h}_k \mathbf{h}_k^\mathsf{H} \mathbf{\hat{R}}_k}{\mathbf{h}_k^\mathsf{H} \mathbf{\hat{R}}_k \mathbf{h}_k }, \quad \forall k
\end{align}
are also optimal. To prove optimality, we need to show that $\mathbf{\tilde{R}}_1, \ldots, \mathbf{\tilde{R}}_K$ satisfy~\eqref{eq:SDR_const_a}-\eqref{eq:SDR_const_b}.
The constraint~\eqref{eq:SDR_const_a} is satisfied since $\sum_k \mathbf{\tilde{R}}_k \preccurlyeq \sum_k \mathbf{\hat{R}}_k \preccurlyeq \mathbf{\hat{R}}$, which follows from the inequality in~\eqref{eq:rank_one_conditions2}.  The constraint~\eqref{eq:SDR_const_b} 
is satisfied due to the equality relation in~\eqref{eq:rank_one_conditions2}.  This shows that the SDR is tight. Strong duality follows by noting that problem~\eqref{prob:max_min_pw} and its SDR give rise to the same dual problem, which is explicitly written down below:
\begin{subequations}\label{prob:dual_pw_max_min}
    \begin{align}
    \underset{\mathbf{Z}, \mu_1, \ldots, \mu_K}{\mathrm{minimize}} ~~ &\Tr\left( \mathbf{Z} \mathbf{\hat R}  \right) \\
        \mathrm{subject \; to} ~~&\mathbf{Z} - \mu_k w_k \mathbf{h}_k \mathbf{h}_k^\mathsf{H} \succcurlyeq \mathbf{0}, ~~ \forall k, \label{eq:dual_const_b} \\
        &\mathbf{Z} \succcurlyeq \mathbf{0}, \quad \mu_k \geq 0, \quad \sum_k \mu_k = 1, \label{eq:dual_const_c}
    \end{align}
\end{subequations}
where $\mathbf{Z}$ and $\mu_k$ are the dual variables associated with~\eqref{eq:SDR_const_a} (or~\eqref{eq:max_min_const_a}) and~\eqref{eq:SDR_const_b}, respectively. The dual problem is convex with nonempty interior, which implies that strong duality holds for the SDR problem. Since the SDR is tight, problem~\eqref{prob:max_min_pw} also enjoys strong duality.

Next, we examine the optimality conditions of problem~\eqref{prob:max_min_pw} to prove the existence of some $\mathbf{V}_\text{s}'$ orthogonal to $\mathbf{h}_1, \ldots, \mathbf{h}_K$. Let $\mathbf{Z}'$ and $\mu'_1, \ldots, \mu'_K$ be the optimal dual optimal variables associated with some primal optimal solution $\mathbf{v}_1', \ldots, \mathbf{v}_K'$. 
By strong duality, the following complementary slackness condition must be satisfied at the optimal solution
\begin{equation}
    \Tr\left(\mathbf{Z}' \left(\mathbf{\hat{R}} - \sum_k \mathbf{v}_k' {\mathbf{v}_k'}^\mathsf{H} \right)\right) = 0.
\end{equation}
By noting that $\mathbf{V}_\text{s}'$ is obtained by decomposing the difference between $\mathbf{\hat R}$ and the sum of communication covariances, this condition can be equivalently expressed in the following way
\begin{equation}\label{eq:CS_cond}
 \Tr\left(\mathbf{Z}' \mathbf{V}'_\text{s} {\mathbf{V}'_\text{s}}^\mathsf{H}\right) = 0.
\end{equation}
This essentially shows that the column space of the optimal $\mathbf{Z}'$ is orthogonal to the sensing beamformers. 

Now, the communication channels and $\mathbf{Z'}$ are related by~\eqref{eq:dual_const_b}, i.e., we have $\mathbf{Z}' - \mu_k' w_k \mathbf{h}_k \mathbf{h}_k^\mathsf{H} \succcurlyeq \mathbf{0},  \forall k.$ Multiplying by $\mathbf{V}'_\text{s} {\mathbf{V}'_\text{s}}^\mathsf{H}$ and applying the trace operation, we obtain  
\begin{equation}
    \Tr\left( \mathbf{Z}'\mathbf{V}'_\text{s} {\mathbf{V}'_\text{s}}^\mathsf{H} \right) - \mu_k' w_k \Tr\left(\mathbf{h}_k \mathbf{h}_k^\mathsf{H}\mathbf{V}'_\text{s} {\mathbf{V}'_\text{s}}^\mathsf{H}\right) \geq 0, ~~~\forall k.
\end{equation}
The first term is zero by~\eqref{eq:CS_cond}, and the second term is nonnegative. Thus, the above relation holds if and only if
\begin{equation}
    \mu_k' \mathbf{h}_k^\mathsf{H} \mathbf{V}_\text{s}' {\mathbf{V}_\text{s}'}^\mathsf{H} \mathbf{h}_k = 0, \quad \forall k. \label{eq:zero_condition}
\end{equation}
Suppose for now that $\mu_1', \ldots, \mu_K'$ are all strictly positive. Then~\eqref{eq:zero_condition} implies that 
$\mathbf{h}_k^\mathsf{H}\mathbf{V}_\text{s}'= \mathbf{0}$ for all $k$. This shows that the columns of $\mathbf{V}_\text{s}'$ are orthogonal to $\mathbf{h}_1, \ldots, \mathbf{h}_K$ in this case.

The above case where $\mu_k' > 0, \forall k$ corresponds to when all of the optimal $\mathbf{v}_1', \ldots, \mathbf{v}_K'$ attain $\min_k w_k |\mathbf{h}_k^\mathsf{H} \mathbf{v}_k'|^2$ with equality. Indeed, this is readily verified by noting that $\mu_k$ is the dual variable associated with~\eqref{eq:SDR_const_b}. By the complementary slackness condition $\mu_k' \left(t -  w_k |\mathbf{h}_k^\mathsf{H} \mathbf{v}_k'|^2 \right) = 0$ (recall that the SDR always has a rank-one solution), since $\mu_k' > 0$, we must have $w_k |\mathbf{h}_k^\mathsf{H} \mathbf{v}_k'|^2 = t$, for all $k$. That is, the optimal $\mathbf{v}_1', \ldots, \mathbf{v}_K'$ all attain the same value.

The analysis becomes more involved when 
some of $\mu'_k = 0$, which occurs when the corresponding optimal $\mathbf{v}_k'$ does not necessarily attain $\min_i w_i |\mathbf{h}_i^\mathsf{H} \mathbf{v}_i'|^2$, but a potentially higher value. Typically, this means that there is some freedom in choosing such optimal communication beamformers, with each choice possibly resulting in a different $\mathbf{V}_\text{s}'$. In these cases, not all such $\mathbf{V}_\text{s}'$ are orthogonal to $\mathbf{h}_1, \ldots, \mathbf{h}_K$.

Despite the above, it is still possible to prove the existence of at least one $\mathbf{V}_\text{s}'$ satisfying $\mathbf{h}_k^\mathsf{H} \mathbf{V}_\text{s}' = \mathbf{0}, \forall k$. The way to obtain such a solution is to consider a perturbed version of the dual problem~\eqref{prob:dual_pw_max_min} with a penalty term in the objective that prevents $\mu_k$ from being zero.  For $\epsilon > 0$, define $\mathcal{D}_\epsilon$ as
\begin{subequations}\label{prob:dual_pw_max_min_perturbed}
    \begin{align}
   \mathcal{D}_\epsilon: \underset{\mathbf{Z}, \mu_1, \ldots, \mu_K}{\mathrm{minimize}} ~~ &\Tr\left( \mathbf{Z} \mathbf{\hat R}  \right) + \epsilon \sum_{k = 1}^K \frac{1}{\mu_k} \\
        \mathrm{subject \; to} ~~&\mathbf{Z} - \mu_k w_k \mathbf{h}_k \mathbf{h}_k^\mathsf{H} \succcurlyeq \mathbf{0}, ~~ \forall k,  \\
        &\mathbf{Z} \succcurlyeq \mathbf{0}, \quad \mu_k \geq 0, \quad \sum_k \mu_k = 1.
    \end{align}
\end{subequations}
 Problem $\mathcal{D}_\epsilon$ differs from the dual problem~\eqref{prob:dual_pw_max_min} in the penalty term $\frac{1}{\mu_k}$. 
Note that we could have used any penalty function that goes to infinity as $\mu_k \rightarrow 0$ from the above. 

Now, for every $\epsilon > 0$, it is readily verified that $\mathcal{D}_\epsilon$ is a strictly feasible convex problem. Thus, like the unperturbed dual problem~\eqref{prob:dual_pw_max_min}, $\mathcal{D}_\epsilon$ has strong duality for every $\epsilon > 0$. The dual of such a problem is the perturbed SDR:
\begin{subequations} \label{prob:max_min_pw_SDR_perturbed}
\begin{align}
\mathcal{R}_\epsilon: \underset{{\mathbf{R}_1 \succcurlyeq \mathbf{0}, \ldots, \mathbf{R}_K \succcurlyeq \mathbf{0}, t, \boldsymbol{\lambda}}}{\mathrm{maximize}}  & t + \sqrt{\epsilon} \sum_{k = 1}^K \sqrt{\lambda_k} \\
    \mathrm{subject \; to} ~~~& \sum_k \mathbf{R}_k  \preccurlyeq \mathbf{\hat R}, \\
    &  t \leq  w_k \Tr\left(\mathbf{R}_k \mathbf{h}_k \mathbf{h}_k^\mathsf{H}\right) - \lambda_k, \forall k \\
    & \boldsymbol{\lambda} \geq 0,
\end{align}
\end{subequations}
where $\boldsymbol{\lambda} \triangleq \left[ \lambda_1, \ldots, \lambda_K\right] \in \mathbb{R}^K$. Note that when $\epsilon = 0$, $\mathcal{R}_\epsilon$ reduces to the SDR \eqref{prob:max_min_pw_SDR}. 


Furthermore, like the SDR problem~\eqref{prob:max_min_pw_SDR}, $\mathcal{R}_\epsilon$ has a rank-one solution for the SDP variables $\mathbf{R}_1, \ldots, \mathbf{R}_K$. This is easily verified using the construction in~\eqref{eq:rank_one_solution2}.

Now, fix any $\epsilon > 0$, let $\mathbf{R}_{1, \epsilon}', \ldots, \mathbf{R}_{K, \epsilon}', t_\epsilon', \lambda_\epsilon'$ be the solution of $\mathcal{R}_\epsilon$, with the matrices $\mathbf{R}_{k, \epsilon}' = \mathbf{v'}_{k, \epsilon} \mathbf{v'}_{k, \epsilon}^\mathsf{H}$ being rank-one. Let $\mathbf{Z}_\epsilon', \mu_{1,\epsilon}', \ldots, \mu_{K,\epsilon}'$ be the corresponding optimal dual solution of $\mathcal{D}_\epsilon$. The following conditions must hold:
\begin{subequations} \label{eq:opt_cond_pertub}
\begin{align}
     \Tr\left(\mathbf{Z}_\epsilon' \left(\mathbf{\hat{R}} - \mathbf{V'}_{\text{c}, \epsilon} \mathbf{V'}_{\text{c}, \epsilon}^\mathsf{H} \right)\right) = 0,& \\
     \mathbf{Z}_\epsilon' - \mu_{k,\epsilon}' w_k \mathbf{h}_k \mathbf{h}_k^\mathsf{H} \succcurlyeq \mathbf{0},&  \quad \forall k,
     \\
     \mu_{k,\epsilon}' > 0&,
\end{align}
\end{subequations}
where $\mathbf{V}'_{\text{c},\epsilon} \triangleq \left[ \mathbf{v'}_{1, \epsilon}, \ldots, \mathbf{v'}_{K, \epsilon}\right]$ is the optimal communication beamforming matrix. In~\eqref{eq:opt_cond_pertub}, the first two conditions are optimality conditions similar to those obtained for the unperturbed case. The last condition is due to the penalty term in the objective of~\eqref{prob:dual_pw_max_min_perturbed}. Following similar arguments as before for the case of unperturbed problems, we obtain the following condition: 
\begin{equation}\label{eq:perturbed_cond}
    \mathbf{h}_k^\mathsf{H} \left(\mathbf{\hat{R}} - \mathbf{V'}_{\text{c}, \epsilon} \mathbf{V'}_{\text{c}, \epsilon}^\mathsf{H} \right) \mathbf{h}_k = 0, \quad \forall k
\end{equation}
i.e., the sensing beamformers are orthogonal to the communication channels.

Since the above relationship is true for all $\epsilon$, it must be true for $\epsilon \rightarrow 0$. Consider the sequence of communication beamformers $\{ \mathbf{V}'_{\text{c}, \epsilon}\}$ resulting from solving a sequence of \eqref{prob:max_min_pw_SDR_perturbed} as $\epsilon \rightarrow 0$. 
Since $\{ \mathbf{V}'_{\text{c}, \epsilon}\}$ lies in a compact set, it must have a limit point. Let $\mathbf{V}_\text{c}'$ be such a limit point. 
At the limit point, since $\epsilon \rightarrow 0$, $\mathbf{V}_\text{c}'$ must achieve the same optimal value as
the optimal solution of \eqref{prob:max_min_pw_SDR}, (or equivalently \eqref{prob:max_min_pw}, because the SDR is tight). 
Further, due to \eqref{eq:perturbed_cond}, we must have 
\begin{align}
	\mathbf{h}_k^\mathsf{H} \left(\mathbf{\hat{R}} - \mathbf{V}'_{\text{c}} {\mathbf{V}'_\text{c}}^\mathsf{H} \right) \mathbf{h}_k = 
    0.
\end{align}
This shows that there exists an optimal solution to \eqref{prob:max_min_pw} 
with $\mathbf{V}_\text{s}' {\mathbf{V}_\text{s}'}^\mathsf{H} = \mathbf{\hat{R}} - \mathbf{V}'_{\text{c}} {\mathbf{V}'_{\text{c}}}^\mathsf{H}$ satisfying $\mathbf{h}_k^\mathsf{H}\mathbf{V}_\text{s}' = \mathbf{0}$ for all $k$.

\section{Proof of Lemma~\ref{lem:quad_eq_sol_NIC}}
\label{appen:quad_eq_sol_NIC}

First, we prove the existence of $\mathbf{U}$ that conforms to~\eqref{eq:lin_transform_compact} and gives rise to some $\mathbf{V}'$ satisfying~\eqref{eq:BFIM_SINR_NIC}, if $N$ satisfies~\eqref{eq:main_cond_NIC}. To do this, we argue that the SINR constraints are automatically satisfied regardless of the choice of $\mathbf{U}$ in~\eqref{eq:lin_transform_compact}. 
To see why, we note that for any $i, k \in \left\{1, \ldots, K \right\}$, we have
\begin{equation}
|\mathbf{h}_i^\mathsf{H} \mathbf{v}'_k|^2 = |\mathbf{h}_i^\mathsf{H} (\mathbf{\hat{v}}_k + \mathbf{\hat{V}}_\text{s}\mathbf{u}_k)|^2 = |\mathbf{h}_i^\mathsf{H} \mathbf{\hat{v}}_k|^2,
\end{equation}
which holds for all $\mathbf{u}_k$ since $\mathbf{\hat{V}}_\text{s}$ is chosen to be orthogonal to $\mathbf{h}_1, \ldots, \mathbf{h}_K$ as given by Lemma~\ref{lem:Ortho_Vs_NIC}. Likewise, it is easy to verify that $\mathbf{h}_k^\mathsf{H} \mathbf{V}_\text{s}'{\mathbf{V}'_\text{s}}^\mathsf{H} \mathbf{h}_K = \mathbf{h}_k^\mathsf{H} \mathbf{\hat{V}}_\text{s}\mathbf{\hat{V}}_\text{s}^\mathsf{H}  \mathbf{h}_K = 0$, for all $\mathbf{U}_\text{s}$. Therefore, any $\mathbf{U}_\text{c}$ and $\mathbf{U}_\text{s}$ (or $\mathbf{U}$ in~\eqref{eq:lin_transform_compact}), the SINR for the $k$-th user remains the same, i.e.,
\begin{subequations}
\begin{align}
\text{SINR}^\text{NIC}_{k, \mathbf{V}'} &= \frac{|\mathbf{h}_k^\mathsf{H} \mathbf{v}'_k|^2}{\sum_{i \neq k} |\mathbf{h}_k^\mathsf{H} \mathbf{v}'_i|^2 + \mathbf{h}_k^\mathsf{H} \mathbf{V}'_\text{s} {\mathbf{V}'_\text{s}}^\mathsf{H} \mathbf{h}_k + \sigma^2} \\
 &=  \frac{|\mathbf{h}_k^\mathsf{H} \mathbf{\hat{v}}_k|^2}{\sum_{i \neq k} |\mathbf{h}_k^\mathsf{H} \mathbf{\hat{v}}_i|^2 + \mathbf{h}_k^\mathsf{H} \mathbf{\hat{V}}_\text{s} {\mathbf{\hat{V}}_\text{s}}^\mathsf{H} \mathbf{h}_k + \sigma^2},  \\
 &=  \text{SINR}^\text{NIC}_{k, \mathbf{\hat{V}}} = \gamma_k'.
\end{align}
\end{subequations}
Next, we focus on the BFIM. For the new $\mathbf{V}'$ to satisfy $\mathbf{J}_{\mathbf{V}'} = \mathbf{J} = \mathbf{J}_{\mathbf{\hat{V}}}$, we must have
\begin{align}
\Tr\left(\mathbf{\tilde{G}}_{ij} \mathbf{V}' {\mathbf{V}'}^\mathsf{H} \right) = \Tr\left(\mathbf{\tilde{G}}_{ij} \mathbf{\hat{V}} \mathbf{\hat{V}}^\mathsf{H} \right),
\end{align}
which is equivalent to
\begin{align}
\Tr\left(\mathbf{\tilde{G}}_{ij} \mathbf{\hat{V}} \left(\mathbf{I}_{N} - \mathbf{U} \mathbf{U}^\mathsf{H} \right) {\mathbf{\hat{V}}}^\mathsf{H} \right) = 0 \label{eq:quad_system_NIC} 
\end{align}
for all $1 \leq i \leq j \leq L$. The system~\eqref{eq:quad_system_NIC} consists of $L (L + 1)/2$ quadratic equations in $\mathbf{U}$. It can be transformed into a linear system as follows. Let $\mathbf{M} \triangleq \mathbf{I}_{N} - \mathbf{U} \mathbf{U}^\mathsf{H}$. We can  write~\eqref{eq:quad_system_NIC} as 
\begin{equation}
\Tr\left(\mathbf{\tilde{G}}_{ij} \mathbf{\hat{V}} \mathbf{M}{\mathbf{\hat{V}}}^\mathsf{H} \right) = 0, \quad 1 \leq i \leq j \leq L. \label{eq:def_M}
\end{equation} 
This is a linear system in $\mathbf{M}$. Additionally, $\mathbf{M}$ must satisfy the following 
\begin{equation}\label{eq:extra_cond}
    \mathbf{I}_{N} -  \mathbf{M} \succcurlyeq \mathbf{0}, \quad \mathbf{I}_{N} - \mathbf{M}~ \text{singular}, ~~~ \mathbf{M} = \left[\begin{matrix} 
    \mathbf{0}_{K \times K} & \mathbf{M}_{12}\\
\mathbf{M}_{12}^\mathsf{H} & \mathbf{M}_{22}\end{matrix} \right]
\end{equation}
for some arbitrary 
 $\mathbf{M}_{12}$ and $\mathbf{M}_{22}$. The first two conditions are similar to the previous case in Lemma~\ref{lem:quad_eq_sol_IC}. The last condition is needed to ensures that $\mathbf{U}$ has the form in~\eqref{eq:lin_transform_compact}. This condition ensures that $\mathbf{U} \mathbf{U}^\mathsf{H}$ has an identity matrix in the $K \times K$ top-left block, which translates to an all-zero matrix in the $K \times K$ top-left block of $\mathbf{M}$.

Given $\mathbf{M}$ that satisfies~\eqref{eq:extra_cond}, we can always find the desired $\mathbf{U}$. Indeed, since $\mathbf{I}_{N} - \mathbf{M}$ is PSD and singular, it must admit a compact decomposition $\mathbf{\tilde{U}} \mathbf{\tilde{U}}^\mathsf{H}$ for a tall matrix $\mathbf{\tilde{U}} \in \mathbb{C}^{\hat{N} \times N'}$. Since the $K \times K$ top-left block of $\mathbf{I}_{N} - \mathbf{M}$ is the identity matrix, the first $K$ rows of $\mathbf{\tilde{U}}$ must be orthonormal, i.e.,
\begin{equation}
    \mathbf{\tilde{U}} = \left[\begin{matrix}
        \mathbf{Q}_1 \\ \mathbf{Q}_2
    \end{matrix} \right], \quad \mathbf{Q}_1 \in \mathbb{C}^{K \times N'}, \quad \mathbf{Q}_1 \mathbf{Q}_1^\mathsf{H} = \mathbf{I}_K.
\end{equation}
It follows that there exists some $\mathbf{Q}_\perp \in \mathbb{C}^{(N' - K) \times N'}$ with $N' - K$ orthonormal rows such that $\mathbf{Q}_1 \mathbf{Q}_\perp^\mathsf{H} = \mathbf{0}$. We can then recover $\mathbf{U}$ as follows:
\begin{equation}
    \mathbf{U} = \mathbf{\tilde{U}} [\begin{matrix}
\mathbf{Q}_1^\mathsf{H} & \mathbf{Q}_\perp^\mathsf{H} \end{matrix}] = \left[ \begin{matrix}
    \mathbf{I}_K & \mathbf{0} \\
    \mathbf{Q}_2\mathbf{Q}_1^\mathsf{H} &  \mathbf{Q}_2\mathbf{Q}_\perp^\mathsf{H}
\end{matrix} \right]. 
\end{equation}
In particular, $\mathbf{U}_\text{c} = \mathbf{Q}_2\mathbf{Q}_1^\mathsf{H}$ and $\mathbf{U}_\text{s} = \mathbf{Q}_2\mathbf{Q}_\perp^\mathsf{H}$.

Now provided that~\eqref{eq:main_cond_NIC} holds, it is always possible to find $\mathbf{M}$ that satisfies both~\eqref{eq:def_M} and~\eqref{eq:extra_cond}. This is because \eqref{eq:def_M} is a system of $L (L + 1)/2$ linear equations in $\mathbf{M}$, where $\mathbf{M}$ has $N^2 - K^2$ real unknowns, since it is a $N \times N$ Hermitian matrix with a fixed $K \times K$ top-left block. Provided that~\eqref{eq:main_cond_NIC} holds, there are more real unknowns than equations, so we can find a nonzero solution $\mathbf{M}'$. Such solution can then be scaled appropriately 
\begin{equation}
    \mathbf{M} = \frac{1}{\delta} \mathbf{M}', \quad |\delta| = \max\{|\delta_1'|, \ldots, |\delta_{\hat{m}}'|\},
\end{equation}
to ensure that $\mathbf{I}_{N} - \mathbf{M}$ is PSD and singular. 

The rest of the proof 
is similar to that of Lemma~\ref{lem:quad_eq_sol_IC}. 
Note that $\mathbf{V}_\text{s}'$ is orthogonal to $\mathbf{h}_1, \ldots, \mathbf{h}_K$ since it is given by a linear map of $\mathbf{\hat{V}}_\text{s}$. 

\section{Computation of $\Tr\left(\mathbf{J}_\mathbf{V}^{-1} \right)$ in \eqref{eq:obj_full_ch}} 
\label{app:Jv_computation}
We first prove the following preliminary result.
Let $\mathbf{\bar J}$ be an invertible $N_\text{T} \times N_\text{T}$ complex matrix defined as $\mathbf{\bar J}  \triangleq \mathbf{B}_1 + \jmath \mathbf{B}_2$, 
where $\mathbf{B}_1 \in  \mathbb{R}^{N_\text{T} \times N_\text{T},}$ and $\mathbf{B}_2 \in  \mathbb{R}^{N_\text{T} \times N_\text{T}}$. 
Then,
\begin{equation}\label{eq:J}
    \mathbf{J} = \left[ \begin{matrix}
        \mathbf{B}_1 & \mathbf{B}_2 \\
        -\mathbf{B}_2  & \mathbf{B}_1
    \end{matrix} \right] \in \mathbb{R}^{2N_\text{T} \times 2N_\text{T}},
\end{equation}
is an invertible  $2N_\text{T} \times 2N_\text{T}$ real matrix, and its inverse is given by
\begin{equation}\label{eq:J_in}
    \mathbf{J}^{-1} = \left[ \begin{matrix}
        \mathbf{B}_3 & \mathbf{B}_4 \\
        -\mathbf{B}_4  & \mathbf{B}_3
    \end{matrix} \right] \in \mathbb{R}^{2N_\text{T} \times 2N_\text{T}},
\end{equation}
where $\mathbf{B}_3  \triangleq \Re\{ \mathbf{\bar J}^{-1} \}$, $\mathbf{B}_4  \triangleq \Im\{ \mathbf{\bar J}^{-1} \}$. 

To show that $\mathbf{J}^{-1}$ indeed has the structure given in~\eqref{eq:J_in}, we note that $\mathbf{\bar J}^{-1} = \mathbf{B}_3 + \jmath \mathbf{B}_4$ since $\mathbf{B}_3$ and $\mathbf{B}_4$ are the real and imaginary part of $\mathbf{\bar J}^{-1}$ by definition. Now, since $\mathbf{I}_{N_\text{T}} = \mathbf{\bar J}  \mathbf{\bar J}^{-1} = \left( \mathbf{B}_1 + \jmath \mathbf{B}_2 \right) \left( \mathbf{B}_3 + \jmath \mathbf{B}_4 \right)$, we must have:
\begin{subequations}\label{eq:B_relation}
\begin{align}
    \mathbf{B}_1 \mathbf{B}_3 - \mathbf{B}_2 \mathbf{B}_4 &= \mathbf{I}_{N_\text{T}} \\
    \mathbf{B}_1 \mathbf{B}_4 + \mathbf{B}_2 \mathbf{B}_3 &= \mathbf{0}_{N_\text{T} \times N_\text{T}}.
\end{align}
\end{subequations}
Now, multiplying $\mathbf{J}$ by the matrix on the right-hand side in~\eqref{eq:J_in} reveals that
\begin{subequations}
\begin{align}
    &\left[ \begin{matrix}
        \mathbf{B}_1 & \mathbf{B}_2 \\
        -\mathbf{B}_2  & \mathbf{B}_1
    \end{matrix} \right] \left[ \begin{matrix}
        \mathbf{B}_3 & \mathbf{B}_4 \\
        -\mathbf{B}_4  & \mathbf{B}_3
    \end{matrix} \right] \nonumber \\ &= \left[ \begin{matrix}
        \mathbf{B}_1 \mathbf{B}_3 - \mathbf{B}_2 \mathbf{B}_4 & \mathbf{B}_1 \mathbf{B}_4 + \mathbf{B}_2 \mathbf{B}_3 \\
        -\mathbf{B}_1 \mathbf{B}_4 - \mathbf{B}_2 \mathbf{B}_3 & \mathbf{B}_1 \mathbf{B}_3 - \mathbf{B}_2 \mathbf{B}_4
    \end{matrix} \right] \\ &= \mathbf{I}_{2N_\text{T} \times 2N_\text{T}}.
\end{align}
\end{subequations}
This shows that $\mathbf{J}^{-1}$ has the structure in~\eqref{eq:J_in}.

We now use the above result to compute $\Tr\left( \mathbf{J}_\mathbf{V}^{-1} \right)$ in~\eqref{eq:obj_full_ch}. To do this, set the above $\mathbf{J}$ equal to $\mathbf{J}_\mathbf{V}$ given in~\eqref{eq:Jv_sensing_full_channel} for a prior matrix $\mathbf{C} = \tfrac{1}{\sigma_0^2} \mathbf{I}_{N_\text{T}}$. Then,  \editrev{$\mathbf{B}_1 = \tfrac{1}{\sigma_0^2} \mathbf{I}_{N_\text{T}} + \tfrac{\Upsilon}{\sigma^2} \Re\{ \mathbf{V} \mathbf{V}^\mathsf{H} \}$,  $\mathbf{B}_2 = \tfrac{\Upsilon}{\sigma^2} \Im\{ \mathbf{V} \mathbf{V}^\mathsf{H} \}$}, and $\mathbf{\bar J} = \tfrac{1}{\sigma_0^2} \mathbf{I}_{N_\text{T}} + \tfrac{\Upsilon}{\sigma^2} \mathbf{V} \mathbf{V}^\mathsf{H}$. In addition, we have
\begin{subequations}\label{eq:B3B4}
    \begin{align}
        \mathbf{B}_3 &= \Re\{ \mathbf{\bar J}^{-1} \} =  \Re\left\{ \left( \frac{1}{\sigma_0^2} \mathbf{I}_{N_\text{T}} + \frac{\Upsilon}{\sigma^2} \mathbf{V} \mathbf{V}^\mathsf{H}\right)^{-1} \right\}, \\
        \mathbf{B}_4 &= \Im\{ \mathbf{\bar J}^{-1} \} = \Im\left\{ \left( \frac{1}{\sigma_0^2} \mathbf{I}_{N_\text{T}} + \frac{\Upsilon}{\sigma^2} \mathbf{V} \mathbf{V}^\mathsf{H}\right)^{-1} \right\},
    \end{align}
\end{subequations}
and $\mathbf{J}_\mathbf{V}^{-1}$ is given in~\eqref{eq:J_in}. Based on this, it is easy to verify that:
\begin{subequations}
\begin{align}
    \Tr\left( \mathbf{J}_\mathbf{V}^{-1} \right) &= \Tr\left( \left[ \begin{matrix}
        \mathbf{B}_3 & \mathbf{B}_4 \\
        -\mathbf{B}_4  & \mathbf{B}_3
    \end{matrix} \right] \right) \\
    &= 2 \Tr \left( \Re\left\{ \left( \frac{1}{\sigma_0^2} \mathbf{I}_{N_\text{T}} + \frac{\Upsilon}{\sigma^2} \mathbf{V} \mathbf{V}^\mathsf{H}\right)^{-1} \right\} \right) \\
    &= 2 \Tr \left( \left( \frac{1}{\sigma_0^2} \mathbf{I}_{N_\text{T}} + \frac{\Upsilon}{\sigma^2} \mathbf{V} \mathbf{V}^\mathsf{H}\right)^{-1} \right),
\end{align}
\end{subequations}
where the second line follows from~\eqref{eq:B3B4} and the third line uses the fact that $\left( \tfrac{1}{\sigma_0^2} \mathbf{I}_{N_\text{T}} + \tfrac{\Upsilon}{\sigma^2} \mathbf{V} \mathbf{V}^\mathsf{H}\right)^{-1}$ is a Hermitian matrix.

\ifCLASSOPTIONcaptionsoff
  \newpage
\fi




\bibliographystyle{IEEEtran}
\bibliography{IEEEabrv,ref}

\begin{thebibliography}{10}
\providecommand{\url}[1]{#1}
\csname url@samestyle\endcsname
\providecommand{\newblock}{\relax}
\providecommand{\bibinfo}[2]{#2}
\providecommand{\BIBentrySTDinterwordspacing}{\spaceskip=0pt\relax}
\providecommand{\BIBentryALTinterwordstretchfactor}{4}
\providecommand{\BIBentryALTinterwordspacing}{\spaceskip=\fontdimen2\font plus
\BIBentryALTinterwordstretchfactor\fontdimen3\font minus
  \fontdimen4\font\relax}
\providecommand{\BIBforeignlanguage}[2]{{%
\expandafter\ifx\csname l@#1\endcsname\relax
\typeout{** WARNING: IEEEtran.bst: No hyphenation pattern has been}%
\typeout{** loaded for the language `#1'. Using the pattern for}%
\typeout{** the default language instead.}%
\else
\language=\csname l@#1\endcsname
\fi
#2}}
\providecommand{\BIBdecl}{\relax}
\BIBdecl

\bibitem{attiahbounds2024}
K.~M. Attiah and W.~Yu, ``Bounds on the minimum number of beamformers for
  integrated sensing and communications,'' in \emph{Proc. Asilomar Conf.
  Signals, Sys. Comput.}, Oct. 2024.

\bibitem{liutcom202}
F.~Liu, C.~Masouros, A.~P. Petropulu, H.~Griffiths, and L.~Hanzo, ``Joint radar
  and communication design: Applications, state-of-the-art, and the road
  ahead,'' \emph{IEEE Trans. Commun.}, vol.~68, pp. 3834--3862, Feb. 2020.

\bibitem{liujsac2022}
F.~Liu, Y.~Cui, C.~Masouros, J.~Xu, T.~X. Han, Y.~C. Eldar, and S.~Buzzi,
  ``Integrated sensing and communications: Toward dual-functional wireless
  networks for {6G} and beyond,'' \emph{IEEE J. Sel. Areas Commun.}, vol.~40,
  no.~6, pp. 1728--1767, June 2022.

\bibitem{Liu2020joint}
X.~Liu, T.~Huang, N.~Shlezinger, Y.~Liu, J.~Zhou, and Y.~C. Eldar, ``Joint
  transmit beamforming for multiuser {MIMO} communications and {MIMO} radar,''
  \emph{{IEEE} Trans. Signal Process.}, vol.~68, pp. 3929--3944, June 2020.

\bibitem{LiuFCRB2022}
F.~Liu, Y.-F. Liu, A.~Li, C.~Masouros, and Y.~C. Eldar, ``{Cramér-Rao} bound
  optimization for joint radar-communication beamforming,'' \emph{{IEEE} Trans.
  Signal Process.}, vol.~70, pp. 240--253, Dec. 2022.

\bibitem{yang}
Y.~Yang and R.~S. Blum, ``{MIMO} radar waveform design based on mutual
  information and minimum mean-square error estimation,'' \emph{{IEEE} Trans.
  Aerosp. Electron. Syst.}, vol.~43, no.~1, pp. 330--343, Jan. 2007.

\bibitem{Li2008range}
J.~Li, L.~Xu, P.~Stoica, K.~W. Forsythe, and D.~W. Bliss, ``Range compression
  and waveform optimization for {MIMO} radar: A {C}ramér–{R}ao bound based
  study,'' \emph{{IEEE} Trans. Signal Process.}, vol.~56, no.~1, pp. 218--232,
  Jan. 2008.

\bibitem{chanisit2024}
C.~Xu and S.~Zhang, ``Integrated sensing and communication exploiting prior
  information: How many sensing beams are needed?'' in \emph{IEEE Int. Symp.
  Inf. Theory (ISIT)}, July 2024.

\bibitem{mateen2023}
M.~Ashraf, B.~Tan, D.~Moltchanov, J.~S. Thompson, and M.~Valkama, ``Joint
  optimization of radar and communications performance in {6G} cellular
  systems,'' \emph{IEEE Trans. Green Commun. Netw.}, vol.~7, no.~1, pp.
  522--536, Jan. 2023.

\bibitem{wen2023}
C.~Wen, Y.~Huang, and T.~N. Davidson, ``Efficient transceiver design for {MIMO}
  dual-function radar-communication systems,'' \emph{{IEEE} Trans. Signal
  Process.}, vol.~71, pp. 1786--1801, May 2023.

\bibitem{soticaprobing2007}
P.~Stoica, J.~Li, and Y.~Xie, ``On probing signal design for {MIMO} radar,''
  \emph{{IEEE} Trans. Signal Process.}, vol.~55, no.~8, pp. 4151--4161, Aug.
  2007.

\bibitem{HuaOptimal2023}
H.~Hua, J.~Xu, and T.~X. Han, ``Optimal transmit beamforming for integrated
  sensing and communication,'' \emph{{IEEE} Trans. Veh. Technol.}, vol.~72,
  no.~8, pp. 10\,588--10\,603, Mar. 2023.

\bibitem{attiahactive2023}
K.~M. Attiah and W.~Yu, ``Active beamforming for integrated sensing and
  communication,'' in \emph{IEEE Int. Conf. Commun. (ICC) Workshops}, Rome,
  Italy, May 2023.

\bibitem{salman}
M.~B. Salman, O.~T. Demir, and E.~Bj\"{o}rnson, ``When are sensing symbols
  required for {ISAC?}'' \emph{{IEEE} Trans. Veh. Technol.}, vol.~73, no.~10,
  pp. 15\,709--15\,714, 2024.

\bibitem{attiah2024ULDL}
K.~M. Attiah and W.~Yu, ``Beamforming design for integrated sensing and
  communications using uplink-downlink duality,'' in \emph{IEEE Int. Symp. Inf.
  Theory (ISIT)}, 2024, pp. 2808--2813.

\bibitem{LiuF2022conf}
F.~Liu, Y.-F. Liu, C.~Masouros, A.~Li, and Y.~C. Eldar, ``A joint
  radar-communication precoding design based on {Cramér-Rao} bound
  optimization,'' in \emph{IEEE Radar Conf.}, New York, USA, May 2022.

\bibitem{hua2025_twc}
H.~Hua, J.~Xu, and R.~Zhang, ``Near-field integrated sensing and communication
  with extremely large-scale antenna array,'' \emph{{IEEE} Trans. Wireless
  Commun.}, vol.~24, no.~12, pp. 9962--9977, June 2025.

\bibitem{yao2025optimal}
J.~Yao and S.~Zhang, ``Optimal beamforming for multi-target multi-user {ISAC}
  exploiting prior information: How many sensing beams are needed?''
  \emph{arXiv preprint arXiv:2503.03560}, 2025.

\bibitem{huangrank2010}
Y.~Huang and D.~P. Palomar, ``Rank-constrained separable semidefinite
  programming with applications to optimal beamforming,'' \emph{{IEEE} Trans.
  Signal Process.}, vol.~58, no.~2, pp. 664--678, Feb. 2010.

\bibitem{pataki1998rank}
G.~Pataki, ``On the rank of extreme matrices in semidefinite programs and the
  multiplicity of optimal eigenvalues,'' \emph{Mathematics of operations
  research}, vol.~23, no.~2, pp. 339--358, May 1998.

\bibitem{Sabhinband2014}
A.~Sabharwal, P.~Schniter, D.~Guo, D.~W. Bliss, S.~Rangarajan, and R.~Wichman,
  ``In-band full-duplex wireless: Challenges and opportunities,'' \emph{{IEEE}
  J. Sel. Areas Commun.}, vol.~32, no.~9, pp. 1637--1652, June 2014.

\bibitem{Eun2018}
H.~Kim and K.~H. Kim, ``Random phase code for automotive {MIMO} radars using
  combined frequency shift keying-linear {FMCW} waveform,'' \emph{IET Radar,
  Sonar, Navigation}, vol.~12, pp. 1090--1095, July 2018.

\bibitem{stoica2005spectral}
P.~Stoica and R.~L. Moses, \emph{Spectral analysis of signals}.\hskip 1em plus
  0.5em minus 0.4em\relax Prentice-Hall, 2005.

\bibitem{Liu2023nearfield}
Y.~Liu, Z.~Wang, J.~Xu, C.~Ouyang, X.~Mu, and R.~Schober, ``Near-field
  communications: A tutorial review,'' \emph{IEEE Open J. Commun. Soc.},
  vol.~4, Aug. 2023.

\bibitem{li2008mimobook}
J.~Li and P.~Stoica, \emph{{MIMO} radar signal processing}.\hskip 1em plus
  0.5em minus 0.4em\relax John Wiley \& Sons, 2008.

\bibitem{Chiriyath2015}
A.~R. Chiriyath and D.~W. Bliss, ``Effect of clutter on joint
  radar-communications system performance inner bounds,'' in \emph{Proc.
  Asilomar Conf. Signals Syst. Comput.}, Nov. 2015, pp. 1379--1383.

\bibitem{Hua_twc_2024}
H.~Hua, T.~X. Han, and J.~Xu, ``{MIMO} integrated sensing and communication:
  {CRB}-rate tradeoff,'' \emph{{IEEE} Trans. Wireless Commun.}, vol.~23, no.~4,
  pp. 2839--2854, Aug. 2024.

\bibitem{van2004detection}
H.~L. Van~Trees, \emph{Detection, Estimation, and Modulation Theory, Part
  I}.\hskip 1em plus 0.5em minus 0.4em\relax John Wiley \& Sons, 1968.

\bibitem{Xiong2023}
Y.~Xiong, F.~Liu, Y.~Cui, W.~Yuan, T.~X. Han, and G.~Caire, ``On the
  fundamental tradeoff of integrated sensing and communications under gaussian
  channels,'' \emph{{IEEE} Trans. Inf. Theory}, vol.~69, no.~9, pp. 5723--5751,
  June 2023.

\bibitem{miller1978}
R.~Miller and C.~Chang, ``A modified {C}ramér-{R}ao bound and its
  applications,'' \emph{{IEEE} Trans. Inf. Theory}, vol.~24, no.~3, pp.
  398--400, 1978.

\bibitem{Fried20212on}
B.~Friedlander, ``On transmit beamforming for {MIMO} radar,'' \emph{{IEEE}
  Trans. Aerosp. Electron. Syst.}, vol.~48, no.~4, pp. 3376--3388, Oct. 2012.

\bibitem{huleihel2013tsp}
W.~Huleihel, J.~Tabrikian, and R.~Shavit, ``Optimal adaptive waveform design
  for cognitive {MIMO} radar,'' \emph{{IEEE} Trans. Signal Process.}, vol.~61,
  no.~20, pp. 5075--5089, Oct. 2013.

\bibitem{bekkerTarget2006}
I.~Bekkerman and J.~Tabrikian, ``Target detection and localization using {MIMO}
  radars and sonars,'' \emph{{IEEE} Trans. Signal Process.}, vol.~54, no.~10,
  pp. 3873--3883, Sept. 2006.

\bibitem{active_sensing_1}
F.~Sohrabi, Z.~Chen, and W.~Yu, ``Deep active learning approach to adaptive
  beamforming for mmwave initial alignment,'' \emph{{IEEE} J. Sel. Areas
  Commun.}, vol.~39, no.~8, pp. 2347--2360, 2021.

\bibitem{active_sensing_2}
F.~Sohrabi, T.~Jiang, W.~Cui, and W.~Yu, ``Active sensing for communications by
  learning,'' \emph{{IEEE} J. Sel. Areas Commun.}, vol.~40, no.~6, pp.
  1780--1794, 2022.

\bibitem{xiewc2022}
L.~Xie, P.~Wang, S.~Song, and K.~B. Letaief, ``Perceptive mobile network with
  distributed target monitoring terminals: Leaking communication energy for
  sensing,'' \emph{{IEEE} Trans. Wireless Commun.}, vol.~21, no.~12, pp.
  10\,193--10\,207, Dec. 2022.

\bibitem{kay1993fundamentals}
S.~M. Kay, \emph{Fundamentals of Statistical Signal Processing: Estimation
  Theory}.\hskip 1em plus 0.5em minus 0.4em\relax Prentice Hall, 1993.

\end{thebibliography}
%



%





\end{document}